\documentclass[12pt,preprint]{aastex}

\usepackage{graphicx}
\usepackage{epstopdf}
\usepackage[multiple]{footmisc}
\usepackage{pdflscape}

\pdfoutput=1

\newcommand{\kms}{km~s$^{-1}$}
\newcommand{\Msun}{$M_{\sun}$}

\newcommand{\footHST}{Based on observations made with the NASA/ESA Hubble Space Telescope, obtained at the Space Telescope Science Institute, which is operated by the Association of Universities for Research in Astronomy, Inc., under NASA contract NAS 5-26555. These observations are associated with program \# 8zt.}
\newcommand{\footMag}{This paper includes data gathered with the 6.5 meter Magellan Telescopes located at Las Campanas Observatory, Chile.}
\newcommand{\footSOAR}{Based on observations obtained at the Southern Astrophysical Research (SOAR) telescope, which is a joint project of the Minist\'{e}rio da Ci\^{e}ncia, Tecnologia, e Inova\c{c}\~{a}o (MCTI) da Rep\'{u}blica Federativa do Brasil, the U.S. National Optical Astronomy Observatory (NOAO), the University of North Carolina at Chapel Hill (UNC), and Michigan State University (MSU).}

\slugcomment{Accepted on 2012-02-02 to appear in \aj.}

\shorttitle{Distance to Westerlund 2}
\shortauthors{Vargas \'{A}lvarez \& Kobulnicky}

\begin{document}

\title{THE DISTANCE TO THE MASSIVE GALACTIC CLUSTER WESTERLUND 2 FROM A SPECTROSCOPIC AND \emph{HST}  PHOTOMETRIC STUDY\footnote{\footHST}\footnote{\footMag}\footnote{\footSOAR}}

\author{Carlos A. Vargas \'{A}lvarez, \& Henry A. Kobulnicky}
\email{cvargasa@uwyo.edu, chipk@uwyo.edu}
\affil{Department of Physics \& Astronomy\\University of Wyoming\\ Dept. 3905\\Laramie, WY 82071}

\author{David R. Bradley, Sheila J. Kannappan, \& Mark A. Norris}
\email{davidbradley512@gmail.com, sheila@physics.unc.edu, manorris@physics.unc.edu}
\affil{Department of Physics and Astronomy\\University of North Carolina\\ Chapel Hill, CB 3255, Phillips Hall\\Chapel Hill, NC 27599-3255}

\author{Richard J. Cool}
\email{rcool@obs.carnegiescience.edu}
\affil{The Observatories of the Carnegie Institution of Washington\\813 Santa Barbara Street\\Pasadena, CA 91101\\Carnegie-Princeton Fellow}

\and 

\author{Brendan P. Miller}
\email{mbrendan@umich.edu}
\affil{Department of Astronomy\\University of Michigan\\745 Dennison Building, 500 Church St.\\Ann Arbor, MI 48109}


\begin{abstract}
We present a spectroscopic and photometric determination of the distance to the young Galactic open cluster Westerlund 2 using \emph{WFPC2} imaging from the \emph{Hubble Space Telescope} and ground-based optical spectroscopy.  \emph{HST} imaging in the F336W, F439W, F555W, and F814W filters resolved many sources previously undetected in ground-based observations and yielded photometry for 1136 stars.  We identified fifteen new O-type stars, along with two probable binary systems, including MSP~188 (O3 + O5.5).  We fit reddened SEDs based on the Padova isochrones to the photometric data to determine individual reddening parameters $R_{V}$ and $A_{V}$ for O-type stars in Wd2.  We find average values $\langle R_{V} \rangle = 3.77 \pm 0.09$ and $\langle A_{V} \rangle = 6.51 \pm 0.38$ mag, which result in a smaller distance than most other spectroscopic and photometric studies.  After a statistical distance correction accounting for close unresolved binaries (factor of 1.08), our spectroscopic and photometric data on 29 O-type stars yield that Westerlund 2 has a distance $\langle d \rangle = 4.16 \pm 0.07$ (random) $+0.26$ (systematic) kpc.  The cluster's age remains poorly constrained, with an upper limit of 3 Myr.  Finally, we report evidence of a faint mid-IR PAH ring surrounding the well-known binary candidate MSP~18, which appears to lie at the center of a secondary stellar grouping within Westerlund 2.
\end{abstract}



\section{Introduction}

\begin{figure}
\epsscale{0.8}
\plotone{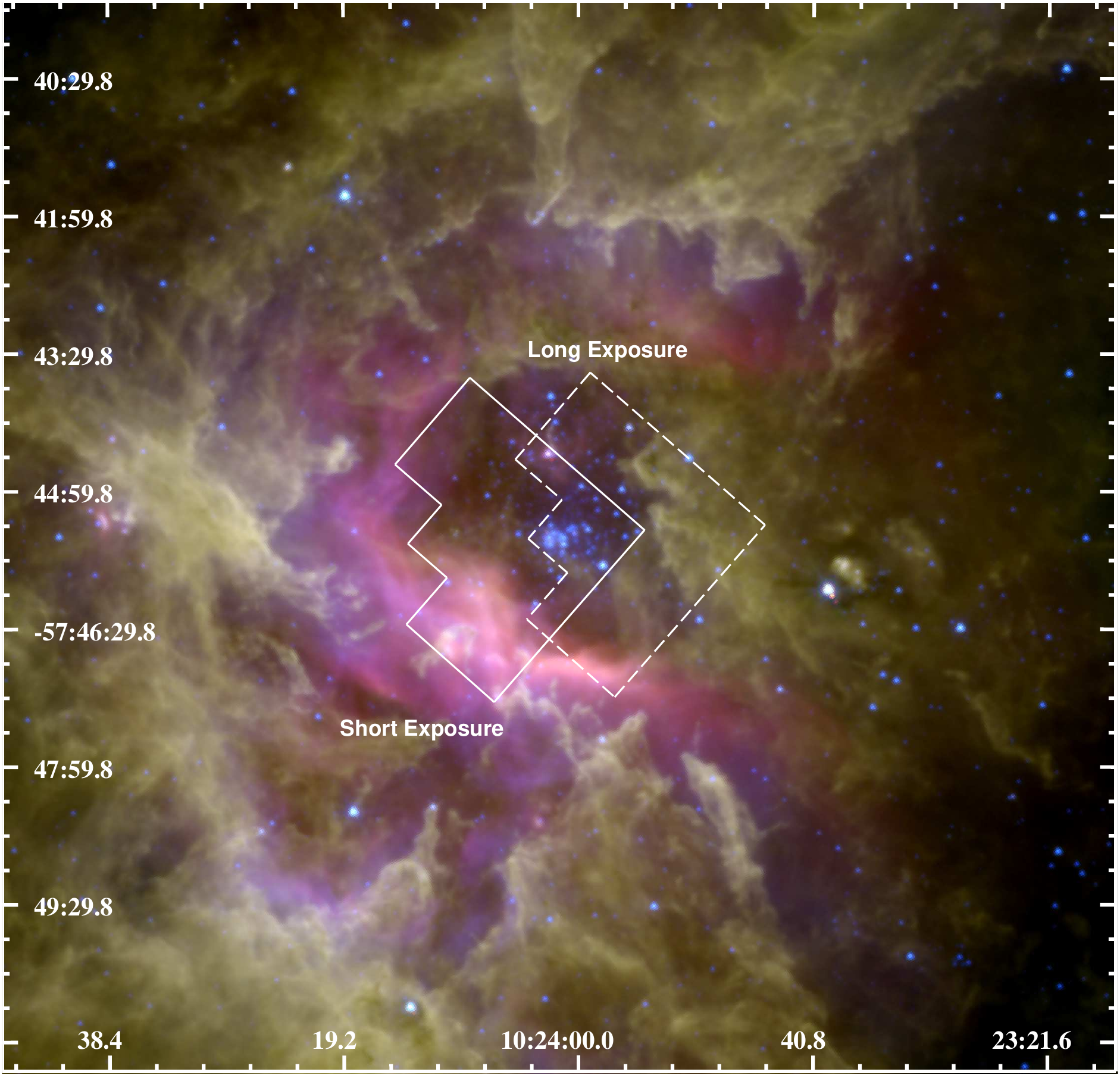}
\caption{\emph{Spitzer} three-color image of RCW49 and the Wd2 cluster with 4.5 $\mu m$ in blue, 5.8 $\mu m$ in green, and 8.0 $\mu m$ in red.  The dashed and solid outlines depict the fields covered by the ``long'' (Position 1) and ``short'' (Position 2) \emph{HST WFPC2} pointing, respectively.\label{glimpsergb}}
\end{figure}

The open cluster Westerlund 2 \citep[hereafter Wd2,][]{west}, at 1 -- 3 Myr, is one of the youngest massive stellar clusters known, similar in many respects to better-known clusters such as the Arches \citep{figer05}, Quintuplet \citep{figer99}, and NGC~3603 \citep{sto04, sto06}.   Located in the Carina arm at $(\alpha,~\delta) = (10^h~24^m~01\fs1, ~-57\degr~45\arcmin~32\arcsec)$, $(l,~b) = (284\fdg3,-0\fdg3)$, it powers the surrounding giant \ion{H}{2} region RCW~49 \citep{rodger}.  Figure~\ref{glimpsergb} shows a three-color image of Wd2 and RCW~49 as seen in the {\it Spitzer Space Telescope} mid-IR \emph{Galactic Legacy MidPlane Survey Extraordinaire} mosaic images \citep[GLIMPSE,][]{benjamin}.  Blue represents the \emph{Infrared Array Camera} \citep[IRAC,][]{fazio} 4.5 $\mu m$ band which highlights the reddened ($A_{V} \simeq6$) stellar cluster, while green (5.8 $\mu m$) and red (8.0 $\mu m$) highlight the photodissociation region (PDR) where PAHs excited by soft UV photons re-radiate stellar energy at mid-IR wavelengths. The seminal imaging study on Wd2 by \citet{moffat} (hereafter MSP91) suggested that the cluster may contain more than 80 O-type stars.  \citet{churchwell} and  \citet{whitney} used \emph{Spitzer} mid-IR images to discover a complex network of dust pillars, filaments, and young stellar objects (YSOs) within RCW~49, suggesting that star formation may be ongoing, or even triggered in the surrounding clouds by winds and ionizing radiation from the central cluster.  

\citet{dame} and \citet{furu} used millimeter-wave CO spectroscopy to conclude that the mass of the molecular cloud associated with Wd2 is $7.5 \times 10^{5}$ \Msun~and $1.7 \pm 0.8 \times 10^{5}$ \Msun, respectively, sufficient to produce the massive stellar cluster Wd2.  The only study that has tried to estimate Wd2's stellar mass is \citet{asc}.  They found that the cluster's mass function is consistent with a \citet{salpeter} power law with an index of $-1.20 \pm 0.16$.  They estimated a total stellar mass of 7000 \Msun, assuming a distance of 2.8 kpc, but the mass may be larger if greater distances are adopted.  If this is so, then Wd2 is among the most massive clusters in the Galaxy,  similar to the Quintuplet cluster \citep[$6.3 \times 10^{3}$ \Msun,][]{figer99}, Arches \citep[$2 \times 10^{4}$ \Msun,][]{figer05}, or NGC~3603 \citep[$7 \times 10^{3}$ \Msun,][]{sto04, sto06}.

Wd2 and RCW~49 have attracted considerable attention because of ongoing star formation \citep{whitney, churchwell}, and the possible association with two very-high-energy (VHE) $\gamma$-ray source HESS J1023--575 and HESS J1023--5746, detected in the \emph{High Energy Stereoscopic System} \citep[HESS,][]{aha, Reimer, hess}, whose positions are coincident with the pulsars PSR~J1023--5746, and PSR~J1028--5819, respectively \citep{saz, ack, abdo}.  This discovery has suggested possible new mechanisms for VHE emissions, such as $\gamma$-ray production in the colliding wind zone of WR 20a, collective effects of stellar winds from the Wd2 cluster, shocks and magneto-hydrodynamic (MHD) turbulent motion inside the hot \ion{H}{2} bubble, or shocks driven by supernova explosions into the interstellar medium (ISM). This cluster is rich in X-ray sources, the most luminous ones being the very massive eclipsing binary Wolf-Rayet stars WR~20a and WR~20b \citep{tsu, naze}.  Many of the brightest X-ray detections correspond to O-type stars, implying the possibility of many close binaries in this cluster \citep{naze}.

Since the discovery of Wd2 there has been much controversy over the distance to this cluster.  MSP91 used \emph{UBV} charge-coupled device photometry and spectroscopy to obtain a distance of $7.9^{+1.2}_{-1.0}$ kpc. \citet{piatti} obtained spatially integrated optical spectra of the cluster and reanalyzed a subset of cluster O stars from MSP91 to infer a revised distance of $d = 5.7 \pm 0.3$ kpc.  \citet{carraro} analyzed new \emph{UBVRI} photometry for the cluster, finding $d = 6.4 \pm 0.4$ kpc and an age $\leqslant 2$ Myr, based on isochrone fitting.  Using \emph{GLIMPSE} photometry and spectra for a single O4 star, MSP~18, \citet{brian} found a distance of $d = 3.23^{+0.54}_{-0.53}$ kpc.  Analyzing optical spectra, light curves, and published photometry of the massive eclipsing binary system WR~20a (WN6ha+WnNha), along with stellar atmosphere models, \citet{rauw05} concluded that this system is a member of Wd2, giving it $d = 7.9 \pm 0.6$ kpc.  The apparent association between Wd2 and nearby molecular and atomic clouds led \citet{dame} to derive a kinematic distance of $d = 6.0 \pm 1.0$ kpc.  \citet{tsu} used \emph{Chandra X-Ray Observatory} and near-IR imaging to identify a population of $\sim$2 \Msun~T-Tauri stars that serve as X-ray ``standard candles''; they obtained a loose distance constraint of $d = $2 -- 5~kpc.  Obtaining new photometry and new spectral classification of 12 O-type stars from MSP91 and the binary WR~20a, \citet{rauw} (hereafter R07) found $d = 8.0 \pm 1.4$ kpc.  Comparing deep near-IR photometry in the $J$, $H$, and $K_{s}$ bands with main-sequence and pre-main-sequence isochrones, \citet{asc} found $d\simeq2.8$ kpc and an age of $2.0\pm0.3$ Myr.  In one of the most recent works, \citet{furu} used CO($J = 2 - 1$) observations to infer a kinematic distance of $5.4^{+1.1}_{-1.4}$ kpc.  In summary, Wd2 has been placed at distances between $\sim$2 and $\sim$8 kpc.

In view of this range of distances for Wd2, we have undertaken a program of optical photometry with the \emph{Hubble Space Telescope} (\emph{HST}) and spectral classification of additional MSP91 sources with the \emph{Magellan} and the \emph{Southern Astrophysical Research} (\emph{SOAR}) telescopes in an attempt to resolve this discrepancy.  In \S \ref{data-red} we describe the reduction of the \emph{HST WFPC2} imaging, photometry comparison with previous works, astrometry, and long-slit spectral data.  In \S \ref{data-ana} we analyze the color-magnitude diagrams (CMD), describe our spectral classification method, and individually comment on each observed source.  In \S \ref{results} we analyze the  \emph{HST} photometry in conjunction with model isochrones and stellar spectroscopy to infer the reddening, distance, and age of the cluster.


\section{WFPC2 Imaging, Photometry, \& Optical Spectroscopy}\label{data-red}

\subsection{Observations}

\subsubsection{\emph{HST}}

The cluster Wd2 was observed using the \emph{Wide-Field Planetary Camera 2} (\emph{WFPC2}) on the \emph{HST}.  The observations were performed during Cycle 13, proposal ID 10276, PI H. Kobulnicky, and \emph{HST} program \emph{8zt}.  A two-orbit \emph{HST} imaging program was performed on 2005 May 19 \& 20 using the standard three-point line dither pattern with $1/3$ pixel spacing. On the first orbit the cluster was centered on the PC1 chip, and a series of three exposures, dubbed the ``long exposures", were obtained in each of the filters F336W (160 sec), F439W (40 sec), F555W (8 sec), and F814W (20 sec).  On the second orbit the cluster was centered on the WF3 chip, and a series of three exposures, dubbed the ``short exposures", were obtained in the same four filters with exposure times of 50 sec, 12 sec, 2 sec, and 2 sec, respectively.   Table~\ref{obs-pro} summarizes the observations. The solid and dashed polygons in Figure~\ref{glimpsergb} depict the locations of the ``short'' and ``long'' exposures, respectively.  The exposure times achieve a $5\sigma$ detection at $V = 20$ on the ``short'' exposure images and at $V = 21.5$ on the ``long'' exposure images.  The \emph{UBV} magnitudes of MSP91 were used to derive the exposure times to avoid saturation and bleeding of bright stars, but at the same time go as deep as possible down the mass function (MF) of the cluster.

\subsubsection{Optical Spectroscopy}

Longslit spectra were obtained for 10 MSP91 sources with the 6.5 m \emph{Magellan I Walter Baade Telescope} and the 4.1 m \emph{SOAR} telescope, using the \emph{Inamori-Magellan Areal Camera \& Spectrograph} (IMACS) and the \emph{Goodman High Throughput Spectrograph} (GHTS), respectively. Owing to the crowding in the Wd2 core, other sources were caught serendipitously on the slit and augmented the number of observed sources to 34.

Spectra for six stars (\#528, \#640, \#714, \#738, \#857, and \#906 in our numeration system, introduced in \S~{2.2.1}) were obtained on 2011 April 6/7 with the \emph{Magellan Telescope}.  Each source was exposed for 300 sec along with a 5 sec exposure of NeHeAr calibration lamps with a $0\farcs9$ wide slit and a grating of 300~l~mm$^{-1}$ which yields a reciprocal dispersion of 1.34~\AA~pixel$^{-1}$.  The typical seeing was 1$\arcsec$ FWHM.  The observations used 1$\times$1 pixel readout modes yielding a scale of $0\farcs2$ pixel$^{-1}$.  The \emph{Magellan} observations provided a wavelength coverage 4640~\AA~$\le \lambda \le$ 10,113~\AA.  Ultimately, only the spectral range from 4640~\AA~-- 6581~\AA~had a sufficient signal-to-noise ratio (S/N) to be usable.

The sources \#137, \#549, \#1004, and \#1031 were observed on 2011 June 9 \& 10 with \emph{SOAR}.  Typical seeing was $1\farcs5$ FWHM at \emph{SOAR} on June 9/10, and $2\farcs2$ FWHM on June 10/11, as measured from the cross-dispersion width of the stellar profile.  The charge-coupled device readout mode employed 2$\times$8 pixel binning yielding a cross-dispersion scale of $1\farcs2$~pixel$^{-1}$. Hence, the spectra are undersampled in the cross-dispersion direction.   Three exposures of 60 sec each were obtained on each source and calibrated with 15 sec FeAr lamps.  The $1\farcs68$ wide slit and the KOSI 600~l~mm$^{-1}$ grating yielded a reciprocal dispersion of 1.3~\AA~pixel$^{-1}$.  The wavelength coverage includes the range 4600~\AA~$\le \lambda \le$ 7300~\AA.  

We obtained spectra of stars \#714 and \#738 with the \emph{Magellan Inamori Kyocera Echelle} (MIKE) spectrograph on the \emph{Magellan II Clay} telescope on 2011 November 14 using a $1\arcsec \times 5\arcsec$ slit in $\sim$1$\farcs2$ seeing.  Exposure times were $2 \times 600$ sec for \#714 and $1 \times 900$ sec for \#738.  The usable wavelength range covered 3900~\AA~-- 5050~\AA~in 27 orders at a nominal spectral resolution of 0.09~\AA~FWHM at 5000~\AA.

\subsection{Reduction}

\subsubsection{\emph{HST} Photometry}
We performed photometry on the long and short exposures in each filter using the PSF-fitting stellar photometry package \emph{HSTphot} \citep{dolphin} that was specially designed for the under-sampled PSF of the \emph{WFPC2}.  \emph{HSTphot} provides both instrumental and standard \emph{UBVI} magnitudes, using the standard reduction recipe and transformations of \citet{holtz06} as updated by \citet{dolphin09}.  This procedure includes masking out the bad pixels, generating sky images, masking cosmic rays, and cleaning hot pixels.  \emph{HSTphot} performs a two-pass search for stars in each \emph{WFPC2} chip to locate brightness peaks, performs PSF-fitting photometry in an iterative manner, and produces a ``refined" photometric measurement and position for each star with an astrometric precision of 0.03 pixels.  The code matches positions of stars detected in more than one filter to provide source-matched multi-color photometry.  The initial run of \emph{HSTphot} yielded 2251 and 577 candidate detections in at least one filter for the ``long" and ``short" images, respectively.   

It is a known characteristic of \emph{HST} \emph{WFPC2} images that spurious sources arising from, for example, diffraction spikes near bright stars, are included in this list. In an attempt to remove false sources we applied a series of additional criteria to the initial source list.  \emph{HSTPhot} reports a series of ``global'' parameters for each source, based on the combined detections in one or more exposures and one or more filters.  These include the source's roundness, sharpness, $\chi$ (as a goodness of fit), object type (point source or extended), and S/N. We determined a threshold roundness parameter, used to discriminate between point-like and elongated objects, by fitting a one-sided Lorentzian profile to the distribution of roundness values and retaining only sources having roundness parameter less than $1 \sigma$ above zero, corresponding to roundness $< 0.37$ and $< 0.19$ for the ``long'' and ``short'' exposures, respectively.  This retains 64\% and 79\% of the sources for the ``long"/``short" exposures, respectively.  The rejected stars are invariably those with the lowest S/N below $\sim$10.   Next, as per the \emph{HSTphot} manual suggestion, we retained only sources having $|$sharpness$| \leq 0.5$.  This criterion removes only a handful of stars that have not already been rejected by the roundness cut.   Finally, we retained only sources with global $S/N \geq 10$ in order to ensure that the final catalog contains only highly reliable sources.   Applying these criteria, the final list contains 967 and 360 stars in at least one filter for the ``long" and ``short" images, respectively:  194/171 in F336W, 227/185 in F439W, 348/286 in F555W, and 918/343 in F814W.  We performed a visual inspection of the images to confirm that these selection criteria retained real stars while rejecting obvious spurious sources such as diffraction spikes. 

\begin{figure}
\epsscale{0.8}
\plotone{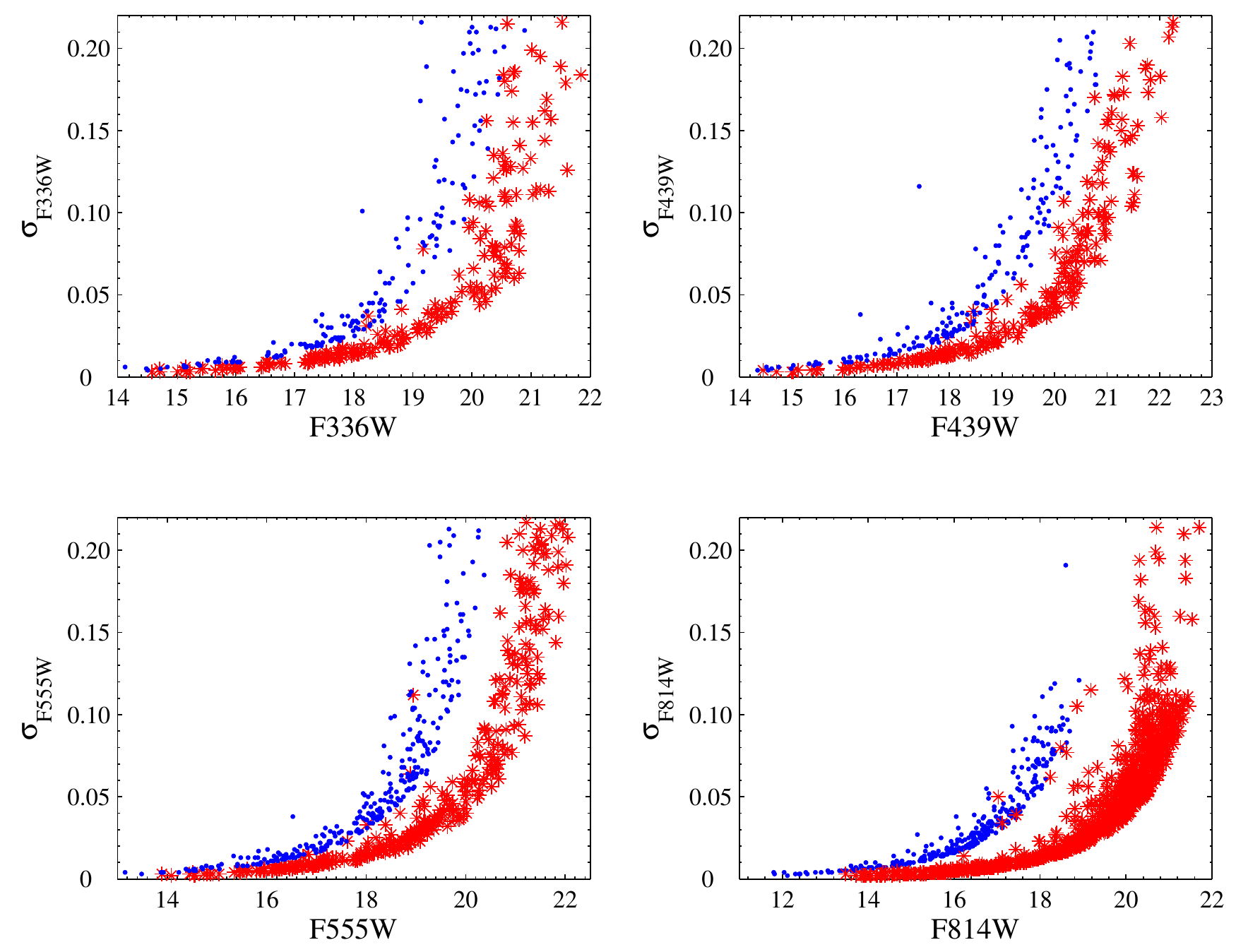}
\caption{Typical photometric uncertainty as a function of instrumental magnitude for each of the \emph{HST} filters.  Asterisks denote the ``long'' exposures (red in the electronic edition) and dots denote the ``short'' exposures (blue in the electronic edition).\label{maguncer}}
\end{figure}

Figure~\ref{maguncer} shows the photometric uncertainty as a function of instrumental magnitude for the short exposures (dots; colored blue in the electronic edition) and long exposures (asterisks; colored red in the electronic edition) in each filter.  Typical uncertainties are $< 0.01$ mag for stars $< 15$ mag in both the long and the short exposures for all filters.  Uncertainties rise to 0.2 mag at the faint end.  This occurs for stars with F555W $\simeq 20$ in the short exposures and F555W$\simeq 21.5$ in the long exposures.  Stars with uncertainties exceeding 0.1 mag appear in these plots because, on a per-filter basis, the S/N may be greater than the threshold global S/N of 10:1.  The sudden drop in source density in the F814W panel (lower right) above uncertainties of $\sim$0.1 mag arises because of the global S/N criterion imposed.  Since F814W is least affected by reddening, most sources are detected more strongly in this band and, therefore, the F814W photon statistics drive the global S/N values.      
 
\begin{figure}
\epsscale{0.8}
\plotone{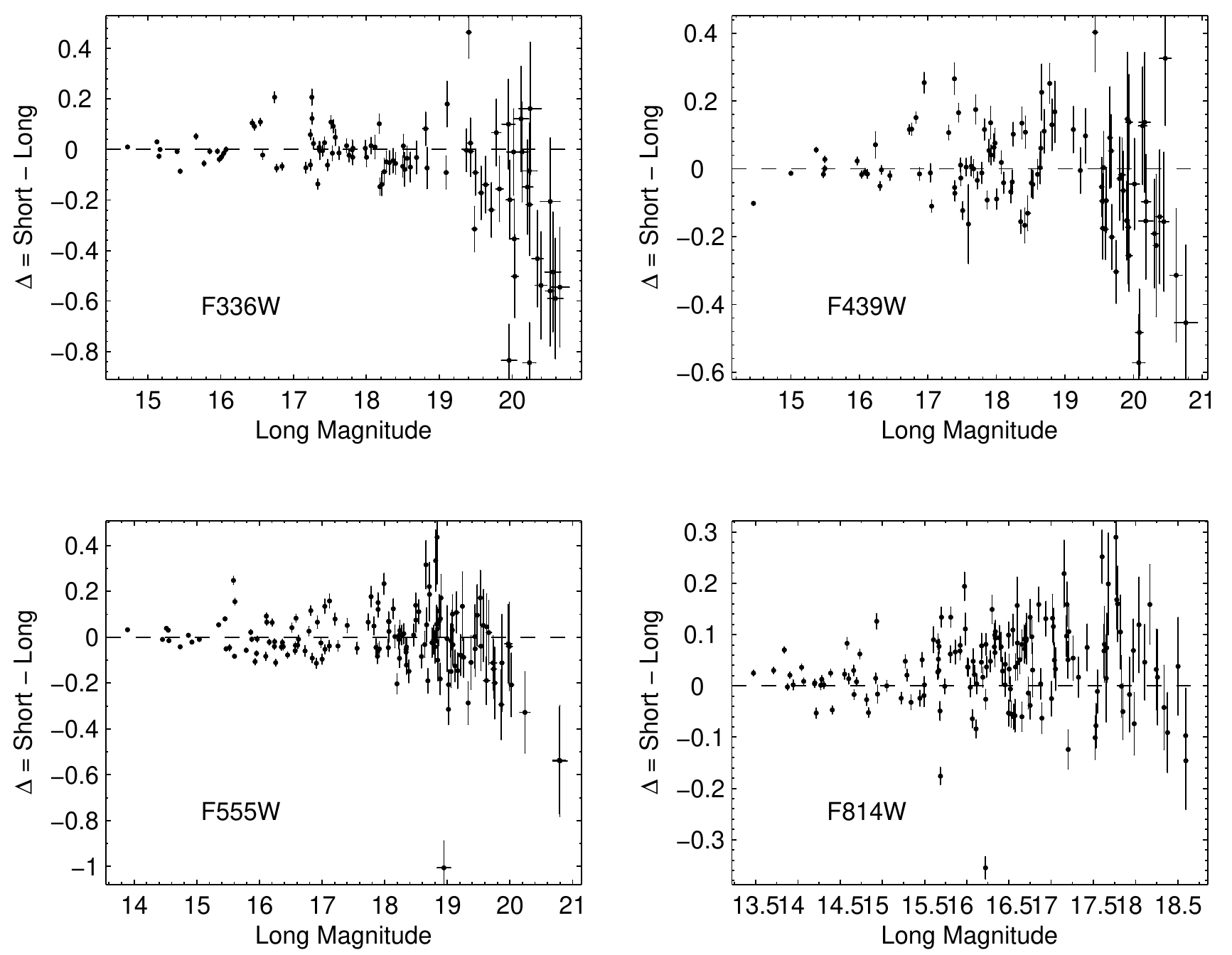}
\caption{Difference between the ``short" and ``long" exposure magnitudes versus long exposure magnitude for stars detected on both sets of images for each of the \emph{HST} filters.  The mean differences are consistent with zero, within the noise.  The rms values are, 0.2162,  0.1567, 0.1577, and 0.0887 magnitudes, respectively, slightly larger at fainter magnitudes.\label{HSTresidual}}
\end{figure}

For the 190 sources detected on both the ``short'' and ``long'' exposure images, we compared the photometry in each \emph{HST} filter to assess any systematic differences between the two orbits observations.  Figure~\ref{HSTresidual} plots the differences between the short- and long-exposure instrumental magnitudes for each \emph{HST} filter as a function of long exposure magnitude.  This figure shows that there are no systematic differences between exposures, except at the faintest levels where the asymmetric nature of Poisson noise produces a systematically brighter magnitude measured in the short exposures relative to the long exposures.  However, the rms deviation from zero is much larger than the typical photometric uncertainty.  The reduced $\chi^{2}$ exceeds 8 in all cases, indicating that there are additional sources of photometric uncertainty that are not included in the errors calculated by \emph{HSTphot}.  This is not an effect of chip position or crowding, but rather is a known limitation of existing \emph{HST} photometry packages when dealing with very short exposures \citep{dolphin}. Therefore, we adopt photometry from the long-exposure images, when available.  For stars appearing on only the short-exposure images (138 of the 1136 stars), we adopt a larger photometric uncertainty that is the quadrature sum of the uncertainty estimated by \emph{HSTphot} and an uncertainty determined empirically from the dispersion in Figure~\ref{HSTresidual} using 1-magnitude wide intervals.  This procedure yields $\chi^{2}_{\nu}$ values of less than two in all of the filters.  Stars detected on the ``short'' or ``long'' exposure images are identified as such in the last column in Table~\ref{mach-tab}.

Coordinates for each star were obtained using the task \emph{METRIC} in the Space Telescope Science Data Analysis System (STSDAS) package.  For a subset of 120 bright isolated stars we identified counterparts in the \emph{2MASS Point Source Catalog} (PSC) and used the mean differences between positions to correct the headers of the \emph{HST} images.  The position header parameters obtained from the guide stars were corrected by $(\Delta \alpha, \Delta \delta) = (0\farcs27182, -0\farcs57479)$ for the ``short" exposure images, while the ``long" exposure images required a correction of $(\Delta \alpha, \Delta \delta) = (-0\farcs01137, -0\farcs52219)$.  After astrometric correction, the rms deviation between positions measured on the \emph{HST} images and the 2MASS coordinates is $0\farcs1$ in R.A. and Dec.  This is consistent with the limitations imposed by the \emph{WFPC2} pixel size coupled with the precision of the \emph{PSC} astrometry.    

Table~\ref{mach-tab} lists the final band- and exposure-merged photometry in each filter, totaling 1136 unique source IDs ordered by right ascension.  A few lines appear in the printed journal as a sample of the table format.  The entire content is available as a machine-readable table in the electronic edition.  Column 1 gives the ID number from this work.  Column 2 provides a cross-identification with the nomenclature of MSP91.  In many instances the angular resolution of \emph{HST} reveals several sources within a $\sim2\arcsec$ radius of the nominal MSP91 star.  These are denoted by the MSP91 (MSP) designation with an alphanumeric suffix ``a'' for the brightest component, ``b'' for the next brightest component, as measured in the F555W band, etc.   Columns 3 and 4 give the (J2000) Right Ascension and Declination in degrees.  Columns 5 through 12 list the instrumental magnitudes and uncertainties in the \emph{HST} filter system. Columns 13 through 16 list the transformed magnitudes in the standard \emph{UBVI}.  Column 17 lists the global S/N value.  Column 18 lists whether the photometry is from the ``long'' exposure image (l) or the ``short'' exposure image (s).

We caution that the transformations from the \emph{HST} to Johnson system used by \emph{HSTphot} are based on stars having a limited range of color, and that many of the Wd2 stars fall at the extreme red end of those calibrations.   These reddest stars are late-type, low-gravity giants rather than heavily reddened early type stars with high surface gravity that typify the Wd2 membership. \citet{holtz95} further caution that the transformation are known to depend on metallicty and surface gravity.  We conducted our own synthetic photometry on reddened O-star spectra to assess the extent to which the \emph{HSTphot} transformations are appropriate to our targets ($(B - V) \simeq 1.4$; $(U - B) \simeq 0.3$.) We found that, for the $B$ and $V$ bands, the transformations are appropriate (within 0.02 mag) for reddened O stars.  In the $U$ and $I$ bands, systematic uncertainties exceeding 0.05 mag may be present, as noted by \citet{holtz95}.  Hence, where high levels of photometric accuracy are required, the \emph{HST} instrumental magnitudes are preferred over the \emph{UBVI} magnitudes.  We use the \emph{HST} instrumental magnitudes hereafter in our analysis, but we provide the \emph{URVI} magnitudes as a convenience and compare them to the MSP91 values in the next subsection.   

Transformations from F336W to the standard $U$-band are particularly problematic.  One complication is the red leak in the F336 filter, which becomes more significant for redder stars.   However, for the great majority of Wd2 stars in the range, $-0.5 \la ($F336W -- F439W$) \la 0.5$, or $(U-B)\simeq0.3$ (MSP91), this only amounts to $\sim$2--3\%, as shown by Figure~3 of \citet{holtz95} and confirmed by our own synthetic photometry of reddened early type stars.  Accordingly, we correct all of the reported $U$ and F336W magnitudes by +0.02 mag, acknowledging that this is only a zeroth-order correction appropriate to the mean Wd2 stars.  More problematic is that the standard Johnson $U$ band straddles the Balmer discontinuity while the F336W filter lies short ward of it.  \citet{holtz95} provides a discussion of difficulties in transforming the F336W magnitudes into standard $U$-band magnitudes, notes that the F336W zero points and $U$-band transformations are based on stars having $(U - B) < -0.1$ and $(V - I) < 1.0$, a regime that excludes the reddened O stars of Wd2.  Accordingly, we urge caution in any interpretation of the transformed $U$-band values.  Nevertheless, our U-band photometry agrees well with that of MSP91. 

\begin{figure}
\epsscale{0.8}
\plotone{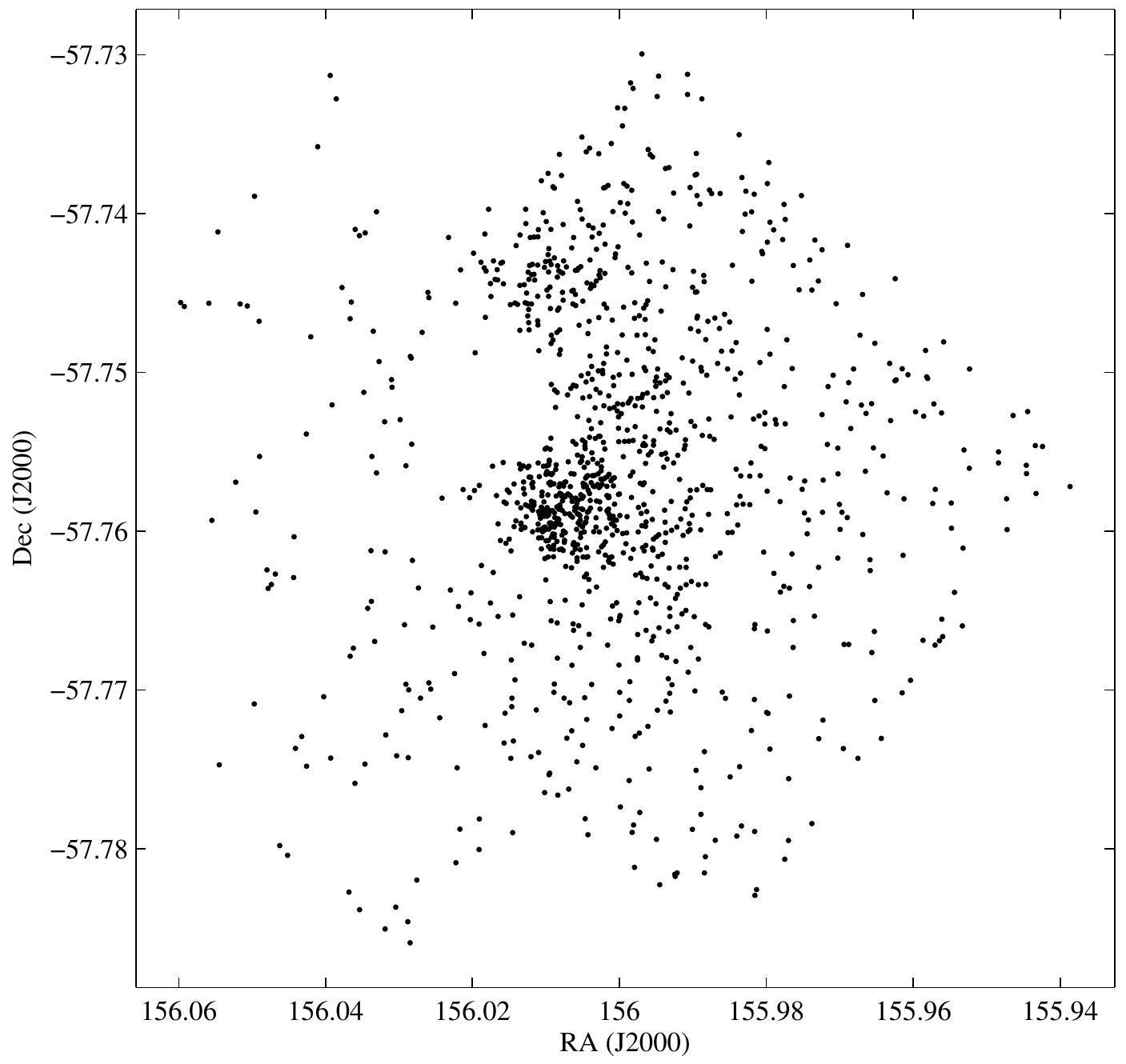}
\caption{Position map of the 1136 observed stars in Wd2.\label{coord}}
\end{figure}

Figure~\ref{coord} displays the positions of the 1136 detected stars in equatorial coordinates.  The concentration of cluster stars located on the PC1 chip of the ``long" exposure images and in the WF3 chip of the ``short" exposure images, is the dominant feature of the cluster, a secondary concentration of stars about $45\arcsec$ to the north is also apparent.  Equally striking is the near-total absence of stars over a $\sim$25$\arcsec$ diameter area to the northeast of the Wd2 core.  This corresponds to a minimum in the mid-IR [5.8] and [8.0] emission in Figure~\ref{glimpsergb}, suggesting the presence of an infrared dark cloud (IRDC) having extremely high extinction.  Our examination of the CO molecular maps of \citet{furu} reveals no obvious molecular clouds at this location, although the angular resolution of those data is $1\farcm5$, several times the size of the region of interest.  The \emph{JHK$_{s}$} infrared images of \citet{asc} show no obvious dearth of stars at this location, lending credence to the hypothesis of a localized region of high extinction that affects the optical wavelength data most significantly. 

\begin{figure}
\epsscale{0.8}
\plotone{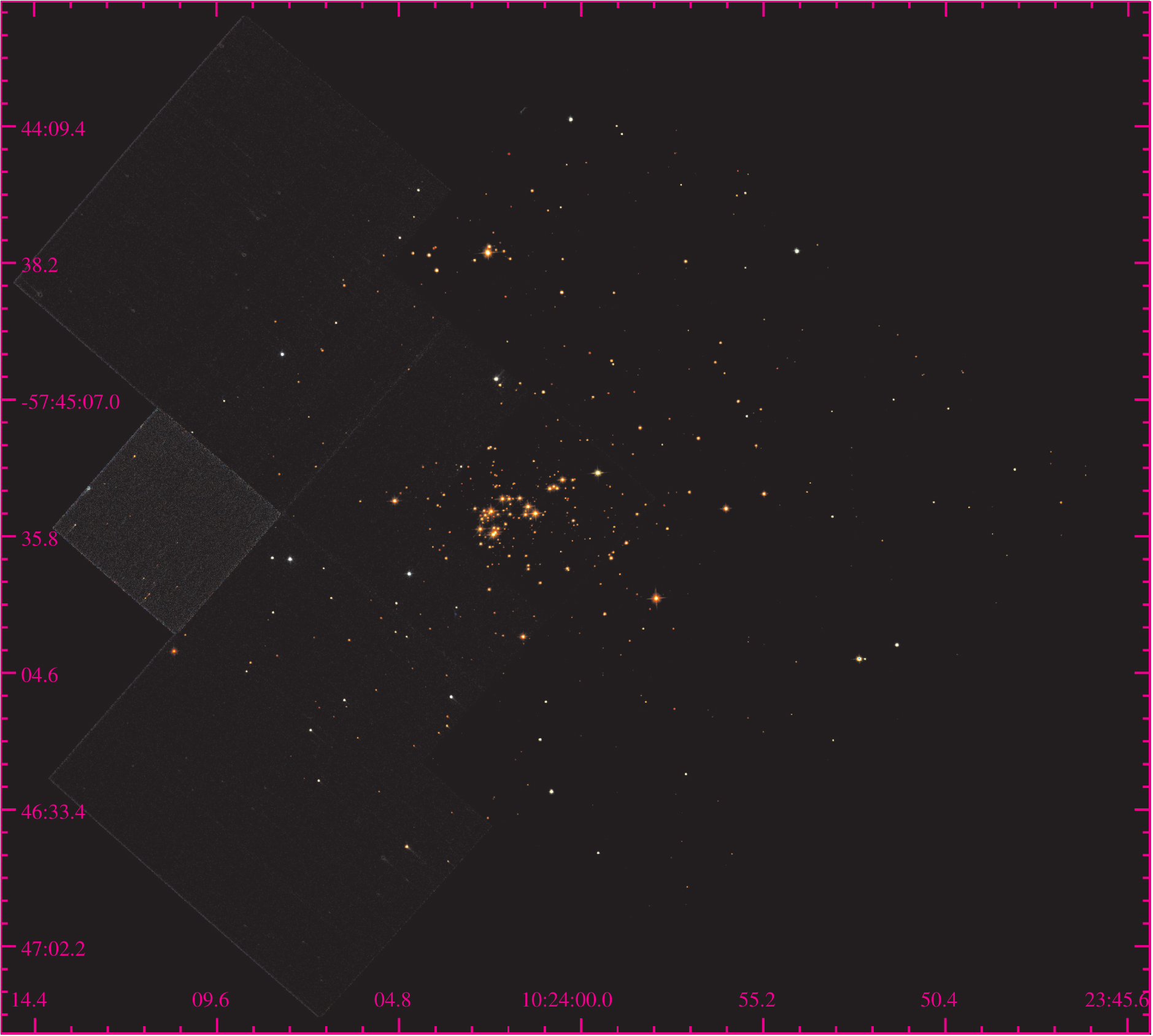}
\caption{RGB image composite of Wd2 using the F439W (blue), F555W (green), and F814W (red) \emph{HST} filters.  Note that the probable member stars near the cluster core appear similarly red, while probable foreground stars having less reddening appear white.\label{hstrgb}}
\end{figure}

Figures~\ref{hstrgb} -- \ref{hstrgb3} show three-color \emph{WFPC2} images of Wd2 with F439W in blue, F555W in green, and F814W in red.  The star ID numbers on the figures denote stars of interest to this study according to the numeration of Table~\ref{mach-tab}.  The color scale reveals that the core cluster stars are red compared to other bright foreground field stars that appear white, such as MSP~158 (\#436), confirmed as a foreground object by R07.  The excellent angular resolution reveals that many of the bright core stars identified by MSP91 have close neighbors that are blended or  unresolved at ground-based resolutions.

\subsubsection{Photometry Comparison}\label{photcomp}

We compared the photometry obtained with \emph{HSTphot} to that presented by MSP91 and R07 for the 75 and 11 stars in common with those works, respectively.\footnote{Unfortunately, \emph{WR~20a} (MSP~240) is saturated on the F555W and F814W images so no photometry is available in those bands.} The catalog provided by MSP91 does not contain positions, so it was necessary to do a visual match between our images and their finding chart.  Since the MSP91 observations were performed with ground-based telescopes that do not possess the resolution of the \emph{HST}, some of the MSP stars are actually blends of several stars that are now resolved in our data.  For such cases we choose the brightest star in the \emph{HST} image to be the one that was originally identified by MSP91.  In 63 of the 75 cases, either the MSP star is isolated and unambiguous, or we can identify the MSP source with the brightest of several close sources in our photometry.  In 8 cases, we found two or more stars of roughly equal brightness within a radius of $2 \arcsec$, so the cross identifications with MSP91 are ambiguous.  

Figures~\ref{moffatphotcomp} and \ref{rauwphotcomp} show differences between our photometry and the works of MSP91 and R07 ($\Delta m = m_{HST} - m_{them}$) versus \emph{HST} magnitudes transformed to the Johnson system.  Numbered points indicate the nomenclature of MSP91 in both Figures.  Where significant photometric differences exist, they are seen in both the $V$ and $B$ bands.  The median differences compared with MSP91 are 0.06, 0.19 and 0.15 mag, for the $U$, $B$ and $V$ bands, respectively.  Compared with R07 the median differences are 0.22 and 0.12 mag for the $B$ and $V$ bands respectively. Both figures show that our measurements are systematically fainter than theirs in both the $BV$ bands, particularly for the brightest stars which lie in the most crowded regions.  Part of this effect can be explained as a result of ground-based photometry that inevitably includes close neighbors in the measurement of the central star.  These neighbors are well-resolved and individually measured in the \emph{HST} data.

To further investigate this systematic magnitude offset we performed a series of tests on the \emph{HST} images.  First, we verified that the images are processed using the latest calibration pipeline routine, which in this case is the OPUS version 2009\_2k with \emph{calwp2} version 2.5.5 (Apr 17, 2009)\footnote{An older version of the pipeline, OPUS version 2008\_5c with \emph{calwp2} version 2.5.3 (Sept 4, 2008) produces the same photometry when using \emph{HSTphot}.}.  This represents the final and best calibration of the WFPC2.  Next, we performed aperture photometry using the Image Reduction and Analysis Facility (IRAF)\footnote{IRAF is distributed by the National Optical Astronomy Observatories, which are operated by the Association of Universities for Research in Astronomy, Inc., under cooperative agreement with the National Science Foundation.} package and compared it with the PSF results of \emph{HSTphot}.  To this photometry we applied the corresponding filter's zeropoint and Charge Transfer Efficiency (CTE) corrections according to \citet{dolphin09} and subsequent updates from the \emph{WFPC2 Calibration and CTE Corrections} webpage\footnote{http://purcell.as.arizona.edu/wfpc2\_calib/}.  We found excellent agreement between the results of the aperture and PSF photometry.  Furthermore, the \emph{HST} zeropoints are well calibrated and their uncertainties range between 0.02 -- 0.04 magnitudes, depending on the filter used \citep{zpt}.

Another possible source of discrepancy with ground-based observations could be the color transformations applied to the \emph{HST} flight system filters.  The color transformation of \citet{holtz95} was made using unreddened standard stars and stars from the cluster $\omega$ Cen rather than reddened O stars like those that dominate the population of Wd2.  We used the \emph{IRAF/STSDAS} {\bf synphot} package to create synthetic photometry of reddened O stars and compare it with the color transformations of \citet{holtz95}.  We found the transformations for reddened O stars to be compatible with the color transformations based on late-type stars, to within a few percent.  In most cases the color transformations are indeed small and the \emph{HST}'s flight system filters can be used as a proxy for standard filters, given the \citet{holtz95} transformations.  For example, the differences between the F555W and the standard $V$-band are on the order of $\sim0.01$ mag for a wide range of stellar colors \cite[see Figure 4 in][]{holtz95}.

We also investigated the known long-short exposure difference \citep{longshort} and the low background level \citep{backillu} problems that affect the \emph{WFPC2} camera.  Both effects are well-calibrated by \emph{HSTphot}, and any residuals in the current CTE corrections are small ($\sim$0.15 mmag/row for images with low-to-moderate background signal of $< 50~e^{-}$) and cannot account for the systematic difference between the ground-based and \emph{WFPC2} observations.  

Others have also noticed discrepancies between \emph{HST} and ground-based observations.  For example, \citet{dotter} reported offsets in the $V$ and $I$ bands of 0.068 and 0.007 mag, respectively.  \citet{turner12} also reports offsets with $UBV$ photometry.  From his Figure~2 we estimated an offset of $\sim 0.2$ magnitudes.  At the moment we can only conclude that there is an offset of unknown origin between the values reported here and those of MSP91 and R07.

\subsubsection{Spectral Reductions}

Optical spectra from \emph{Magellan} and \emph{SOAR} were bias-subtracted, flat-fielded, wavelength-calibrated and normalized using the standard IRAF procedures.   The rms of the wavelength calibration was 0.09~\AA~for the \emph{Magellan} data and 0.28~\AA~for the \emph{SOAR} data.  The spectral resolution determined from the width of arc lamp exposures was 4.1~\AA~FWHM, measured at 6172~\AA, and 7.6~\AA~FWHM, measured at 6172~\AA, for the \emph{Magellan} and \emph{SOAR} spectra, respectively.  Signal-to-noise ratios of the reduced spectra range between 15:1 and 91:1 pixel$^{-1}$ at 6000~\AA~for the \emph{Magellan} data and 11:1 to 80:1 pixel$^{-1}$ for the \emph{SOAR} data.  For purposes of spectral classification, we used a subsection of the observed spectral range from 4600~\AA~-- 6600\AA~having good S/N  and a number of temperature-sensitive \ion{He}{1} and \ion{He}{2} lines. The blue spectral range ($<$ 4600~\AA) commonly used for spectral classification was either not observed (\emph{Magellan}) or unusable (\emph{SOAR}) because of low S/N.  

The MIKE spectra of \#714 and \#738 were reduced using flat fields from the internal quartz lamps with a diffuser screen.  Spectra were extracted for each star and wavelength calibrated to an RMS of 0.004~\AA~using ThAr lamps. Multiple spectra and spectral orders were combined, discarding data of low S/N within 10~\AA~of the edge of each order, and continuum normalized using a 7th order Legendre polynomial before being Doppler corrected to the Heliocentric velocity frame.  S/N ratios were 23 and 11 per pixel for \#714 and \#738, respectively, at 4500~\AA~with 0.04~\AA~pixels, but we smoothed the spectra to $\sim$1~\AA~resolutions, yielding S/N ratios of 85 and 50, respectively.


\section{Data Analysis}\label{data-ana}

\subsection{Color Magnitude Diagrams}

Figure~\ref{cmdccd} shows the CMD for Wd2 in the \emph{HST}'s standard filters.  The CMD is the F555W magnitude vs. the (F439W -- F555W) color.  Detections with MSP91 counterparts, 81 stars, are identified by an asterisk (colored red in the electronic edition).  Stars of relevance to this work are labeled using the numeration system of this work.  The CMD clearly show shows two distinct populations.  On the CMD the blue ``arm" represents the field stars, while the redder ``arm" clearly shows the cluster's members.  Two stars, MSP~91 and 158 (\#832 and 436, respectively), fall on the field ``arm" and are likely foreground objects, consistent with the conclusions of R07.

We attempted to clean the CMD of field stars by using the 3D field star decontamination algorithm developed by \citet{bona07} and then improved in \citet{bica} and \citet{bona09}.  Unfortunately, the limited \emph{WFPC2} field of view means that we lack off-source observations of field stars needed to make a proper field star decontamination.  However, given the richness of the cluster core in the center of the small \emph{WFPC2} field, these data are dominated by cluster stars and, at the brightest magnitudes, suffer little field star contamination in the red arm.   

\subsection{Spectral Classification}\label{classification}

The vast majority of the new spectra for Wd2 exhibit lines of neutral and/or ionized He, indicative of O or early-B type stars.  Four sources exhibit H$\alpha$ in emission.  Figures~\ref{spec1} and \ref{spec2} show the normalized spectra for early-type stars labeled according to the IDs of this work and organized by decreasing value of temperature/spectral type from top to bottom.  Figure~\ref{spec3} shows the spectra for stars having spectral type consistent with late-O/early-B, undetermined, or field stars.  The main stellar absorption lines, interstellar lines (mainly Na~I~D lines $\lambda\lambda$5889,5895), and Diffuse Interstellar Bands (DIB) are labeled.  We classified stars having new spectra using the ratio of the equivalent width (EW) of \ion{He}{2} ($\lambda$5411) to the EW of \ion{He}{1} ($\lambda$5876) using the diagram and analytic fit from \citet{kobulnicky12}:
\begin{equation}
\frac{EW(\lambda5411)}{EW(\lambda5876)} =  1.16208 \times 10^{-12} T^{3} - 1.19205 \times 10^{-7} T^{2} + 4.22137 \times 10^{-3} T - 50.5093. \label{equ}
\end{equation}
\noindent This calibration is based on model spectra from the \citet{lanz} Tlusty models and hot stars from the \citet{jacoby} spectral atlas.  Because we have not observed spectral standards using our spectral setup, this may introduce additional uncertainty into our procedure for placing our targets on the MK system.  We used the robust curve fitting package MPFIT \citep{mark} as implemented in IDL to measure the EW of the \ion{He}{2} $\lambda$5411 and  \ion{He}{1} $\lambda$5876 lines and their uncertainties.  To improve the robustness of the fit for weak lines, we constrained the Gaussian width to that measured for the stronger of the two lines.  Table~\ref{spec-tab} list the stars having spectral types determined using the above procedure.  Columns 1 and 2 are the stars' identification numbers according to this work and MPS91, where available.  Columns 3 and 4 are the photometry, $V$ mag and $B - V$ color, for such stars.  Columns 5 and 6 are the measured EW for \ion{He}{2} and \ion{He}{1} lines having detections above $2 \sigma$; negative values correspond to emission lines, and upper limits are indicated.  Column 7 is the ratio of the \ion{He}{2}/\ion{He}{1} EW and column 8 is the stellar effective temperature corresponding to the ratio as determined by the relation in equation \ref{equ}.  Column 9 is the spectral type determined from the temperature approximation to one subtype \citep{martins05}.  Because our SOAR/IMACS data do not cover the blue portion of the optical spectrum typically used for classification, we are not able to constrain the luminosity class, thus we adopt a default of dwarf (V).  This is likely to be correct in most cases, since our targets are not among the brightest cluster members, assigned V or III by R07.   A constraint on the luminosity is obtained when analyzing the \emph{MIKE} spectra for stars \#714, and \#738.  The lack of \ion{N}{3} $\lambda \lambda$4634,4640,4662 in emission in spectra of Figures~\ref{mike714}, and \ref{mike738} confirms that these stars are not evolved.  Remarkably, there are no known supergiants in the cluster, consistent with  it being very young.    

\begin{figure}
\epsscale{0.8}
\plotone{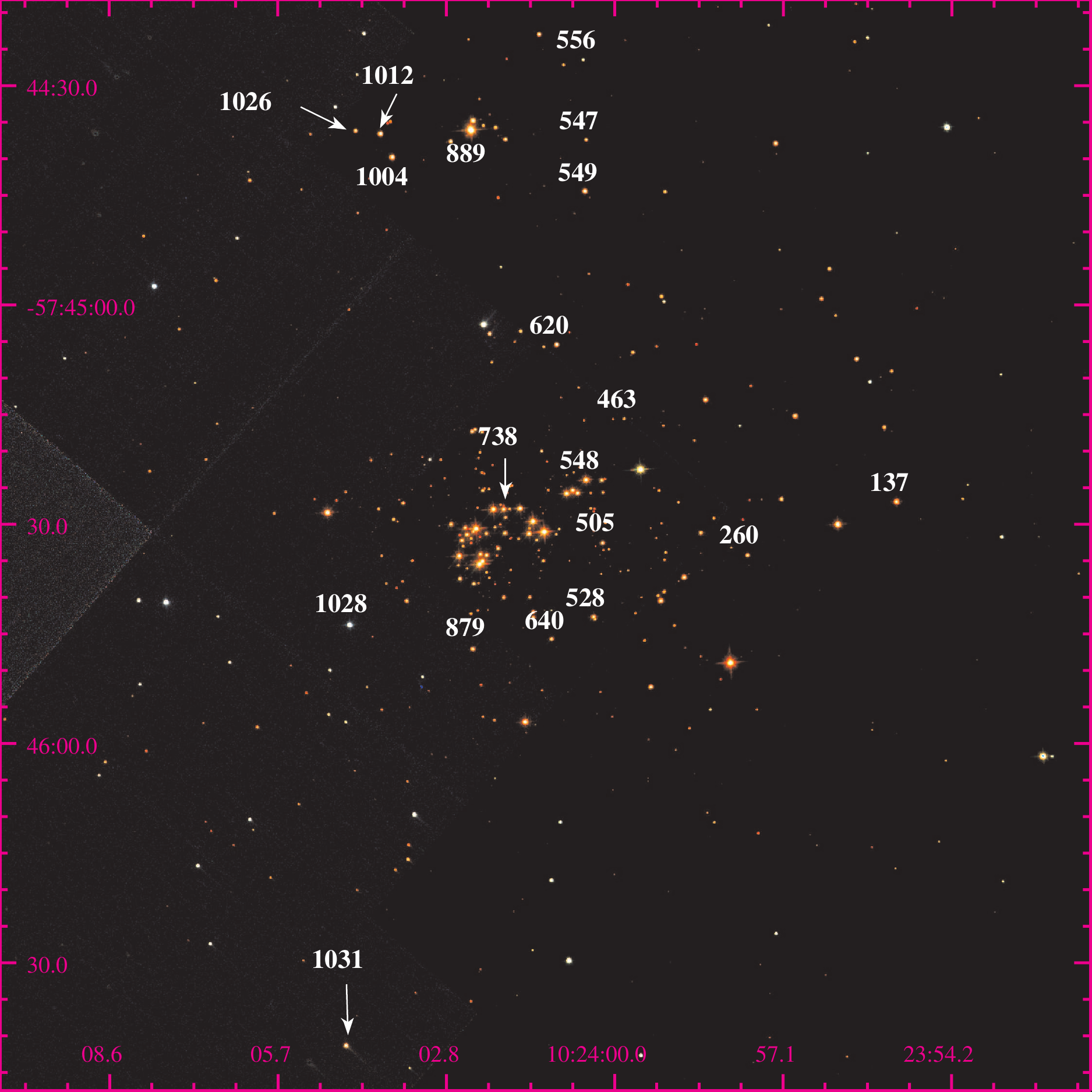}
\caption{Same as figure~\ref{hstrgb} for a zoomed view of the cluster core.  Key stars to this investigation are labeled according to the numeration system created for this work.\label{hstrgb2}}
\end{figure}

\begin{figure}
\epsscale{0.8}
\plotone{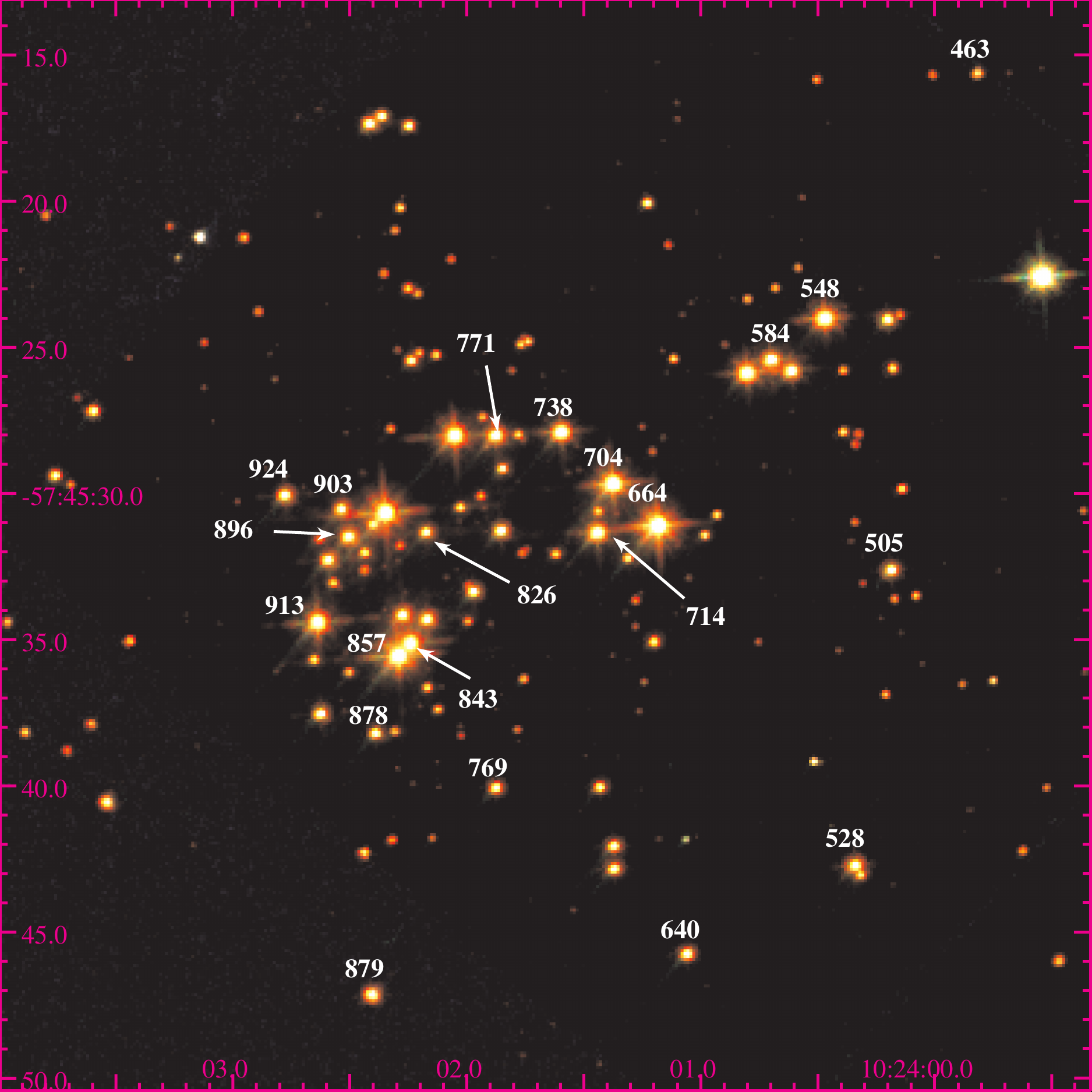}
\caption{Same as Figure~\ref{hstrgb}, showing a close-up view of the cluster core.  Note the bright star in the upper right (not labeled) that appears less red than other stars, marking it as a foreground object.\label{hstrgb3}}
\end{figure}

\begin{figure}
\epsscale{0.8}
\plotone{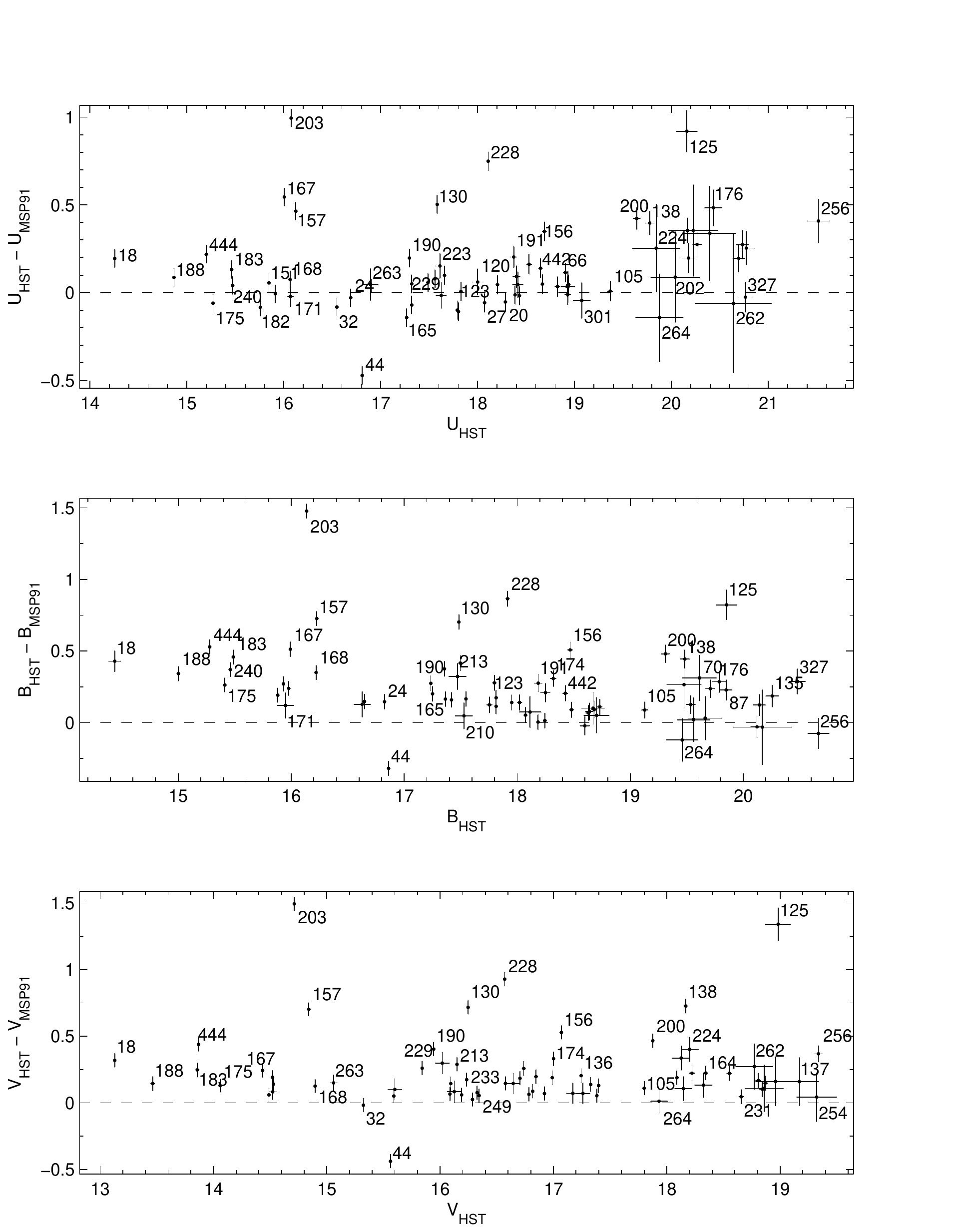}
\caption{Photometric differences between MSP91 and this work.  The \emph{HST} photometry is fainter than the ground-based observations.  The numeration follows the identification system of MSP91.\label{moffatphotcomp}}
\end{figure}

\begin{figure}
\epsscale{0.8}
\plotone{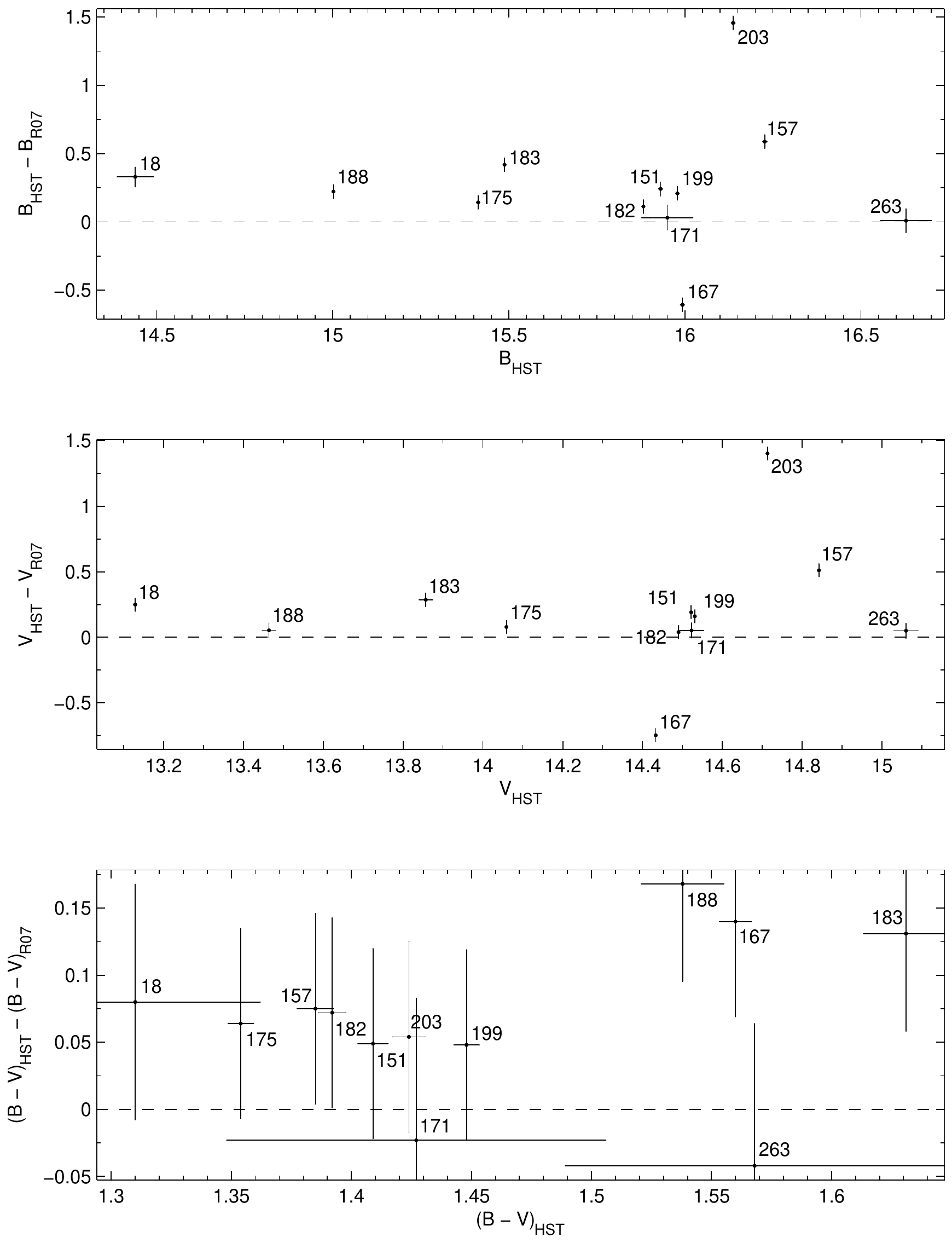}
\caption{Photometric differences between R07 and this work. Once again the \emph{HST} photometry is fainter than the ground-based observations.  The numeration follows the systems of MSP91 and R07.\label{rauwphotcomp}}
\end{figure}

\begin{figure}
\epsscale{0.8}
\plotone{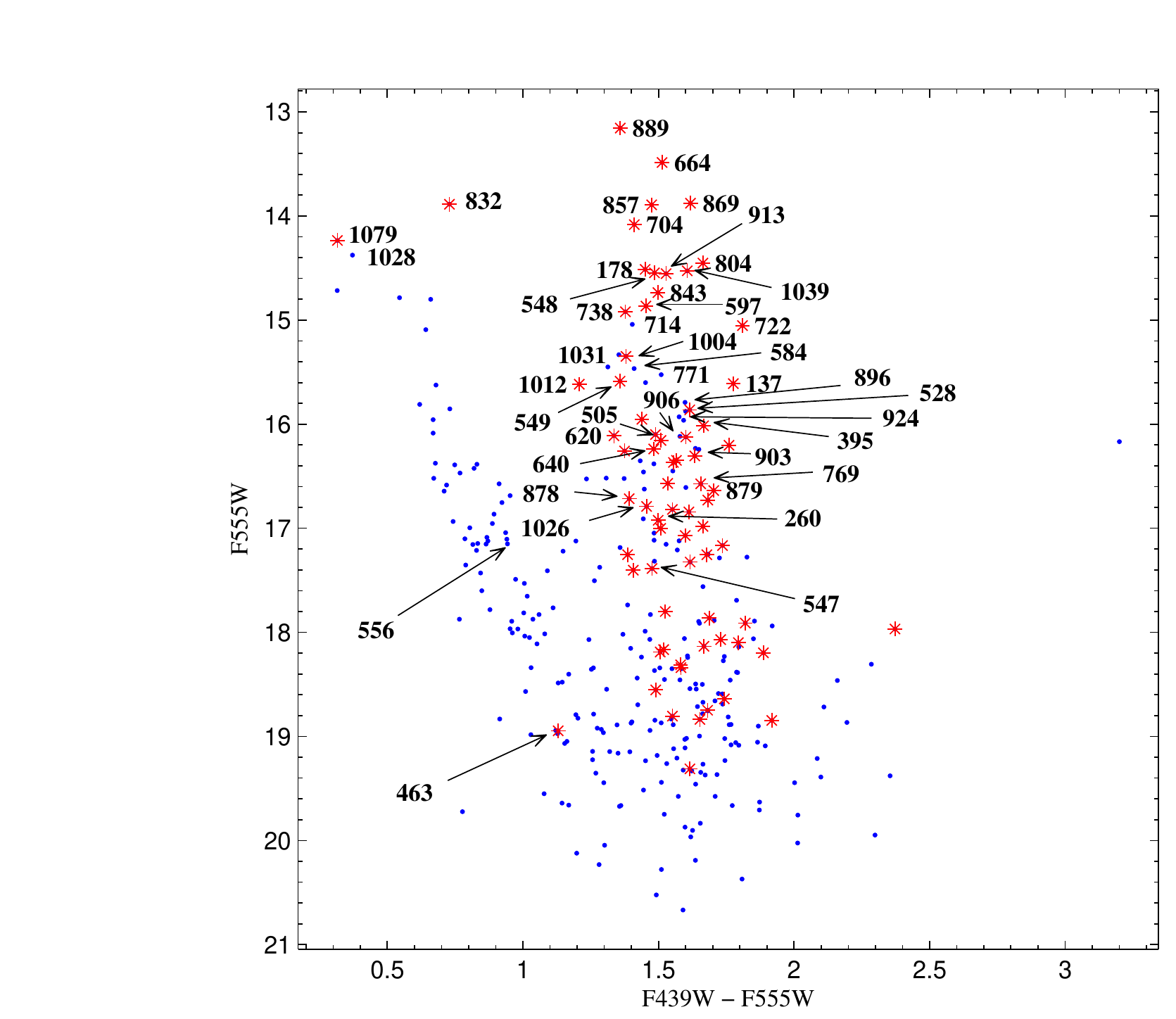}
\caption{CMD of Wd2 using the HST instrumental filters.  Asterisks (red colored in the online edition) represent the stars that were previously observed by MSP91.  Key stars are labeled according to the numeration system created for this work.\label{cmdccd}}
\end{figure}

{\it \#137 (MSP~165)}--- We tentatively classify this star as O4 on the basis of the relatively large \ion{He}{2} EW (0.97 $\pm$0.09~\AA) and relatively small \ion{He}{1} EW ($0.30 \pm 0.07$~\AA) in the \emph{SOAR} spectrum (S/N = 50).  This ratio indicates a very hot star with $\sim$43,000 K, very close to the maximum range calibrated in the \citet{kobulnicky12} diagnostic diagram.  However, the detection of \ion{He}{1} precludes classification as a true O3 given the \citet{walborn71} criteria requiring its absence.  This star is among the brightest in the sample at $V = 15.5$ and also one of the reddest with $B - V = 1.66$.  It does not have detectable H$\alpha$ emission.  Its location on the cluster outskirts near a region of diffuse IR emission  (see Figures~\ref{glimpsergb} \& \ref{hstrgb2}) is consistent with higher dust extinction at a location that has not yet been fully evacuated by the powerful winds of the central cluster.

{\it \#260 (MSP~201)}--- Located on the cluster outskirts not far from \#137 (see Figure~\ref{hstrgb2}), we classify as a late O or early B because of the lack of a \ion{He}{2} line from our \emph{SOAR} spectrum (S/N = 21).  H$\alpha$ appears in emission, along with [\ion{O}{3}] $\lambda\lambda$4959,5007, despite our best attempts at background subtraction, indicating the presence of small-scale variations in nebular emission near the star.  

{\it \#463 (MSP~125)}--- This star, together with \#483 at $1\farcs5$ to the E, constitute MSP~125.  Our \emph{SOAR} spectrum of this star (S/N = 11) has insufficient S/N for spectral classification. With $(B - V)$ = 0.87,  this star is unusually blue for a cluster member.

{\it \#505 (MSP~196)}--- This particular spectrum from \emph{SOAR} (S/N = 27) is dominated by light from \#505 but may contain contributions from \#495 and \#502 located within about 1\arcsec.  Hence, MSP~196 is a blend of at least three stars.  The dominant component, \#505, has $(B - V) = 1.3$ and spectral type O8.5, consistent with cluster membership.

{\it \#528 (MSP~229)}--- This source and its companion \#523, separated by $\sim$$0\farcs4$, constitute MSP~229.  Our two spectra of this object, one from \emph{Magellan} (S/N = 42) and one from \emph{SOAR} (S/N = 53),  yield O8.5 and O8, respectively, so we adopt O8 as a weighted mean of the brighter component which dominates the spectrum.

{\it \#547 (MSP~28)}--- This $V = 17.4$ source has [\ion{O}{3}] emission and H$\alpha$ emission.  \ion{He}{2} is a non-detection, and \ion{He}{1} is marginally detected in our S/N = 22 \emph{SOAR} spectrum, making this a late-O or early-B star.  The position of this star is interesting since it is located near the secondary cluster, centered on MSP~18, and it lies upon a ring of diffuse mid-IR emission which surrounds the secondary cluster.  Figure~\ref{ring} is a \emph{Spitzer} 3-color image of the region surrounding MSP~18 showing this $\sim36\arcsec$ diameter ring-like structure, which is present in all IRAC bands but best seen at [8.0], consistent with PAH emission from a photo-dissociation region.  We believe that this is the first recognition of this feature, which possibly demarcates the boundary of the \ion{H}{2} region surrounding MSP~18, although MSP~18 does not lie at its geometric center.

{\it \#548 (MSP~151)}---  Although this star is located $\sim$13\arcsec NW of the cluster core in an area that may be affected by small-scale dust extinction.  Observed with \emph{SOAR} (S/N = 80), a spectral type of O4 was determined.  This classification is at odds with MSP91 and R07 classification of O6 -- 7III.  The \ion{He}{2} is very strong with EW $= 0.94 \pm 0.04$~\AA~and \ion{He}{1} is rather weak with  EW $= 0.31 \pm 0.06$~\AA.  Nebular lines are well subtracted, so we do not have an explanation for the earlier spectral type compared to R07.
  
{\it \#549 (MSP~44)}--- Star \#549, along with the faint companion \#538 constitute the source MSP~44.  Indeed, \citet[hereafter R11,][]{rauw11} identify this system as a P = 5.176 d eclipsing binary ($\Delta V = 0.6$ mag) composed of an B1V + pre-main-sequence star, or possibly even a triple system.  We obtained two spectra for this source, one as target (S/N = 59), and the second being serendipitous (S/N = 28).  Both observations, obtained with the \emph{SOAR} telescope, yield consistent spectra where the He lines are very weak or non-detections. For this star we also detected emission at [\ion{O}{3}] and H$\alpha$.  The presence of emission lines may indicate a localized knot of nebular emission.  The residual \ion{He}{1} emission may affect the stellar EW measurement, rendering stellar classification unreliable, but consistent with a very late-O or early-B star, in agreement with R11.  R11 reports nebular emission associated with MSP~18, MSP~223, and several other stars.  Evidently, the nebular structure in the Wd2 region is complex both spatially and, as indicated by R11, in velocity as well.  Like \#547, which also exhibits nebular emission, \#549 lies near the  ring of mid-IR emission encircling MSP~18.  Mid-IR PAH emission usually traces the UV-heated surfaces of molecular clouds.  Such a location could indicate that the MSP~44 system is very young, consistent with the interpretation of R11 that the secondary star is a pre-main-sequence object.

{\it \#556}---  Located $\sim$66\arcsec\ to the north of the cluster core, \#556 ($V = 17.1$) seems to be surrounded by diffuse IR emission, very likely coming from the ring surrounding MSP~18. \ion{He}{1} is detected in our S/N = 23 \emph{SOAR} spectrum with an EW of $0.53 \pm 0.11$~\AA, but \ion{He}{2} is not detected, yielding only an upper limit on the spectral type of late O or early B.

{\it \#584 (MSP~157b)}---  Located at $\sim$10\arcsec\ from the cluster core where blending becomes problematic,  MSP~157 contains contributions from as many as seven stars, the brightest three of which are stars \#597, \#584, and \#568 located within $\sim1\farcs2$ radius.  Our \emph{Magellan} spectrum (S/N = 58) and classification implies an O8.  This is ``slightly" later type than the O5.5 -- 6.5V by R07, but their spectrum is likely dominated by the brighter component, MSP~157a (\#597).

{\it \#620 (MSP~96)}--- Located in a  relatively isolated region at $\sim$26\arcsec\ to the north of the main cluster, \#620 along with \#663, constitute MSP~96, and its photometry is consistent with the measurement from MSP91.  Both He lines are non-detections in our S/N = 43 \emph{SOAR} spectrum.  This limits our classification to a very late O or early B star.

{\it \#640 (MSP~233)}--- This $V = 16.2$ source is relatively isolated and lies close to diffuse IR emission at $\sim14\arcsec$ to the SW of the main cluster.  The clear detections of the He lines in our \emph{Magellan} spectrum (S/N = 36) provide a classification of about O9.5.

{\it \#664 (MSP~188)}--- This bright star has two faint ($\Delta V = 4.3$ mag) neighbors at $\sim$$1\farcs5$.  The very strong \ion{He}{2} and the weak \ion{He}{1} in our S/N = 91 \emph{Magellan} spectrum yield an EW ratio of the He lines of $8.99 \pm 4.20$, placing this star at the upper limit of our classification scheme; therefore, we type it as an O4 or earlier.  This is in agreement with R07 who revised the MSP91 spectral type to O4.  As portrayed in the velocity plot of the \ion{He}{2} and \ion{He}{1} lines in Figure~\ref{664vel} the spectrum shows that both He lines are double-peaked, with \ion{He}{1} showing two approximately equal-depth absorption lines, and \ion{He}{2} displaying different strength absorption lines. We measure the velocity difference between the peaks for \ion{He}{1} to be $\Delta v \approxeq 360.5 \pm 103.$ \kms and for \ion{He}{2} to be $\Delta v \approxeq 489.6 \pm 36.8$ \kms.  The He lines were deblended to calculate individual EW and EW ratios, so as to determine the spectral types of what we infer to be a binary system.  The two components have EW ratios of $5.70 \pm 3.32$ and $2.04 \pm 1.18$, implying spectral types of at least O4 and O5.5, respectively.  The binary nature of this system partially explains why \#664 is so luminous compared to other very early members of Wd2.  

{\it \#704 (MSP~175)}--- Star \#704, along with the faint and close neighbor \#713 (lying $\sim$$0\farcs44$ away, midway between \#704 and \#714), forms MSP~175.  It shows very strong \ion{He}{2} in our S/N = 47 \emph{Magellan} spectrum, while \ion{He}{1} is very weak (EW $= 0.15 \pm 0.05$), implying a spectral type near O4.  This is slightly earlier than the O5V -- O6V type assigned by R07.  R07 notes that the He lines in MSP~175 are broad making classification uncertain; He II lines appear asymmetric with a blue wing in Figure~3 of R07.  Our \emph{Magellan} spectrum shows asymmetry with a red wing in both the He II and He I lines, suggesting a possible binary system.

{\it \#714}--- Star \#714, located at $\sim$3\arcsec to the W of the main cluster core, is in a region of heavy crowding.  Stars \#664 (MSP~188), and \#704 (MSP~175) lie in a crowded region near the cluster core within $1\farcs5$ of several other stars, including \#714 (no MSP identification) and \#713.  Figure~\ref{mike714} shows the MIKE spectrum of \#714 (S/N = 91), smoothed to a resolution of about 1~\AA~with key spectral features of hot stars labeled.  This spectrum is consistent with the lower resolution IMACS spectrum in that  \ion{He}{1} $\lambda$4471 is extremely weak (not detected here), while the \ion{He}{2} lines are very strong, indicating an extremely hot star.  By comparison with the O-star atlases of \citet{wal} and \citet{walborn02} we assign this star an O3 spectral type and dwarf (V) luminosity class.   Signatures of O2 spectral types, such as \ion{N}{5} $\lambda\lambda$4604,4620 absorption or strong emission lines of \ion{N}{4} $\lambda$4058 and \ion{N}{3} $\lambda\lambda$4634,4640,4642 are not seen, although these latter N features are weakly present.  The lack of He emission lines and lack of strong N emission lines further supports a dwarf luminosity classification.  Nebular [\ion{O}{3}] $\lambda\lambda$4959,5007 and H$\beta$ lines also appear in this spectrum, but we consider these to originate from the  diffuse ionized gas that pervades the RCW49 \ion{H}{2} region; these emission lines remain after background subtraction, presumably because of small scale variations in the nebular emission along the slit.  In our single spectrum we find no evidence of binarity. However, all of the principle spectral stellar features in \#714 are redshifted by about 1~\AA~(67 \kms) compared to expected rest wavelengths, and compared to molecular gas associated with the cluster which spans the range -11 to 11 \kms~LSR \citep{ohama}. This redshift may constitute evidence of being a single-lined spectroscopic binary or may indicate that \#714 has a peculiar velocity sufficient to unbind it from the cluster. 

{\it  \#738 (MSP~168)}--- Figure~\ref{mike738} shows our single MIKE spectrum of \#738 ($V = 14.9$) boxcar smoothed to about 1~\AA~resolution (S/N = 48).  \ion{He}{1} $\lambda4471$, with an equivalent width of 0.29~\AA, appears in this spectrum, making \#738 later than O3.  The EW ratio EW($\lambda4686$)/EW($\lambda4471$) is 1.86, similar to O5 -- O6 stars from the \citet{wal} atlas.  The radial velocity, as judged from the mean of  key spectral features, is consistent with zero \kms~LSR,  making it consistent with the molecular gas surrounding the cluster \citep{ohama}. We adopt a classification of O5.5.

{\it \#769 (MSP~219)}--- This star is located at $\sim9\arcsec$ to the S of the main cluster core in a relatively isolated area.  \ion{He}{2} is a marginal detection in the Magellan spectrum (S/N = 31) but \ion{He}{1} is strong with EW $= 0.85 \pm 0.05$.  This  suggests that \#769 is O9.5 or later.

{\it \#771 (MSP~167)}--- MSP~167 is a blend of the stars \#804, \#771, \#765, \#777, and \#772.  A broad \ion{He}{2} line and a well-defined \ion{He}{1} line produced an EW ratio of $1.00 \pm 0.13$ implying that this star is O8.

{\it \#857 (MSP~444)}--- MSP~444 is one of those hard-to-identify stars owing to its position in the cluster core and its close proximity to \#843 (MSP~203; $\sim$$0\farcs6$).  With  \emph{HST} MSP~444 was resolved into three components, \#857\, and  \#895, and \#824.  The Magellan spectrum (S/N = 78) of \#857 possesses a strong and well-defined \ion{He}{2} line while the \ion{He}{1} is broader and weaker than \ion{He}{2}, implying a hot star of type O4.5, consistent with O4 -- O5 by R11.  With $(B - V) = 1.4$, its color and magnitude are about right to be an early type cluster member.

{\it \#878}---  Located at $\sim$8\arcsec to the SW at the outskirts of the main cluster, the two Magellan spectra obtained for this faint star ($V = 16.6$) have insufficient S/N (S/N = 15) to perform a reliable classification.  \ion{He}{2} is not detected, and \ion{He}{1} is weak, hinting at a late O or early B.

{\it \#879 (MSP~235)}--- This star is in relative isolation at $\sim$16\arcsec\ to the SE of the main cluster.  The \emph{Magellan} observations, with a S/N = 36, show strong \ion{He}{1} but minimal \ion{He}{2}, implying a spectral type of O9.5.

{\it \#896 \& \#903 (MSP~183 subcomponents)}--- These stars are located on the outskirts of the main cluster core where heavy blending in ground-based observations is inevitable.  Within a radius of $\sim$$1\farcs5$ there are seven stars that could contribute to the brightness of MSP~183, namely, \#869, \#896, \#903, \#826, \#880, \#887, and \#856.  These two stars, separated by $\sim1\farcs5$, became partially blended on the longslit observations with the \emph{Magellan Telescope} (S/N = 43 for \#896 \& S/N = 39 for \#903).  The strong He lines in both stars resulted in an EW ratio of $0.83 \pm 0.13$ for \#896 and $0.90 \pm 0.14$ for \#903, implying an O8.5 type for both.  

{\it \#1004 (MSP~32), \#1012 (MSP~24), and \#1026 (MSP~20)}--- This stellar trio, observed spectroscopically with SOAR, is located on the diffuse mid-IR ring that surrounds the secondary cluster, at about $\sim12\arcsec$ to the W of MSP~18.  Star \#1004's spectrum (S/N = 53) possesses strong He lines, although \ion{He}{2} is noisy and broad, implying a classification of O9.5.  The spectra of both \#1012 (S/N = 36) and \#1026 (S/N = 14) display [\ion{O}{3}] and H$\alpha$ in emission and have insufficient S/N for reliable classification.

{\it \#1028}--- With $V = 14.3$, and $(B - V) = 0.37$, this star is clearly a field star in the vicinity of the cluster core.  The serendipitous \emph{SOAR} spectrum (S/N = 62) shows only strong Balmer lines with EW(H$\alpha$) = 7.67~\AA, making it a probable A star.

{\it \#1031}--- Greatly isolated from the cluster core at $\sim$73\arcsec\ to the SE it lies projected against a bright mid-IR rim illuminated by the central cluster. The S/N = 32 SOAR spectrum is heavily contaminated by emission lines, and no spectral  type determination is possible.  Interestingly, as shown in Figure~\ref{spec3}, \#1031 shows \ion{Mg}{2} $\lambda5173$ \AA\ absorption lines.


\section{Results}\label{results}

\subsection{De-Reddening}\label{dered}

We measured the extinction, $A_{V}$, and the ratio of selective to total extinction $R_{V} = A_{V}/E(B - V)$, for each star by individually fitting a spectral energy distribution (SED) based on the known spectral type/temperature of the star and the theoretical absolute magnitudes intrinsic colors of the Padova stellar evolution tracks and isochrones (CMD v.2.2)\footnote{Available from the website: http://stev.oapd.inaf.it/cgi-bin/cmd\_2.2} \citep{girardi00, marigo, girardi02}.  Each SED was reddened by each of three reddening laws \citep{cardelli, F04, FM07} (hereafter CCM89, F04, and FM07 respectively) and compared, after appropriate normalization, with the \emph{HST}\footnote{\citet{holtz95} caution that reddening determinations should be made in the \emph{HST} filter system rather than the transformed Landolt \emph{UBVI} system.} and \citet{asc} \emph{JHK$_{s}$} photometry to find a global chi-squared minimum over all plausible values for $R_{V}$, and $A_{V}$.  We note here that, while CCM89 provide an analytical approximation of the mean interstellar extinction with $R_{V}$ and $A_{V}$ as free parameters, FM04 tabulates 38 reddening parameterizations as a function of $R_{V}$, and FM07 tabulates parameterizations of 328 sightlines, creating 328 discrete reddening laws.  As an additional method to estimate $R_{V}$ we also used the method of \citet[hereafter FM09][]{FM09} which is based on an average relation between \emph{BVK$_{s}$} photometry in their Figure~7.  The fitted methods are sightly dependent on the anchor point used to normalize the Padova magnitudes to the data.  To eliminate this dependency the fitting was done using each of the seven photometric values, when available, as anchors points and the resultant values of $R_{V}$, $A_{V}$ and $A_{\lambda}$ are weighted averages of all seven fits.  Figure~\ref{sed-plot} shows examples of the best fits for stars \#137, 528, 704, and 1039.  Each subplot displays the \emph{HST} photometry as squares and the \citet{asc} IR photometry as circles.  The lines are the best fit to the photometry after reddening the predicted magnitudes for the appropriate star's spectral type.  For comparison purposes the best fit for each method was included: the solid lines is the CCM89, the dash line is the F04, and the dotted line is the FM07 reddening laws (each one is colored in blue, cyan and magenta in the electronic edition).

Table~\ref{reddening-compilation} compiles the fit results and shows that best-fitting values range between $5.71< A_{V} < 7.54$ and $3.44 < R_{V} < 4.31$.  Column 1 lists the star's ID according to this work, column 2 lists the adopted spectral type, columns 3 -- 6 are the best-fitting $R_{V}$ using the methods of CCM89, F04, FM07, and FM09 respectively, and columns 7 -- 10 are the the best-fitting $A_{V}$ values.  The results indicate that $R_{V}$ is significantly higher than the mean interstellar value of $R_{V} = 3.1$, meaning that the obscuring dust produces more extinction for a given amount of reddening, often interpreted as a result of a population of larger ``gray'' dust grains.  On the other hand  FM07 and FM09 argue that there is no universal reddening law or extinction curves and that extinction cannot be parameterized in terms of a single variable such as $R_{V}$.  The so-called ``anomalous" dust would, in reality, be ``normal" for the Wd2 region and may be different from any other dust in the Galaxy.

Stars \#664, 869, 889 are not included in Table~\ref{reddening-compilation} since these only have photometry in three bands (F336W, F439W, F555W) because of saturation or blending.  We regard that the adopted fit procedure provides unreliable results and we adopt, for these stars, the average $R_{V}$ and $A_{V}$ from the 29 other stars.  Excluding these three stars, the average values for the cluster are $\langle R_{V} \rangle = 3.77 \pm 0.09$ and $\langle A_{V} \rangle = 6.51 \pm 0.38$ mag where the uncertainties are the dispersions of the sample.   We conclude that the mean $R_{V}$ is substantially larger than mean interstellar values, and that all stars are consistent, within the uncertainties, of this mean.  On the other hand, there appears to be a significant scatter in $A_{V}$, consistent with appreciable internal extinction variations. 

Table~\ref{reddening-table} compiles the corresponding $A_{\lambda}$ for the \emph{HST} filters corresponding to the best model fit in Table~\ref{reddening-compilation}.  Column 1 lists the star's ID according to this work, column 2 lists the adopted spectral type,  columns 3, 4 \& 5 are the $A_{F336W}$ values for CCM89, F04 and F07, and column 6 contains the weighted average of columns 3, 4, \& 5.  The same pattern is followed by columns 5 thru 18 for filters F439W, F555W, and F814W.  We use the values of Table~\ref{reddening-table} to deredden the listed stars in the HST filter system and correct for the heavy, patchy extinction toward Wd2 which produces considerable dispersion in the raw CMD (Figure~\ref{cmdccd}).

The (F336W -- F439W) vs (F439W -- F555W) color-color diagram (CCD) in Figure~\ref{ccd-iso-fit} shows a portion of a reddened Zero-Age Main Sequence (ZAMS) Padova isochrone\footnote{The Padova isochrones are based on the non-overshoot models of ATLAS9 by \citet{castelli}, which in turn recommends using \citet{martins05} observational values for O stars to transform effective temperatures and surface gravities into spectral types. We adopt the \citet{martins05} observational calibration.} using the global $A_{\lambda}$ averages from Tables~\ref{reddening-compilation} \& \ref{reddening-table} ($R_{V} = 3.77$, and $A_{V} = 6.51$ with corresponding color excess of $E$(F336W -- F439W) = 1.47 and $E$(F439W -- F555W) = 1.81).  The dashed line (magenta in the electronic edition) is part of the reddening vector connecting the unreddened ZAMS O stars and the reddened Wd2 stars.  Asterisks (red in the electronic edition) represent Wd2 stars having spectral types that we used to obtain the reddening parameters.  The reddened isochrone provides a good match to the cluster stars, given that these stars exhibit a dispersion that is parallel to the vector, indicating substantial differential reddening and consistency with the mean reddening parameters found for this sightline.  Additionally, the dotted curve (cyan in the electronic edition), is a reddening track that takes into account small systematic color trends due to the broadband nature of the \emph{HST} filters and the large reddening towards Wd2. The curve was constructed by performing synthetic photometry with the software package {\bf synphot} using a CCM89 reddening law.  As pointed out by \citet{schmidt} and \citet{turner89}, the effective wavelength of a filter changes as a function of spectral type and reddening, leading to curved reddening tracks.  Along this curve we label points that correspond to particular extinction values $A_{V}$.  According to \citet{turner89} the difference between a linear and a curved reddening track is $\sim$2\% per $E$($B - V$).  Although this is a small factor we included it in Figure~\ref{ccd-iso-fit} for illustrative purposes.  When compared to the random and systematic errors of the intrinsic and observed colors that are discussed previously and in the next section, this difference is small so we adopt the straight vector in our reddening analysis.  From the figure it is evident that there exists a systemic offset between the linear vector and the curved track, but this is small when compared to the dispersion in the photometry.  This offset can be explained by the method we used in the creation of the linear vector.  This vector was created using the average $A_{\lambda}$ from the reddening laws of CCM89, FM04, and FM07.  Such values do not necessary have to reproduce the curve created by {\bf synphot} that only uses the CCM89 reddening law.

Support for the ``anomalous'' value of $R_{V}$ toward Wd2 can be found in the studies of other open clusters and their Cepheid stars in the vicinity of the Carina arm.  The open cluster Ruprecht 91, Collinder 236, and Shorlin 1; with projected distances of $\sim$$3\fdg2$, $\sim$$5\fdg6$, and $\sim$$6\fdg4$, respectively, away from Wd2, also have different values from the nominal $R_{V} = 3.1$; Ruprecht 91 with $R_{V} = 3.82 \pm 0.13$ \citep{turner05}, Collinder 236 with $R_{V} = 3.82 \pm 0.13$ \citep[the $R_{V}$ was an adopted value because of similar reddening law with Ruprecht 91;][]{turner09}, and Shorlin 1 with $R_{V} = 4.0 \pm 0.1$ \citep{turner12}.  Furthermore \citet{turner12} made a compilation showing that various clusters in the Carina complex have values greater than 3.1 in general (see his Table 1).

\subsection{Distance Determination}\label{distance}

We adopt the Padova isochrones to determine the cluster's distance modulus ($DM$).  Isochrones with ages between 1 -- 4 Myr and solar metallicity ($Z = 0.019$) were chosen to match the expected properties of Wd2.  The Padova database provides absolute magnitudes (and therefore intrinsic colors) for stars in the \emph{HST} filter system which we use to compute a spectroscopic and photometric distance for each star, following the prescription used by \citet{hanson} in finding the distance to Cygnus OB2.  We adopt either our own spectral classifications or those of R07 and R11 (based on the \citet{wal} calibration) in order to calculate the appropriate $M_{\mbox{F555W}}$ and (F439W -- F555W)$_{0}$ values and their associated uncertainties by interpolating the Padova data tables.  For stars typed using our procedure described above, we estimate absolute magnitude and color uncertainties by propagating errors from our determination of the effective temperature. For stars typed by R07, we adopt temperature uncertainties given in that work.  We deredden each star according to the values in Table~\ref{reddening-table} and then solve for the distance modulus of each.  

Table~\ref{dis-table} displays the results of our distance analysis.    The columns list (1) the star ID numbers according to this work, (2) the MSP91 identification numbers, (3) the star's spectral type, (4) the observed apparent F555W magnitude, (5) the observed (F439W - F555W) color, (6) the absolute F555W magnitude, (7) the intrinsic (F439W -- F555W) color, (8) the extinction $A_{\mbox{F555W}}$, (9) the distance modulus, and (10) the distance, in kpc, obtained for each star.  The weighted mean are $\langle DM \rangle = 12.98 \pm 0.03$, and $\langle d \rangle = 3.85 \pm 0.06$ kpc (the simple mean yields $\langle d \rangle = 4.20 \pm 0.06$ kpc). This weighted mean is taking into account uncertainties in the spectral/luminosity classification.  We exclude from this calculation stars \#664, 869, 889, which, as stated in the previous section, lack complete photometry.   This distance is substantially smaller than the estimate that would result if we were to assume the canonical $R_{V} = 3.1$.  If we were to assume such, along with $\langle A_{V} \rangle = 5.16$ (calculated from the $A_{V}$ values in R07), then the resulting distance modulus becomes $\langle DM \rangle = 14.246 \pm 0.038$  and $\langle d \rangle =7.07 \pm 0.12$ kpc, about twice the distance than when using $\langle R_{V} \rangle = 3.77$.

Figure~\ref{unred-cmd} shows the extinction- and distance-corrected Wd2 CMD in the HST filter system, illustrating the application of the results from Table~\ref{dis-table}, along with ZAMS, 1 -- 4, and 10 Myr isochrones.  Each star was individually dereddened, extinction corrected, and then shifted by the weighted mean $DM$. For easier comparison, key spectral types are labeled at the right edge of the figure.  Each star is labeled with its ID and spectral type.   The dispersion around the ZAMS is minimal, and very few stars have an observed spectral type inconsistent with the spectral type expected along the ZAMS isochrone.  Star \#889 (MSP~18) is not shown in the figure because of its odd position in the CMD after adopting the average $A_{\lambda}$ (overcorrection in color).  Since this star lies to the north of the main cluster, it is very likely that the average reddening determined for the main cluster is not applicable to this star.

The presence of close, unresolved binaries in a stellar population produces brighter apparent magnitudes than if no binaries were present.  Consequently, the derived distances are systematically low compared to the single-star case.  \citet{kiminki} modeled the light contribution from secondary components among massive binaries in Cygnus~OB2 and found that they contribute 16\% of the luminosity, on average.  This translates into a systematic distance error of 8\%, in the sense that the derived distances are too small unless binaries are taken into account.  We apply this statistical correction to find a distance of $4.16 \pm 0.07$ kpc.  

A possible source of systematic error in our distance measurement is the intrinsic colors adopted for massive stars.   There is a difference in the intrinsic colors of stars between the Padova isochrones and \citet{martins} colors.  The \citet{martins} $(B - V)_0$ colors are, on average, 0.04 mag redder than those that result from the Kurucz models or the photometric measurements of \citet{johnson} and \citet{fitzgerald}.  This difference implies a smaller color excess and, therefore, a smaller extinction and larger distance modulus.  \citet{martins} warn that a redder intrinsic color introduces a difference of $\Delta A_{V} \sim$0.124 mag assuming $R_{V} = 3.1$.  For our mean value of $R_{V} = 3.77$, the change in extinction required is $\Delta A_{V} = 0.15$, equivalent to a systematic distance error of $+ 0.26$ kpc.   We conclude that our measurement of the Wd2 distance is $4.16 \pm 0.06$ (random) $ + 0.26$ (systematic) kpc.

The distance that we determined favors a small value, like some in the literature, but not as small as the 2.8 kpc suggested by \citet{asc}.  \citet{asc} measured a distance by fitting deep \emph{JHK$_{s}$} CMDs to pre-main-sequence isochrones after correcting for reddening, assuming a single value for $A_{V}$ derived from the four brightest cluster stars.  Despite adopting an $A_{V}$ smaller than our preferred value, they obtain a smaller distance, which we attribute to uncertainties in the pre-main-sequence models and the inherent ambiguities of fitting such models to the data.  Furthermore, the distance calculation is based on a predetermined age for Wd2, which has not yet been properly constrained.  Our measurement, like that of \citet{brian}, falls in the middle of the $d =$ 2 -- 5 kpc constraints determined in the X-ray study of pre-main-sequence stars in Wd2 \citep{tsu}.  Our result is also very close to the 4.2 kpc kinematic distance adopted by \citet{churchwell} on the basis of 21-cm  absorption arguments \citep{caswell}.  Other kinematic estimates from the CO radial velocity of the associated molecular cloud \citep{dame, furu} are slightly larger, 5 -- 6 kpc --- still broadly consistent with our result but with larger uncertainties owing to the range of molecular cloud velocities along this sightline.  However, our measurement is inconsistent with distances larger than 5 kpc.  The main reason why our preferred distance is smaller than that found by \citet{piatti, carraro}, R05, R07, and R11 is that all of these works adopted a standard $R_{V} = 3.1$ which leads to a substantially larger distance.  Our adopted distance of 4.16 kpc places Wd2 comfortably within the Carina spiral arm and near the solar circle, about 40\% further away than where it is pictured in Figure~3 of the review of massive stellar clusters by \citet{review}.  

Due to the discrepancy between our transformed $UBVI$ photometry and that reported via ground-based observations, as attested in \S~\ref{photcomp}, we consider here what would happen if we brighten our flight system photometry by 0.2 magnitudes to achieve better agreement with those works.  When such is the case, we find that Wd2 has a slightly smaller reddening, $\langle R_{V} \rangle = 3.60 \pm 0.10$ and $\langle A_{V} \rangle = 6.23 \pm 0.39$.  This smaller $A_V$ more than offsets the 0.2 mag photometric correction and yields $\langle DM \rangle = 13.26 \pm 0.03$, and $\langle d \rangle = 4.38 \pm 0.07$ kpc.  After corrections for binary stars this becomes $4.73 \pm 0.08$ (random) $ + 0.26$ (systematic) kpc.  In summary, the photometric offset compared to other works has only a small impact on the final distance determination.   This alternative distance is still much smaller than the ``far'' distances in the literature, substantially larger than some of the ``near'' distances, and still broadly consistent with the kinematic distance estimates.

\subsection{Cluster Age} \label{age}

In Figure~\ref{unred-cmd} we used the values of Table~\ref{reddening-table} to deredden all the available photometry from Figure~\ref{cmdccd}, and we compared the data with 1 -- 4 Myr isochrones in order to constrain Wd2's age.  From the inspection it becomes evident that isochrones with ages 1 -- 3 Myr are degenerate, and it can be ``safely'' assumed the majority of the stars are dwarf class.  The 4 Myr isochrone shows that there should be a population of evolved giant stars that is absent in the data.  At the most we can conclude that Wd2 has an age between 1 and 3 Myr.  Other authors have determined the same age constraint on the basis of integrated spectra, CMD fitting, and the lack of evolved stars within the cluster \citep{piatti, asc}.  

We attempt here a different kind  of age constraint using the size of the cavity carved in the ISM by the winds of massive stars \citep{churchwell}.  \citet{weaver77} derives (Equation 51) the radius of an interstellar bubble as a function of time $t_{6}$ in Myr, ambient density $n_{0}$, and mechanical energy injection $L_{36}$, in units of $10^{36}$ erg s$^{-1}$ .  We invert this expression to find
\begin{equation}
t_{6} = \left[ \left( \frac{R_{2}(pc)}{27} \right)^{5} \frac{n_{0}}{ L_{36}} \right]^{1/3} Myr.
\end{equation}   
 From the angular size of the cavity in Figure~\ref{glimpsergb} \citep[see also][]{churchwell} we adopt a radius of $105\arcsec$ or 2.0 pc at the adopted distance.  We assume $n_{0} = 10^{5}$ cm$^{-3}$, typical of molecular clouds.   Based on the wind speeds and (highly uncertain) mass loss rates for O stars from \citet{mokiem07} we estimate a time-averaged  energy injection rate of  $L_{36} = 1.0$ from the ensemble of massive stars. The mechanical wind luminosity is dominated by the few most massive stars, so the exact number and type of stars later than about O5 is inconsequential.  The resulting expansion timescale of the (still ill-defined) interstellar cavity is $\sim$0.6 Myr, indicating extreme youth, consistent with the abundance of molecular material just outside the cluster core.   We caution that the \citet{weaver77} analysis was developed for interstellar bubbles in uniform media and densities more typical of the diffuse ISM.  The presence of a radio ``blister'' structure on the west side of the Wd2 \ion{H}{2} region \citep{whiteoak77} suggests that the uniform density assumption is not valid and that the bubble may be expanding asymmetrically into a region of lower density on that side.  The derived expansion timescale should be considered a lower limit.


\section{Conclusions}

We have used \emph{Hubble Space Telescope WFPC2} imaging to obtain photometry for 1136 stars in the Wd2 field, many of which are blends of multiple sources that are difficult to distinguish using ground-based observations.  With  \emph{Magellan} and \emph{SOAR} optical spectroscopy we identified fifteen new O-type stars, including two new probable binary systems: \#664 (O3 + O5.5) (MSP~188) and \#714 (O3 + ?).  The photometry and spectral type information allowed us to determine the reddening parameters $R_{V}$ and $A_{V}$ individually for the 15 new O stars and 14 additional O stars from the literature.  The mean $R_{V}$ of 3.77 is larger than the canonical Galactic value and leads to a larger inferred extinction and a smaller distance modulus compared to most previous studies.  Using Padova stellar isochrones we determined a new spectroscopic and photometric distance to Westerlund 2 of $\langle d \rangle = 4.16 \pm 0.06$ (random) $+ 0.26$ (systematic) kpc.   In archival \emph{Spitzer GLIMPSE} images we discovered a faint mid-IR PAH ring that surrounds \#889 (MSP~18) which seems to lie at the center of a secondary cluster located $\sim$55\arcsec\ to the north of the cluster core.

The distance revision of Wd2 calls into question whether the Wolf-Rayet star WR~20a and the TeV $\gamma$-ray sources HESS J1023--575 and HESS J1023--5746 are indeed cluster members or even located at a similar distance.  Our distance determination allows that the HESS objects could be cluster members, but they would be almost twice as luminous when placed at 4.16 kpc instead of 2.4 kpc \citep{ack}.  The distance of 4.16 kpc potentially leaves WR~20a in isolation and calls into question ideas about the formation of massive WR stars.  If WR~20a is at a distance of $\sim$8 kpc like \citet{rauw05, rauw} and \citet{rauw11} conclude, then, did it form in isolation, and its relative position to Wd2 is just coincidence?  It could be argued that neither the WR star nor the $\gamma$-ray sources belong to Wd2.  However, if we use R07's photometry for WR~20a in conjunction with our $R_{V}$ determination, then this star would lie at a distance of $\sim$4.44 kpc, well within the uncertainties of our new distance, thus making its association with Wd2 likely.  Our new distance determination also motivates a revision of Wd2's mass.  A distance of 4.16 kpc will revise its mass downward compared to estimates using larger distances and will make Wd2 less comparable to the more massive clusters of the Milky Way.


\acknowledgments

Acknowledgments: We thank Evan Skillman, Daniel Weiss, and Matthew Povich, Giovanni Carraro for helpful discussions throughout this work, and our referee, Anthony Moffat, whose comments improved this article.  Our deepest gratitude goes to Andrew Dolphin for his support and help in the \emph{WFPC2} data reduction.  The extended \emph{GLIMPSE} team contributed to the scientific motivation for an \emph{HST} study of Wd2.   This publication has made use of SAOImage DS9, developed by Smithsonian Astrophysical Observatory; and data products from the Two Micron All Sky Survey, which is a joint project of the University of Massachusetts and the Infrared Processing and Analysis Center/California Institute of Technology, funded by the National Aeronautics and Space Administration and the National Science Foundation.



\clearpage
\begin{deluxetable}{cccccc}
\tablecolumns{5}
\tabletypesize{\scriptsize}
\tablecaption{Observations\label{obs-pro}}
\tablewidth{0pt}
\tablehead{
\colhead{Position} & \colhead{Number} & \colhead{Filter} & \colhead{Exposure Time} & \colhead{UT} \\
\colhead{} & \colhead{of exposures} & \colhead{} & \colhead{(sec)} & \colhead{Date}
}
\startdata
1 & 3 & F336W & 160 & 2005-05-19 \\
1 & 3 & F439W &  40 & 2005-05-19 \\
1 & 3 & F555W &   8 & 2005-05-19 \\
1 & 3 & F814W &  20 & 2005-05-19 \\
2 & 3 & F336W &  50 & 2005-05-20 \\
2 & 3 & F439W &  12 & 2005-05-20 \\
2 & 3 & F555W &   2 & 2005-05-20 \\
2 & 3 & F814W &   2 & 2005-05-20 \\
\enddata
\end{deluxetable}  

\clearpage
\begin{landscape}
\begin{center}
\begin{deluxetable}{ccrrrrrrrrrrrrrrrc}
\tablecolumns{18}
\tabletypesize{\scriptsize}
\setlength{\tabcolsep}{0.04in} 
\tablecaption{Photometry\label{mach-tab}}
\tablewidth{0pt}
\tablehead{
\colhead{ID} & \colhead{MSP91} & \colhead{$\alpha$ (J2000 $\degr$)} & \colhead{$\delta$ (J2000 $\degr$)} & \colhead{F336W} & \colhead{$\sigma_{F336W}$} & \colhead{F439W} & \colhead{$\sigma_{F439W}$} & \colhead{F555W} & \colhead{$\sigma_{F555W}$} & \colhead{F814W} & \colhead{$\sigma_{F814W}$} & \colhead{U} & \colhead{B} & \colhead{V} & \colhead{I}  & \colhead{S/N} & \colhead{Source}
}
\startdata
     1	&	\nodata	&	155.93865 & -57.75719 & 99.999 & 0.000 & 99.999 & 0.000 & 17.621 & 0.023 & 16.585 & 0.009 & 99.999 & 99.999 & 17.597 & 16.547 & 147.5 & l \\
     2	&	\nodata	&	155.94239 & -57.75466 & 17.757 & 0.016 & 17.640 & 0.014 & 16.687 & 0.009 & 15.521 & 0.004 & 17.780 & 17.567 & 16.661 & 15.478 & 320.0 & l \\
     3	&	\nodata	&	155.94329 & -57.75762 & 99.999 & 0.000 & 99.999 & 0.000 & 20.203 & 0.075 & 17.438 & 0.010 & 99.999 & 99.999 & 20.270 & 17.461 & 109.1 & l \\
\nodata	&	\nodata	&	\nodata	&	\nodata	&	\nodata	&	\nodata	&	\nodata	&	\nodata	&	\nodata	&	\nodata	&	\nodata	&	\nodata	&	\nodata	&	\nodata	&	\nodata	&	\nodata	&	\nodata	&	\nodata	\\
  66	&	\nodata	&	155.96852 & -57.75353 & 99.999 & 0.000 & 99.999 & 0.000 & 99.999 & 0.000 & 20.894 & 0.094 & 99.999 & 99.999 & 99.999 & 99.999 &   11.4 & l \\
  67	&     273b 	&	155.96877 & -57.76713 & 18.633 & 0.018 & 18.534 & 0.016 & 17.529 & 0.013 & 16.454 & 0.006 & 18.656 & 18.462 & 17.503 & 16.411 & 231.7 & l \\
  68	&	\nodata	&	155.96880 & -57.75063 & 99.999 & 0.000 & 99.999 & 0.000 & 20.986 & 0.137 & 18.910 & 0.022 & 99.999 & 99.999 & 20.996 & 18.890 &   49.7 & l \\
  69	&	\nodata	&	155.96893 & -57.74199 & 99.999 & 0.000 & 99.999 & 0.000 & 99.999 & 0.000 & 21.443 & 0.111 & 99.999 & 99.999 & 99.999 & 99.999 &   10.2 & l \\
  70	&	\nodata	&	155.96899 & -57.75914 & 99.999 & 0.000 & 99.999 & 0.000 & 21.375 & 0.176 & 19.061 & 0.023 & 99.999 & 99.999 & 21.401 & 19.053 &   47.8 & l \\
  71	&	\nodata	&	155.96914 & -57.75185 & 99.999 & 0.000 & 99.999 & 0.000 & 99.999 & 0.000 & 20.238 & 0.055 & 99.999 & 99.999 & 99.999 & 99.999 &   19.9 & l \\
  72	&     273a 	&	155.96941 & -57.76712 & 15.801 & 0.004 & 15.118 & 0.004 & 99.999 & 0.000 & 99.999 & 0.000 & 15.438 & 15.059 & 99.999 & 99.999 & 429.6 & l \\
\nodata	&	\nodata	&	\nodata	&	\nodata	&	\nodata	&	\nodata	&	\nodata	&	\nodata	&	\nodata	&	\nodata	&	\nodata	&	\nodata	&	\nodata	&	\nodata	&	\nodata	&	\nodata	&	\nodata	&	\nodata	\\
1134 &	\nodata	&	156.05588 & -57.74563 & 18.930 & 0.105 & 18.966 & 0.127 & 18.005 & 0.081 & 16.861 & 0.077 & 18.991 & 18.895 & 17.979 & 16.819 &   42.2 & s \\
1135 &	\nodata	&	156.05922 & -57.74585 & 15.547 & 0.030 & 15.330 & 0.022 & 14.785 & 0.030 & 14.082 & 0.024 & 15.548 & 15.298 & 14.761 & 14.045 & 264.7 & s \\
1136 &	\nodata	&	156.05973 & -57.74560 & 19.166 & 0.223 & 99.999 & 0.000 & 99.999 & 0.000 & 18.597 & 0.202 & 19.362 & 99.999 & 99.999 & 18.560 &   10.1 & s \\
\enddata
\tablecomments{Table \ref{mach-tab} is published in its entirety in the electronic edition of this journal.  A portion is shown here for guidance regarding its form and content.}
\end{deluxetable}  
\end{center}
\end{landscape}

\clearpage
\begin{center}
\begin{deluxetable}{rrcccccccc}
\tablecolumns{10}
\tabletypesize{\scriptsize}
\tablecaption{New Spectral Classification\label{spec-tab}}
\tablewidth{0pt}
\tablehead{
\colhead{ID} & \colhead{MSP91} & \colhead{V} & \colhead{B - V} & \colhead{EW($\lambda$5410)} & \colhead{EW($\lambda$5876)} & \colhead{EW(5410)/EW(5876)} & \colhead{T} & \colhead{Spectral} & \colhead{Comment} \\
\colhead{} & \colhead{} &\colhead{(mag)} & \colhead{(mag)} & \colhead{(\AA)} & \colhead{(\AA)} & \colhead{} &\colhead{(K)} & \colhead{Type} &\colhead{}
}
\startdata
\cutinhead{Stars with only absorption lines}
  137				&	165		&	15.591	&	1.660	&	0.97	$\pm$	0.09	&	0.30	$\pm$	0.07	&	3.20	$\pm$	0.79	&	43000	$\pm$	2000	&	O4		&	S	\\
  505\tablenotemark{a}	&	196a 	&	16.094	&	1.325	&	0.33	$\pm$	0.11	&	0.37	$\pm$	0.09	&	0.89	$\pm$	0.36	&	34000	$\pm$	2000	&	O8.5	&	S	\\
  528\tablenotemark{a}	&	229a	 	&	15.841	&	1.515	&	0.55	$\pm$	0.10	&	0.53	$\pm$	0.09	&	1.05	$\pm$	0.26	&	35000	$\pm$	2000	&	O8		&	S	\\
  548				&	151		&	14.522	&	1.409	&	0.94	$\pm$	0.04	&	0.31	$\pm$	0.06	&	3.00	$\pm$	0.58	&	43000	$\pm$	1000	&	O4		&	S	\\
  584				&	157b	&	15.442	&	1.346	&	0.87	$\pm$	0.06	&	0.79	$\pm$	0.05	&	1.10	$\pm$	0.10	&	35000	$\pm$	  700	&	O8		&	M	\\
  640				&	233		&	16.234	&	1.312	&	0.66	$\pm$	0.12	&	1.05	$\pm$	0.11	&	0.63	$\pm$	0.13	&	32000	$\pm$	  800	&	O9.5	&	M	\\
  664a				&	188a	 	&	13.464	&	1.538	&	0.12	$\pm$	0.01	&	0.02	$\pm$	0.01	&	5.70	$\pm$	3.32	&	48000	$\pm$	4000	&	O3		&	M	\\
  664b				&	188a 	&	13.464	&	1.538	&	0.05	$\pm$	0.01	&	0.02	$\pm$	0.01	&	2.04	$\pm$	1.18	&	40000	$\pm$	4000	&	O5.5	&	M	\\
  704				&	175a		&	14.059	&	1.354	&	0.95	$\pm$	0.06	&	0.15	$\pm$	0.05	&	6.24	$\pm$	1.92	&	48000	$\pm$	2000	&	O4		&	M	\\
  769				&	219		&	16.576	&	1.445	&	0.42	$\pm$	0.12	&	0.85	$\pm$	0.05	&	0.50	$\pm$	0.14	&	31000	$\pm$	  800	&	O9.5	&	M	\\
  771				&	167b	&	15.501	&	1.433	&	0.63	$\pm$	0.06	&	0.63	$\pm$	0.06	&	1.00	$\pm$	0.13	&	35000	$\pm$	  900	&	O8		&	M	\\
  857				&	444a		&	13.869	&	1.410	&	1.13	$\pm$	0.04	&	0.45	$\pm$	0.05	&	2.52	$\pm$	0.29	&	42000	$\pm$	  800	&	O4.5	&	M	\\
  879				&	235		&	16.645	&	1.469	&	0.59	$\pm$	0.12	&	0.77	$\pm$	0.09	&	0.77	$\pm$	0.18	&	33000	$\pm$	1000	&	O9.5	&	M	\\
  896\tablenotemark{a}	&	183c		&	16.246	&	1.438	&	0.66	$\pm$	0.08	&	0.80	$\pm$	0.07	&	0.83	$\pm$	0.13	&	34000	$\pm$	  900	&	O8.5	&	M	\\
  903				&	183b	&	15.789	&	1.409	&	0.77	$\pm$	0.10	&	0.85	$\pm$	0.08	&	0.90	$\pm$	0.14	&	34000	$\pm$	1000	&	O8.5	&	M	\\
1004				&	32		&	15.323	&	1.325	&	0.53	$\pm$	0.06	&	0.74	$\pm$	0.05	&	0.72	$\pm$	0.09	&	33000	$\pm$	  600	&	O9.5	&	S	\\
\cutinhead{Stars believed to be late O/early B}
  260\tablenotemark{a,b}	&	201		&	16.920	&	1.324	&		$<$	-0.15	&	0.43	$\pm$	0.12	&	\nodata	&	\nodata	&	\nodata	&	S	\\
  547				&	28		&	17.383	&	1.298	&		$<$   0.19	&	0.79	$\pm$	0.14	&	\nodata	&	\nodata	&	\nodata	&	S	\\
  549				&	44a		&	15.562	&	1.300	&		$<$   0.04	&	0.70	$\pm$	0.29	&	\nodata	&	\nodata	&	\nodata	&	S	\\
  556				&	\nodata	&	17.124	&	0.900	&		$<$	-0.21	&	0.53	$\pm$	0.11	&	\nodata	&	\nodata	&	\nodata	&	S	\\
  620				&	96a		&	16.086	&	1.279	&		$<$	-0.07	&	0.18	$\pm$	0.06	&	\nodata	&	\nodata	&	\nodata	&	S	\\
  878				&	\nodata	&	16.615	&	1.291	&	 	$<$   0.85	&	2.17	$\pm$	0.94	&	\nodata	&	\nodata	&	\nodata	&	M	\\
\cutinhead{Field Stars}
1028				&	\nodata	&	14.355	&	0.373	&	0.15	$\pm$	0.08	&	-0.31	$\pm$	0.09	&	\nodata	&	\nodata	&	A	&	S	\\
\cutinhead{Stars unable to be identified}
  463\tablenotemark{a,b}	&	125a		&	18.982	&	0.871	&	\nodata	&			\nodata		&	\nodata	&	\nodata	&	\nodata	&	S	\\
1012\tablenotemark{b}	&	24		&	15.601	&	1.225	&	\nodata	&			\nodata		&	\nodata	&	\nodata	&	\nodata	&	S	\\
1026\tablenotemark{b}	&	20a		&	16.784	&	1.289	&	$<$	0.19	&			\nodata		&	\nodata	&	\nodata	&	\nodata	&	S	\\
1031\tablenotemark{b,c}	&	\nodata	&	15.308	&	1.257	&	$<$	0.11	&	-0.93	$\pm$	0.07	&	\nodata	&	\nodata	&	\nodata	&	S	\\
\enddata
\tablecomments{Comment column lists the source of the spectroscopic data, {\it M} stands for the Magellan Telescope while the {\it S} stands for the SOAR Telescope.  Negative values of EW indicate an emission line.}
\tablenotetext{a}{This spectrum also exhibits H$\alpha$ emission.}
\tablenotetext{b}{This spectrum is a blend of two or more stars. The labeled one is the dominant source on the slit.}
\tablenotetext{c}{This source exhibits a \ion{Mg}{2} absorption line.}
\end{deluxetable}  
\end{center}

\clearpage
\begin{landscape}
\begin{center}
\begin{deluxetable}{rrrrrrrrrrrrrr}
\tablecolumns{13}
\tabletypesize{\scriptsize}
\setlength{\tabcolsep}{0.04in} 
\tablecaption{Reddening Summary\label{reddening-compilation}}
\tablewidth{0pt}
\tablehead{
\colhead{ID} & \colhead{Spectral Type} & \multicolumn{5}{c}{$R_{V}$ (mag)} & \colhead{} & \multicolumn{5}{c}{$A_{V}$ (mag)} \\
\cline{3-7} \cline{9-13} \\
\colhead{} & \colhead{} & \colhead{CCM89} & \colhead{F04} & \colhead{FM07} & \colhead{FM09} & \colhead{Average} & \colhead{} & \colhead{CCM89} & \colhead{F04} & \colhead{FM07} & \colhead{FM09} & \colhead{Average}
}
\startdata
  137 &   O4V &    3.76 $\pm$    0.09 &    3.99 $\pm$    0.07 &    3.84 $\pm$    0.07 &    3.79 $\pm$    0.43 &    3.85 $\pm$    0.10 & &    7.55 $\pm$    0.09 &    7.51 $\pm$    0.21 &    7.41 $\pm$    0.22 &    7.49 $\pm$    0.86 &    7.49 $\pm$    0.06 \\ 
  178 &   O4V &    3.74 $\pm$    0.03 &    3.95 $\pm$    0.02 &    3.93 $\pm$    0.03 &    3.79 $\pm$    0.17 &    3.85 $\pm$    0.10 & &    6.40 $\pm$    0.03 &    6.35 $\pm$    0.05 &    6.38 $\pm$    0.07 &    6.37 $\pm$    0.29 &    6.37 $\pm$    0.02 \\ 
  395 &   O7.5V &    3.73 $\pm$    0.03 &    3.96 $\pm$    0.02 &    3.77 $\pm$    0.03 &    4.12 $\pm$    0.33 &    3.89 $\pm$    0.18 & &    7.06 $\pm$    0.03 &    7.03 $\pm$    0.06 &    6.92 $\pm$    0.07 &    7.19 $\pm$    0.91 &    7.05 $\pm$    0.11 \\ 
  505 &   O8.5V &    3.62 $\pm$    0.06 &    3.86 $\pm$    0.05 &    3.71 $\pm$    0.06 &    4.01 $\pm$    0.35 &    3.80 $\pm$    0.17 & &    6.34 $\pm$    0.06 &    6.33 $\pm$    0.12 &    6.36 $\pm$    0.14 &    6.46 $\pm$    0.57 &    6.38 $\pm$    0.06 \\ 
  528 &   O8V &    3.81 $\pm$    0.06 &    4.04 $\pm$    0.05 &    3.99 $\pm$    0.05 &    3.87 $\pm$    0.32 &    3.93 $\pm$    0.11 & &    7.00 $\pm$    0.06 &    6.97 $\pm$    0.13 &    6.97 $\pm$    0.14 &    6.99 $\pm$    0.58 &    6.98 $\pm$    0.02 \\ 
  548 &   O4V &    3.64 $\pm$    0.05 &    3.90 $\pm$    0.04 &    3.76 $\pm$    0.04 &    3.73 $\pm$    0.25 &    3.76 $\pm$    0.10 & &    6.46 $\pm$    0.04 &    6.46 $\pm$    0.09 &    6.48 $\pm$    0.10 &    6.43 $\pm$    0.43 &    6.46 $\pm$    0.02 \\ 
  549\tablenotemark{a} &   B1.0V &    4.01 $\pm$    0.05 &    4.32 $\pm$    0.03 &    4.01 $\pm$    0.04 &    3.91 $\pm$    0.22 &    4.06 $\pm$    0.18 & &    6.12 $\pm$    0.04 &    6.11 $\pm$    0.07 &    6.09 $\pm$    0.08 &    6.03 $\pm$    0.35 &    6.09 $\pm$    0.04 \\ 
  584 &   O8V &    3.69 $\pm$    0.02 &    3.94 $\pm$    0.01 &    3.73 $\pm$    0.02 &    3.84 $\pm$    0.13 &    3.80 $\pm$    0.11 & &    6.22 $\pm$    0.02 &    6.22 $\pm$    0.04 &    6.19 $\pm$    0.05 &    6.29 $\pm$    0.21 &    6.23 $\pm$    0.04 \\ 
  597 &   O6.5V &    3.70 $\pm$    0.04 &    3.95 $\pm$    0.03 &    3.76 $\pm$    0.03 &    3.80 $\pm$    0.20 &    3.80 $\pm$    0.11 & &    6.41 $\pm$    0.03 &    6.40 $\pm$    0.07 &    6.39 $\pm$    0.08 &    6.42 $\pm$    0.34 &    6.40 $\pm$    0.01 \\ 
  620 &   B1V &    3.70 $\pm$    0.04 &    3.93 $\pm$    0.03 &    3.82 $\pm$    0.04 &    3.76 $\pm$    0.23 &    3.80 $\pm$    0.10 & &    5.76 $\pm$    0.03 &    5.74 $\pm$    0.07 &    5.77 $\pm$    0.08 &    5.71 $\pm$    0.35 &    5.75 $\pm$    0.03 \\ 
  640 &   O9.5V &    3.64 $\pm$    0.02 &    3.87 $\pm$    0.01 &    3.73 $\pm$    0.02 &    4.12 $\pm$    0.15 &    3.84 $\pm$    0.21 & &    6.38 $\pm$    0.02 &    6.35 $\pm$    0.04 &    6.37 $\pm$    0.05 &    6.55 $\pm$    0.25 &    6.41 $\pm$    0.09 \\ 
  704 &   O4V &    3.68 $\pm$    0.14 &    3.92 $\pm$    0.11 &    3.76 $\pm$    0.12 &    3.75 $\pm$    0.72 &    3.78 $\pm$    0.10 & &    6.27 $\pm$    0.12 &    6.27 $\pm$    0.27 &    6.27 $\pm$    0.29 &    6.28 $\pm$    1.20 &    6.27 $\pm$    0.01 \\ 
  714 &   O3V &    3.64 $\pm$    0.06 &    3.83 $\pm$    0.04 &    3.73 $\pm$    0.05 &    3.61 $\pm$    0.29 &    3.70 $\pm$    0.10 & &    6.08 $\pm$    0.05 &    6.04 $\pm$    0.10 &    6.08 $\pm$    0.11 &    6.01 $\pm$    0.49 &    6.05 $\pm$    0.04 \\ 
  722 &   O6V &    3.59 $\pm$    0.02 &    3.79 $\pm$    0.01 &    3.65 $\pm$    0.01 &    3.91 $\pm$    0.22 &    3.74 $\pm$    0.14 & &    7.24 $\pm$    0.02 &    7.19 $\pm$    0.04 &    7.23 $\pm$    0.04 &    7.32 $\pm$    0.71 &    7.25 $\pm$    0.05 \\ 
  738 &   O5.5V &    3.62 $\pm$    0.04 &    3.85 $\pm$    0.03 &    3.73 $\pm$    0.04 &    3.71 $\pm$    0.22 &    3.73 $\pm$    0.09 & &    6.03 $\pm$    0.04 &    6.01 $\pm$    0.07 &    6.02 $\pm$    0.08 &    6.05 $\pm$    0.36 &    6.03 $\pm$    0.02 \\ 
  769 &   O9.5V &    3.62 $\pm$    0.02 &    3.84 $\pm$    0.01 &    3.65 $\pm$    0.02 &    3.94 $\pm$    0.14 &    3.76 $\pm$    0.15 & &    6.66 $\pm$    0.02 &    6.63 $\pm$    0.04 &    6.63 $\pm$    0.06 &    6.76 $\pm$    0.29 &    6.67 $\pm$    0.06 \\ 
  771 &   O8V &    3.72 $\pm$    0.03 &    3.94 $\pm$    0.01 &    3.80 $\pm$    0.03 &    3.86 $\pm$    0.15 &    3.83 $\pm$    0.09 & &    6.64 $\pm$    0.03 &    6.61 $\pm$    0.05 &    6.60 $\pm$    0.06 &    6.65 $\pm$    0.27 &    6.62 $\pm$    0.03 \\ 
  804 &   O6V &    3.58 $\pm$    0.02 &    3.79 $\pm$    0.01 &    3.71 $\pm$    0.01 &    3.66 $\pm$    0.09 &    3.69 $\pm$    0.09 & &    6.84 $\pm$    0.02 &    6.81 $\pm$    0.03 &    6.91 $\pm$    0.04 &    6.82 $\pm$    0.17 &    6.85 $\pm$    0.05 \\ 
  826 &   O9.5V &    3.60 $\pm$    0.03 &    3.77 $\pm$    0.02 &    3.67 $\pm$    0.03 &    3.89 $\pm$    0.17 &    3.73 $\pm$    0.12 & &    6.54 $\pm$    0.03 &    6.45 $\pm$    0.06 &    6.49 $\pm$    0.07 &    6.55 $\pm$    0.32 &    6.51 $\pm$    0.04 \\ 
  843\tablenotemark{b} &   O4.5V &    3.53 $\pm$    0.05 &    3.79 $\pm$    0.04 &    3.64 $\pm$    0.04 &    3.70 $\pm$    0.24 &    3.67 $\pm$    0.11 & &    6.32 $\pm$    0.05 &    6.33 $\pm$    0.10 &    6.32 $\pm$    0.11 &    6.42 $\pm$    0.41 &    6.35 $\pm$    0.05 \\ 
  857 &   O4.5V &    3.47 $\pm$    0.03 &    3.65 $\pm$    0.02 &    3.63 $\pm$    0.03 &    3.55 $\pm$    0.19 &    3.58 $\pm$    0.08 & &    6.14 $\pm$    0.03 &    6.11 $\pm$    0.07 &    6.13 $\pm$    0.08 &    6.11 $\pm$    0.33 &    6.12 $\pm$    0.02 \\ 
  879 &   O9.5V &    3.58 $\pm$    0.03 &    3.79 $\pm$    0.02 &    3.70 $\pm$    0.03 &    4.02 $\pm$    0.35 &    3.77 $\pm$    0.19 & &    6.91 $\pm$    0.03 &    6.87 $\pm$    0.06 &    6.98 $\pm$    0.07 &    7.04 $\pm$    0.99 &    6.95 $\pm$    0.07 \\ 
  896 &   O8.5V &    3.55 $\pm$    0.03 &    3.76 $\pm$    0.01 &    3.69 $\pm$    0.02 &    4.00 $\pm$    0.15 &    3.75 $\pm$    0.19 & &    6.60 $\pm$    0.03 &    6.58 $\pm$    0.05 &    6.58 $\pm$    0.06 &    6.78 $\pm$    0.27 &    6.64 $\pm$    0.10 \\ 
  903 &   O8.5V &    3.58 $\pm$    0.03 &    3.79 $\pm$    0.02 &    3.67 $\pm$    0.03 &    3.97 $\pm$    0.17 &    3.75 $\pm$    0.17 & &    6.71 $\pm$    0.03 &    6.68 $\pm$    0.06 &    6.72 $\pm$    0.07 &    6.85 $\pm$    0.31 &    6.74 $\pm$    0.08 \\ 
  903 &   O6.5V &    3.56 $\pm$    0.05 &    3.79 $\pm$    0.04 &    3.66 $\pm$    0.04 &    3.95 $\pm$    0.27 &    3.74 $\pm$    0.17 & &    6.73 $\pm$    0.05 &    6.72 $\pm$    0.11 &    6.78 $\pm$    0.12 &    6.87 $\pm$    0.49 &    6.77 $\pm$    0.07 \\ 
  906 &   O6.5V &    3.53 $\pm$    0.05 &    3.74 $\pm$    0.04 &    3.64 $\pm$    0.05 &    3.99 $\pm$    0.28 &    3.72 $\pm$    0.19 & &    6.56 $\pm$    0.05 &    6.54 $\pm$    0.11 &    6.51 $\pm$    0.12 &    6.74 $\pm$    0.49 &    6.59 $\pm$    0.11 \\ 
  913 &   O3.5V &    3.56 $\pm$    0.05 &    3.80 $\pm$    0.04 &    3.66 $\pm$    0.04 &    3.61 $\pm$    0.26 &    3.65 $\pm$    0.10 & &    6.41 $\pm$    0.05 &    6.40 $\pm$    0.10 &    6.42 $\pm$    0.11 &    6.36 $\pm$    0.46 &    6.40 $\pm$    0.03 \\ 
  924 &   O8V &    3.51 $\pm$    0.03 &    3.74 $\pm$    0.02 &    3.60 $\pm$    0.03 &    3.88 $\pm$    0.17 &    3.68 $\pm$    0.16 & &    6.47 $\pm$    0.03 &    6.47 $\pm$    0.06 &    6.40 $\pm$    0.07 &    6.59 $\pm$    0.30 &    6.48 $\pm$    0.08 \\ 
 1004\tablenotemark{a} &   O9.5V &    3.65 $\pm$    0.02 &    3.83 $\pm$    0.01 &    3.79 $\pm$    0.02 &    3.62 $\pm$    0.13 &    3.72 $\pm$    0.10 & &    6.00 $\pm$    0.02 &    5.92 $\pm$    0.03 &    5.94 $\pm$    0.04 &    5.83 $\pm$    0.21 &    5.92 $\pm$    0.07 \\ 
 1039 &   O4.5V &    3.44 $\pm$    0.04 &    3.63 $\pm$    0.03 &    3.47 $\pm$    0.04 &    3.69 $\pm$    0.31 &    3.56 $\pm$    0.12 & &    6.43 $\pm$    0.04 &    6.39 $\pm$    0.09 &    6.42 $\pm$    0.10 &    6.42 $\pm$    0.74 &    6.42 $\pm$    0.02 \\ 
\cutinhead{}
      &         &    3.64 $\pm$    0.11 &    3.86 $\pm$    0.13 &    3.73 $\pm$    0.11 &    3.84 $\pm$    0.15 &    3.77 $\pm$    0.09 & &    6.51 $\pm$    0.38 &    6.48 $\pm$    0.37 &    6.49 $\pm$    0.37 &    6.54 $\pm$    0.42 &    6.51 $\pm$    0.38 \\ 
\enddata
\tablenotetext{a}{The IR photometry of these stars comes from the \emph{2MASS PSC} instead of \citet{asc}.}
\tablenotetext{b}{This star lacks $J$ photometry.}
\end{deluxetable}  
\end{center}

\clearpage
\begin{center}
\begin{deluxetable}{rrrrrrcrrrrcrrrrcrrrr}
\tablecolumns{21}
\tabletypesize{\scriptsize}
\setlength{\tabcolsep}{0.04in} 
\tablecaption{Reddening Table for the \emph{HST} Filters\label{reddening-table}}
\tablewidth{0pt}
\tablehead{
\colhead{ID}  & \colhead{Spectral Type}  & \multicolumn{4}{c}{$A_{\mbox{F336W}}$ (mag)}  &  \colhead{} &\multicolumn{4}{c}{$A_{\mbox{F439W}}$ (mag)}  & \colhead{} & \multicolumn{4}{c}{$A_{\mbox{F555W}}$ (mag)}  & \colhead{} & \multicolumn{4}{c}{$A_{\mbox{F814W}}$ (mag)}\\
\cline{3-6} \cline{8-11} \cline{13-16} \cline{18-21} \\
\colhead{}  & \colhead{}  & \colhead{CCM89}  & \colhead{F04}  & \colhead{FM07}  & \colhead{Average}  & \colhead{} & \colhead{CCM89}  & \colhead{F04}  & \colhead{FM07}  & \colhead{Average}  & \colhead{} & \colhead{CCM89}  & \colhead{F04}  & \colhead{FM07}  & \colhead{Average}  & \colhead{} & \colhead{CCM89}  & \colhead{F04}  & \colhead{FM07}  & \colhead{Average}
}
\startdata
  137 &   O4.0V &   11.36 $\pm$    0.03 &   11.17 $\pm$    0.17 &   11.24 $\pm$    0.18 &   11.35 $\pm$    0.03 & &    9.80 $\pm$    0.03 &    9.53 $\pm$    0.17 &    9.54 $\pm$    0.18 &    9.79 $\pm$    0.03 & &    7.68 $\pm$    0.02 &    7.56 $\pm$    0.16 &    7.54 $\pm$    0.17 &    7.68 $\pm$    0.02 & &    4.82 $\pm$    0.01 &    4.57 $\pm$    0.14 &    4.60 $\pm$    0.15 &    4.82 $\pm$    0.01 \\ 
  178 &   O4.0V &    9.65 $\pm$    0.03 &    9.51 $\pm$    0.05 &    9.53 $\pm$    0.06 &    9.61 $\pm$    0.02 & &    8.32 $\pm$    0.03 &    8.10 $\pm$    0.05 &    8.09 $\pm$    0.06 &    8.25 $\pm$    0.02 & &    6.52 $\pm$    0.02 &    6.41 $\pm$    0.04 &    6.43 $\pm$    0.06 &    6.49 $\pm$    0.02 & &    4.09 $\pm$    0.01 &    3.86 $\pm$    0.04 &    3.89 $\pm$    0.05 &    4.05 $\pm$    0.01 \\ 
  395 &   O7.5V &   10.67 $\pm$    0.03 &   10.51 $\pm$    0.05 &   10.45 $\pm$    0.06 &   10.61 $\pm$    0.02 & &    9.18 $\pm$    0.03 &    8.95 $\pm$    0.05 &    8.94 $\pm$    0.06 &    9.11 $\pm$    0.02 & &    7.18 $\pm$    0.02 &    7.09 $\pm$    0.05 &    7.05 $\pm$    0.06 &    7.16 $\pm$    0.02 & &    4.50 $\pm$    0.01 &    4.27 $\pm$    0.04 &    4.20 $\pm$    0.05 &    4.46 $\pm$    0.01 \\ 
  505 &   O8.5V &    9.70 $\pm$    0.03 &    9.55 $\pm$    0.11 &    9.57 $\pm$    0.12 &    9.69 $\pm$    0.03 & &    8.31 $\pm$    0.03 &    8.10 $\pm$    0.11 &    8.12 $\pm$    0.12 &    8.29 $\pm$    0.03 & &    6.46 $\pm$    0.02 &    6.38 $\pm$    0.10 &    6.38 $\pm$    0.11 &    6.45 $\pm$    0.02 & &    4.02 $\pm$    0.01 &    3.81 $\pm$    0.09 &    3.86 $\pm$    0.10 &    4.01 $\pm$    0.01 \\ 
  528 &   O8.0V &   10.49 $\pm$    0.03 &   10.35 $\pm$    0.11 &   10.36 $\pm$    0.12 &   10.48 $\pm$    0.03 & &    9.07 $\pm$    0.03 &    8.84 $\pm$    0.11 &    8.82 $\pm$    0.12 &    9.05 $\pm$    0.03 & &    7.13 $\pm$    0.02 &    7.03 $\pm$    0.10 &    7.03 $\pm$    0.11 &    7.12 $\pm$    0.02 & &    4.49 $\pm$    0.01 &    4.27 $\pm$    0.09 &    4.31 $\pm$    0.10 &    4.48 $\pm$    0.01 \\ 
  548 &   O4.0V &    9.86 $\pm$    0.03 &    9.69 $\pm$    0.08 &    9.74 $\pm$    0.09 &    9.83 $\pm$    0.02 & &    8.45 $\pm$    0.03 &    8.23 $\pm$    0.08 &    8.24 $\pm$    0.09 &    8.42 $\pm$    0.02 & &    6.58 $\pm$    0.02 &    6.50 $\pm$    0.08 &    6.49 $\pm$    0.09 &    6.57 $\pm$    0.02 & &    4.10 $\pm$    0.01 &    3.89 $\pm$    0.07 &    3.96 $\pm$    0.08 &    4.09 $\pm$    0.01 \\ 
  549\tablenotemark{a} &   B1.0V &    8.96 $\pm$    0.03 &    8.84 $\pm$    0.07 &    8.99 $\pm$    0.07 &    8.95 $\pm$    0.02 & &    7.83 $\pm$    0.03 &    7.63 $\pm$    0.07 &    7.67 $\pm$    0.07 &    7.79 $\pm$    0.02 & &    6.23 $\pm$    0.02 &    6.15 $\pm$    0.06 &    6.11 $\pm$    0.07 &    6.22 $\pm$    0.02 & &    3.98 $\pm$    0.01 &    3.82 $\pm$    0.05 &    3.96 $\pm$    0.06 &    3.97 $\pm$    0.01 \\ 
  584 &   O8.0V &    9.43 $\pm$    0.03 &    9.28 $\pm$    0.03 &    9.34 $\pm$    0.04 &    9.37 $\pm$    0.02 & &    8.11 $\pm$    0.03 &    7.89 $\pm$    0.03 &    7.91 $\pm$    0.04 &    8.00 $\pm$    0.02 & &    6.33 $\pm$    0.02 &    6.24 $\pm$    0.03 &    6.23 $\pm$    0.04 &    6.29 $\pm$    0.01 & &    3.96 $\pm$    0.01 &    3.75 $\pm$    0.02 &    3.80 $\pm$    0.04 &    3.89 $\pm$    0.01 \\ 
  597 &   O6.5V &    9.71 $\pm$    0.03 &    9.54 $\pm$    0.06 &    9.60 $\pm$    0.07 &    9.67 $\pm$    0.02 & &    8.35 $\pm$    0.03 &    8.12 $\pm$    0.06 &    8.14 $\pm$    0.07 &    8.29 $\pm$    0.02 & &    6.52 $\pm$    0.02 &    6.43 $\pm$    0.06 &    6.41 $\pm$    0.07 &    6.51 $\pm$    0.02 & &    4.08 $\pm$    0.01 &    3.87 $\pm$    0.05 &    3.91 $\pm$    0.06 &    4.06 $\pm$    0.01 \\ 
  620 &   B1.0V &    8.73 $\pm$    0.03 &    8.59 $\pm$    0.06 &    8.62 $\pm$    0.08 &    8.70 $\pm$    0.02 & &    7.51 $\pm$    0.03 &    7.30 $\pm$    0.06 &    7.30 $\pm$    0.08 &    7.46 $\pm$    0.02 & &    5.87 $\pm$    0.02 &    5.77 $\pm$    0.06 &    5.77 $\pm$    0.07 &    5.85 $\pm$    0.02 & &    3.67 $\pm$    0.01 &    3.47 $\pm$    0.05 &    3.49 $\pm$    0.06 &    3.65 $\pm$    0.01 \\ 
  640 &   O9.5V &    9.74 $\pm$    0.03 &    9.58 $\pm$    0.03 &    9.61 $\pm$    0.05 &    9.67 $\pm$    0.02 & &    8.35 $\pm$    0.03 &    8.13 $\pm$    0.03 &    8.13 $\pm$    0.05 &    8.24 $\pm$    0.02 & &    6.50 $\pm$    0.02 &    6.40 $\pm$    0.03 &    6.40 $\pm$    0.04 &    6.46 $\pm$    0.02 & &    4.05 $\pm$    0.01 &    3.82 $\pm$    0.02 &    3.87 $\pm$    0.04 &    3.98 $\pm$    0.01 \\ 
  704 &   O4.0V &    9.52 $\pm$    0.03 &    9.38 $\pm$    0.24 &    9.41 $\pm$    0.27 &    9.52 $\pm$    0.03 & &    8.18 $\pm$    0.03 &    7.97 $\pm$    0.24 &    7.98 $\pm$    0.27 &    8.17 $\pm$    0.03 & &    6.38 $\pm$    0.02 &    6.30 $\pm$    0.22 &    6.28 $\pm$    0.25 &    6.38 $\pm$    0.02 & &    3.99 $\pm$    0.01 &    3.78 $\pm$    0.20 &    3.83 $\pm$    0.22 &    3.99 $\pm$    0.01 \\ 
  714 &   O3.0V &    9.28 $\pm$    0.03 &    9.13 $\pm$    0.10 &    9.17 $\pm$    0.11 &    9.26 $\pm$    0.02 & &    7.96 $\pm$    0.03 &    7.73 $\pm$    0.10 &    7.76 $\pm$    0.11 &    7.93 $\pm$    0.02 & &    6.19 $\pm$    0.02 &    6.08 $\pm$    0.09 &    6.11 $\pm$    0.10 &    6.18 $\pm$    0.02 & &    3.85 $\pm$    0.01 &    3.62 $\pm$    0.08 &    3.72 $\pm$    0.09 &    3.85 $\pm$    0.01 \\ 
  722 &   O6.0V &   11.11 $\pm$    0.03 &   10.93 $\pm$    0.03 &   10.93 $\pm$    0.04 &   11.00 $\pm$    0.02 & &    9.50 $\pm$    0.03 &    9.25 $\pm$    0.03 &    9.26 $\pm$    0.04 &    9.35 $\pm$    0.02 & &    7.37 $\pm$    0.02 &    7.26 $\pm$    0.02 &    7.25 $\pm$    0.03 &    7.31 $\pm$    0.01 & &    4.57 $\pm$    0.01 &    4.30 $\pm$    0.02 &    4.36 $\pm$    0.03 &    4.45 $\pm$    0.01 \\ 
  738 &   O5.5V &    9.21 $\pm$    0.03 &    9.05 $\pm$    0.07 &    9.09 $\pm$    0.08 &    9.18 $\pm$    0.02 & &    7.89 $\pm$    0.03 &    7.68 $\pm$    0.07 &    7.69 $\pm$    0.08 &    7.84 $\pm$    0.02 & &    6.13 $\pm$    0.02 &    6.04 $\pm$    0.06 &    6.05 $\pm$    0.07 &    6.12 $\pm$    0.02 & &    3.82 $\pm$    0.01 &    3.60 $\pm$    0.05 &    3.66 $\pm$    0.06 &    3.80 $\pm$    0.01 \\ 
  769 &   O9.5V &   10.17 $\pm$    0.03 &   10.00 $\pm$    0.04 &   10.04 $\pm$    0.05 &   10.10 $\pm$    0.02 & &    8.71 $\pm$    0.03 &    8.47 $\pm$    0.04 &    8.51 $\pm$    0.05 &    8.60 $\pm$    0.02 & &    6.77 $\pm$    0.02 &    6.66 $\pm$    0.03 &    6.66 $\pm$    0.04 &    6.73 $\pm$    0.02 & &    4.21 $\pm$    0.01 &    3.97 $\pm$    0.02 &    4.02 $\pm$    0.04 &    4.14 $\pm$    0.01 \\ 
  771 &   O8.0V &   10.04 $\pm$    0.03 &    9.89 $\pm$    0.04 &    9.92 $\pm$    0.06 &    9.98 $\pm$    0.02 & &    8.64 $\pm$    0.03 &    8.42 $\pm$    0.04 &    8.42 $\pm$    0.06 &    8.56 $\pm$    0.02 & &    6.76 $\pm$    0.02 &    6.66 $\pm$    0.04 &    6.64 $\pm$    0.05 &    6.73 $\pm$    0.02 & &    4.23 $\pm$    0.01 &    4.01 $\pm$    0.03 &    4.05 $\pm$    0.05 &    4.19 $\pm$    0.01 \\ 
  804 &   O6.0V &   10.50 $\pm$    0.03 &   10.34 $\pm$    0.02 &   10.40 $\pm$    0.04 &   10.41 $\pm$    0.02 & &    8.98 $\pm$    0.03 &    8.75 $\pm$    0.02 &    8.81 $\pm$    0.04 &    8.84 $\pm$    0.02 & &    6.96 $\pm$    0.02 &    6.87 $\pm$    0.02 &    6.92 $\pm$    0.03 &    6.92 $\pm$    0.01 & &    4.32 $\pm$    0.01 &    4.07 $\pm$    0.01 &    4.23 $\pm$    0.03 &    4.21 $\pm$    0.01 \\ 
  826 &   O9.5V &   10.03 $\pm$    0.03 &    9.85 $\pm$    0.05 &    9.79 $\pm$    0.06 &    9.96 $\pm$    0.02 & &    8.58 $\pm$    0.03 &    8.33 $\pm$    0.05 &    8.32 $\pm$    0.06 &    8.49 $\pm$    0.02 & &    6.66 $\pm$    0.02 &    6.53 $\pm$    0.04 &    6.53 $\pm$    0.06 &    6.63 $\pm$    0.02 & &    4.13 $\pm$    0.01 &    3.85 $\pm$    0.03 &    3.83 $\pm$    0.05 &    4.08 $\pm$    0.01 \\ 
  843\tablenotemark{b} &   O4.5V &    9.76 $\pm$    0.03 &    9.58 $\pm$    0.09 &    9.63 $\pm$    0.10 &    9.73 $\pm$    0.03 & &    8.32 $\pm$    0.03 &    8.10 $\pm$    0.09 &    8.14 $\pm$    0.10 &    8.29 $\pm$    0.03 & &    6.43 $\pm$    0.02 &    6.36 $\pm$    0.08 &    6.37 $\pm$    0.09 &    6.42 $\pm$    0.02 & &    3.98 $\pm$    0.01 &    3.77 $\pm$    0.07 &    3.83 $\pm$    0.08 &    3.96 $\pm$    0.01 \\ 
  857 &   O4.5V &    9.58 $\pm$    0.03 &    9.41 $\pm$    0.06 &    9.37 $\pm$    0.07 &    9.53 $\pm$    0.02 & &    8.14 $\pm$    0.03 &    7.91 $\pm$    0.06 &    7.87 $\pm$    0.07 &    8.08 $\pm$    0.02 & &    6.27 $\pm$    0.02 &    6.15 $\pm$    0.05 &    6.17 $\pm$    0.07 &    6.25 $\pm$    0.02 & &    3.85 $\pm$    0.01 &    3.58 $\pm$    0.04 &    3.58 $\pm$    0.06 &    3.82 $\pm$    0.01 \\ 
  879 &   O9.5V &   10.62 $\pm$    0.03 &   10.44 $\pm$    0.05 &   10.50 $\pm$    0.06 &   10.57 $\pm$    0.02 & &    9.08 $\pm$    0.03 &    8.83 $\pm$    0.05 &    8.89 $\pm$    0.06 &    9.00 $\pm$    0.02 & &    7.04 $\pm$    0.02 &    6.93 $\pm$    0.04 &    6.98 $\pm$    0.06 &    7.01 $\pm$    0.02 & &    4.36 $\pm$    0.01 &    4.10 $\pm$    0.03 &    4.25 $\pm$    0.05 &    4.33 $\pm$    0.01 \\ 
  896 &   O8.5V &   10.18 $\pm$    0.03 &   10.02 $\pm$    0.04 &    9.98 $\pm$    0.06 &   10.11 $\pm$    0.02 & &    8.69 $\pm$    0.03 &    8.46 $\pm$    0.04 &    8.44 $\pm$    0.06 &    8.59 $\pm$    0.02 & &    6.72 $\pm$    0.02 &    6.63 $\pm$    0.04 &    6.63 $\pm$    0.05 &    6.69 $\pm$    0.02 & &    4.16 $\pm$    0.01 &    3.91 $\pm$    0.03 &    3.89 $\pm$    0.05 &    4.10 $\pm$    0.01 \\ 
  903 &   O8.5V &   10.30 $\pm$    0.03 &   10.15 $\pm$    0.05 &   10.16 $\pm$    0.06 &   10.25 $\pm$    0.02 & &    8.80 $\pm$    0.03 &    8.58 $\pm$    0.05 &    8.61 $\pm$    0.06 &    8.73 $\pm$    0.02 & &    6.83 $\pm$    0.02 &    6.74 $\pm$    0.04 &    6.75 $\pm$    0.06 &    6.81 $\pm$    0.02 & &    4.24 $\pm$    0.01 &    3.99 $\pm$    0.03 &    4.06 $\pm$    0.05 &    4.20 $\pm$    0.01 \\ 
  906 &   O6.5V &   10.15 $\pm$    0.03 &    9.98 $\pm$    0.09 &    9.94 $\pm$    0.10 &   10.12 $\pm$    0.03 & &    8.65 $\pm$    0.03 &    8.42 $\pm$    0.09 &    8.39 $\pm$    0.10 &    8.62 $\pm$    0.03 & &    6.69 $\pm$    0.02 &    6.59 $\pm$    0.08 &    6.57 $\pm$    0.10 &    6.68 $\pm$    0.02 & &    4.13 $\pm$    0.01 &    3.88 $\pm$    0.07 &    3.82 $\pm$    0.09 &    4.12 $\pm$    0.01 \\ 
  913 &   O3.5V &    9.87 $\pm$    0.03 &    9.71 $\pm$    0.09 &    9.75 $\pm$    0.10 &    9.85 $\pm$    0.03 & &    8.43 $\pm$    0.03 &    8.21 $\pm$    0.09 &    8.26 $\pm$    0.10 &    8.40 $\pm$    0.03 & &    6.53 $\pm$    0.02 &    6.45 $\pm$    0.08 &    6.47 $\pm$    0.09 &    6.52 $\pm$    0.02 & &    4.04 $\pm$    0.01 &    3.82 $\pm$    0.07 &    3.90 $\pm$    0.08 &    4.03 $\pm$    0.01 \\ 
  924 &   O8.0V &   10.03 $\pm$    0.03 &    9.84 $\pm$    0.05 &    9.85 $\pm$    0.06 &    9.97 $\pm$    0.02 & &    8.54 $\pm$    0.03 &    8.31 $\pm$    0.05 &    8.32 $\pm$    0.06 &    8.46 $\pm$    0.02 & &    6.59 $\pm$    0.02 &    6.49 $\pm$    0.04 &    6.50 $\pm$    0.06 &    6.57 $\pm$    0.02 & &    4.06 $\pm$    0.01 &    3.82 $\pm$    0.03 &    3.82 $\pm$    0.05 &    4.02 $\pm$    0.01 \\ 
 1004\tablenotemark{a} &   O9.5V &    9.13 $\pm$    0.03 &    9.00 $\pm$    0.03 &    8.97 $\pm$    0.04 &    9.05 $\pm$    0.02 & &    7.83 $\pm$    0.03 &    7.63 $\pm$    0.03 &    7.59 $\pm$    0.04 &    7.72 $\pm$    0.02 & &    6.10 $\pm$    0.02 &    6.00 $\pm$    0.03 &    6.00 $\pm$    0.04 &    6.05 $\pm$    0.01 & &    3.80 $\pm$    0.01 &    3.56 $\pm$    0.02 &    3.63 $\pm$    0.04 &    3.72 $\pm$    0.01 \\ 
 1039 &   O4.5V &   10.06 $\pm$    0.03 &    9.87 $\pm$    0.08 &    9.87 $\pm$    0.09 &   10.02 $\pm$    0.03 & &    8.53 $\pm$    0.03 &    8.29 $\pm$    0.08 &    8.32 $\pm$    0.09 &    8.49 $\pm$    0.03 & &    6.55 $\pm$    0.02 &    6.45 $\pm$    0.07 &    6.44 $\pm$    0.08 &    6.54 $\pm$    0.02 & &    4.02 $\pm$    0.01 &    3.75 $\pm$    0.06 &    3.76 $\pm$    0.08 &    4.00 $\pm$    0.01 \\ 
\cutinhead{}
Averages &         &    9.92 $\pm$    0.00 &    9.87 $\pm$    0.01 &    9.85 $\pm$    0.01 &    9.90 $\pm$    0.58 & &    8.50 $\pm$    0.00 &    8.37 $\pm$    0.01 &    8.35 $\pm$    0.01 &    8.46 $\pm$    0.49 & &    6.63 $\pm$    0.00 &    6.59 $\pm$    0.01 &    6.57 $\pm$    0.01 &    6.62 $\pm$    0.38 & &    4.12 $\pm$    0.00 &    3.94 $\pm$    0.01 &    3.97 $\pm$    0.01 &    6.62 $\pm$    0.24 \\ 
\enddata
\tablenotetext{a}{The IR photometry of these stars comes from the \emph{2MASS PSC} instead of \citet{asc}.}
\tablenotetext{b}{This star lacks $J$ photometry.}
\end{deluxetable}  
\end{center}

\clearpage
\begin{center}
\begin{deluxetable}{cccccccccc}
\tablecolumns{10}
\tabletypesize{\scriptsize}
\tablecaption{Distance Calculation.\label{dis-table}}
\tablewidth{0pt}
\tablehead{
\colhead{ID} & \colhead{MSP91} & \colhead{Spectral} & \colhead{F555W} & \colhead{(F439W - F555W)}  & \colhead{$M_{\mbox{F555W}}$} &  \colhead{(F439W - F555W)$_{0}$}  & \colhead{$A_{\mbox{F555W}}$} & \colhead{$DM$}  &  \colhead{d}  \\
\colhead{} & \colhead{} & \colhead{Type} & \colhead{(mag)} & \colhead{(mag)} & \colhead{(mag)} & \colhead{(mag)} & \colhead{(mag)} & \colhead{(mag)} & \colhead{(kpc)}
}
\startdata
  137	&	165		&	O4 V		&	15.610	$\pm$	0.006	&	1.776	$\pm$	0.013	&	-5.171	$\pm$	0.445	&	-0.278	$\pm$	0.003	&	7.595	$\pm$	0.076	&	13.186	$\pm$	0.450	&	4.338	$\pm$	0.900	\\
  178	&	182		&	O4 V-III((f))*	&	14.514	$\pm$	0.004	&	1.451	$\pm$	0.006	&	-3.727	$\pm$	0.149	&	-0.257	$\pm$	0.004	&	6.453	$\pm$	0.054	&	11.788	$\pm$	0.158	&	2.279	$\pm$	0.166	\\
  395	&	223a		&	O7.5 V*		&	16.015	$\pm$	0.062	&	1.666	$\pm$	0.098	&	-3.878	$\pm$	0.153	&	-0.260	$\pm$	0.003	&	7.108	$\pm$	0.067	&	12.785	$\pm$	0.178	&	3.605	$\pm$	0.295	\\
  505	&	196a		&	O8.5 V		&	16.100	$\pm$	0.005	&	1.489	$\pm$	0.011	&	-3.598	$\pm$	0.294	&	-0.254	$\pm$	0.007	&	6.405	$\pm$	0.046	&	13.293	$\pm$	0.297	&	4.556	$\pm$	0.622	\\
  528	&	229a		&	O8 V		&	15.863	$\pm$	0.005	&	1.615	$\pm$	0.013	&	-3.746	$\pm$	0.299	&	-0.257	$\pm$	0.007	&	7.062	$\pm$	0.058	&	12.547	$\pm$	0.304	&	3.231	$\pm$	0.452	\\
  548	&	151		&	O4 V		&	14.546	$\pm$	0.002	&	1.485	$\pm$	0.006	&	-5.171	$\pm$	0.222	&	-0.278	$\pm$	0.002	&	6.525	$\pm$	0.050	&	13.193	$\pm$	0.228	&	4.350	$\pm$	0.456	\\
  549	&	44a		&	B1 V*		&	15.587	$\pm$	0.005	&	1.358	$\pm$	0.009	&	-2.254	$\pm$	0.184	&	-0.213	$\pm$	0.006	&	6.165	$\pm$	0.061	&	11.676	$\pm$	0.193	&	2.164	$\pm$	0.192	\\
  584	&	157b	&	O8 V		&	15.466	$\pm$	0.004	&	1.410	$\pm$	0.009	&	-3.746	$\pm$	0.105	&	-0.257	$\pm$	0.002	&	6.266	$\pm$	0.056	&	12.945	$\pm$	0.119	&	3.882	$\pm$	0.212	\\
  597	&	157a		&	O6.5 V*		&	14.866	$\pm$	0.003	&	1.454	$\pm$	0.008	&	-4.368	$\pm$	0.173	&	-0.269	$\pm$	0.003	&	6.454	$\pm$	0.059	&	12.780	$\pm$	0.183	&	3.597	$\pm$	0.303	\\
  620	&	96a		&	B1 V*		&	16.111	$\pm$	0.006	&	1.336	$\pm$	0.013	&	-2.254	$\pm$	0.184	&	-0.213	$\pm$	0.006	&	5.803	$\pm$	0.056	&	12.561	$\pm$	0.192	&	3.253	$\pm$	0.287	\\
  640	&	233		&	O9.5 V		&	16.239	$\pm$	0.006	&	1.482	$\pm$	0.014	&	-3.306	$\pm$	0.117	&	-0.246	$\pm$	0.003	&	6.433	$\pm$	0.056	&	13.112	$\pm$	0.129	&	4.192	$\pm$	0.249	\\
  664	&	188a		&	O3 V		&	13.487	$\pm$	0.017	&	1.513	$\pm$	0.017	&	-6.499	$\pm$	1.259	&	-0.283	$\pm$	0.002	&	6.561	$\pm$	0.381	&	13.425	$\pm$	1.314	&	4.841	$\pm$	2.929	\\
  704	&	175a		&	O4 V		&	14.084	$\pm$	0.002	&	1.411	$\pm$	0.005	&	-6.499	$\pm$	0.629	&	-0.283	$\pm$	0.001	&	6.323	$\pm$	0.052	&	14.259	$\pm$	0.631	&	7.110	$\pm$	2.065	\\
  714	&	\nodata	&	O3 V		&	15.042	$\pm$	0.003	&	1.403	$\pm$	0.007	&	-5.610	$\pm$	0.252	&	-0.280	$\pm$	0.001	&	6.126	$\pm$	0.058	&	14.526	$\pm$	0.258	&	8.040	$\pm$	0.957	\\
  722	&	263a		&	O6 V*		&	15.053	$\pm$	0.030	&	1.810	$\pm$	0.079	&	-4.274	$\pm$	0.084	&	-0.268	$\pm$	0.001	&	7.291	$\pm$	0.068	&	12.036	$\pm$	0.113	&	2.554	$\pm$	0.132	\\
  738	&	168		&	O5.5 V		&	14.921	$\pm$	0.003	&	1.378	$\pm$	0.007	&	-4.540	$\pm$	0.183	&	-0.272	$\pm$	0.002	&	6.076	$\pm$	0.051	&	13.385	$\pm$	0.189	&	4.753	$\pm$	0.414	\\
  769	&	219		&	O9.5 V		&	16.574	$\pm$	0.008	&	1.656	$\pm$	0.022	&	-3.160	$\pm$	0.118	&	-0.241	$\pm$	0.004	&	6.697	$\pm$	0.060	&	13.037	$\pm$	0.132	&	4.049	$\pm$	0.247	\\
  771	&	167b	&	O8 V		&	15.524	$\pm$	0.005	&	1.510	$\pm$	0.009	&	-3.746	$\pm$	0.134	&	-0.257	$\pm$	0.003	&	6.688	$\pm$	0.062	&	12.581	$\pm$	0.148	&	3.283	$\pm$	0.224	\\
  804	&	167a		&	O6 III*		&	14.453	$\pm$	0.003	&	1.664	$\pm$	0.007	&	-4.249	$\pm$	0.084	&	-0.267	$\pm$	0.001	&	6.917	$\pm$	0.047	&	11.785	$\pm$	0.096	&	2.275	$\pm$	0.100	\\
  826	&	183d	&	O9.5 V*		&	16.609	$\pm$	0.007	&	1.601	$\pm$	0.018	&	-3.289	$\pm$	0.146	&	-0.245	$\pm$	0.004	&	6.572	$\pm$	0.076	&	13.327	$\pm$	0.164	&	4.627	$\pm$	0.350	\\
  843	&	203a		&	O4-5 V*		&	14.736	$\pm$	0.003	&	1.498	$\pm$	0.007	&	-4.927	$\pm$	0.206	&	-0.276	$\pm$	0.002	&	6.385	$\pm$	0.041	&	13.278	$\pm$	0.210	&	4.526	$\pm$	0.438	\\
  857	&	444a		&	O4.5 V		&	13.893	$\pm$	0.003	&	1.475	$\pm$	0.005	&	-4.956	$\pm$	0.167	&	-0.276	$\pm$	0.001	&	6.196	$\pm$	0.061	&	12.653	$\pm$	0.177	&	3.394	$\pm$	0.277	\\
  869	&	183a		&	O3 V((f))*	&	13.878	$\pm$	0.017	&	1.618	$\pm$	0.018	&	-5.610	$\pm$	0.252	&	-0.280	$\pm$	0.001	&	6.561	$\pm$	0.381	&	12.927	$\pm$	0.457	&	3.850	$\pm$	0.810	\\
  879	&	235		&	O9.5 V		&	16.637	$\pm$	0.056	&	1.704	$\pm$	0.111	&	-3.452	$\pm$	0.146	&	-0.250	$\pm$	0.004	&	6.981	$\pm$	0.053	&	13.108	$\pm$	0.164	&	4.184	$\pm$	0.317	\\
  889	&	18a		&	O5 V-III*		&	13.155	$\pm$	0.004	&	1.359	$\pm$	0.052	&	-4.583	$\pm$	0.148	&	-0.272	$\pm$	0.002	&	6.561	$\pm$	0.381	&	11.177	$\pm$	0.408	&	1.720	$\pm$	0.323	\\
  896	&	183b	&	O8.5 V		&	15.790	$\pm$	0.005	&	1.598	$\pm$	0.011	&	-3.598	$\pm$	0.132	&	-0.254	$\pm$	0.003	&	6.660	$\pm$	0.054	&	12.727	$\pm$	0.143	&	3.512	$\pm$	0.230	\\
  903	&	183c		&	O8.5 V		&	16.243	$\pm$	0.006	&	1.649	$\pm$	0.014	&	-3.598	$\pm$	0.147	&	-0.254	$\pm$	0.004	&	6.770	$\pm$	0.050	&	13.071	$\pm$	0.155	&	4.113	$\pm$	0.293	\\
  906	&	\nodata	&	O6-7 V*		&	16.118	$\pm$	0.005	&	1.579	$\pm$	0.013	&	-4.194	$\pm$	0.247	&	-0.266	$\pm$	0.004	&	6.613	$\pm$	0.063	&	13.699	$\pm$	0.255	&	5.493	$\pm$	0.645	\\
  913	&	199a		&	O3-4 V*		&	14.554	$\pm$	0.002	&	1.528	$\pm$	0.005	&	-5.378	$\pm$	0.236	&	-0.279	$\pm$	0.001	&	6.481	$\pm$	0.043	&	13.451	$\pm$	0.240	&	4.900	$\pm$	0.541	\\
  924	&	\nodata	&	O8 V		&	15.963	$\pm$	0.005	&	1.593	$\pm$	0.012	&	-3.727	$\pm$	0.149	&	-0.257	$\pm$	0.004	&	6.526	$\pm$	0.054	&	13.164	$\pm$	0.158	&	4.294	$\pm$	0.313	\\
1004	&	32		&	O9.5 V		&	15.349	$\pm$	0.004	&	1.380	$\pm$	0.008	&	-3.452	$\pm$	0.087	&	-0.250	$\pm$	0.002	&	6.031	$\pm$	0.057	&	12.770	$\pm$	0.104	&	3.580	$\pm$	0.172	\\
1039	&	171		&	O4-5 V*		&	14.528	$\pm$	0.030	&	1.606	$\pm$	0.079	&	-4.935	$\pm$	0.207	&	-0.276	$\pm$	0.002	&	6.479	$\pm$	0.065	&	12.984	$\pm$	0.218	&	3.952	$\pm$	0.398	\\
\enddata
\tablenotetext{*}{These spectral classification were determined by R07 and R11.}
\end{deluxetable}
\end{center}
\end{landscape}


\clearpage

\clearpage

\clearpage

\clearpage

\clearpage

\clearpage

\clearpage

\clearpage

\clearpage

\clearpage

\clearpage

\begin{figure}
\epsscale{0.6}
\plotone{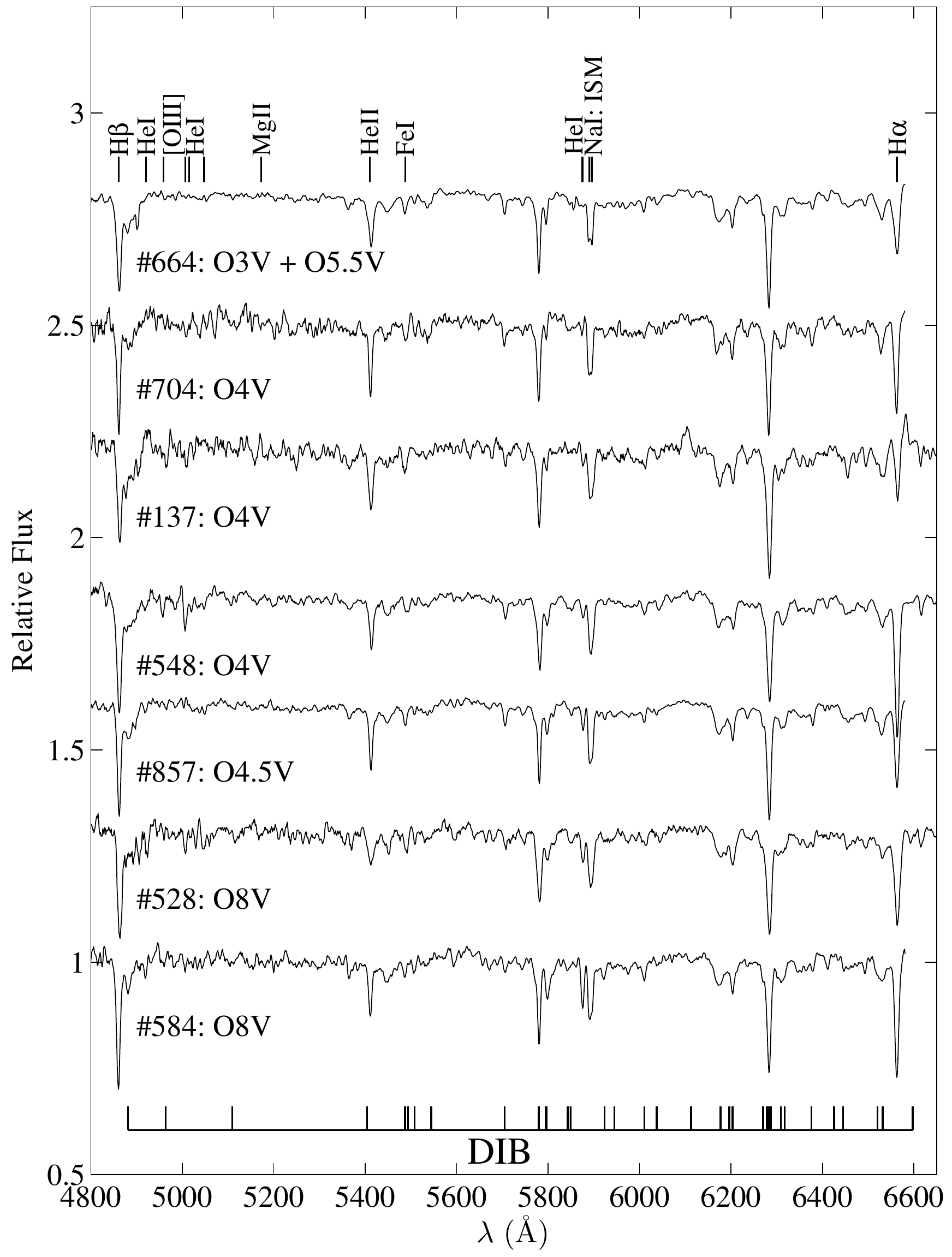}
\caption{Normalized spectra obtained at both the \emph{Magellan} and \emph{SOAR} telescopes.  Each spectrum is offset by a constant value for easier visualization. Under each spectra is the stars identification number assigned in this work along with the spectral type.  Spectra appear in order of descending temperature/spectral type from top to bottom.  The main stellar absorption lines are labeled at the top of the figure (along with the interstellar \ion{Na}{1} D lines), and the strongest diffuse interstellar band (DIB) lines are identified at the bottom the figure.\label{spec1}}
\end{figure}


\clearpage

\begin{figure}
\epsscale{0.6}
\plotone{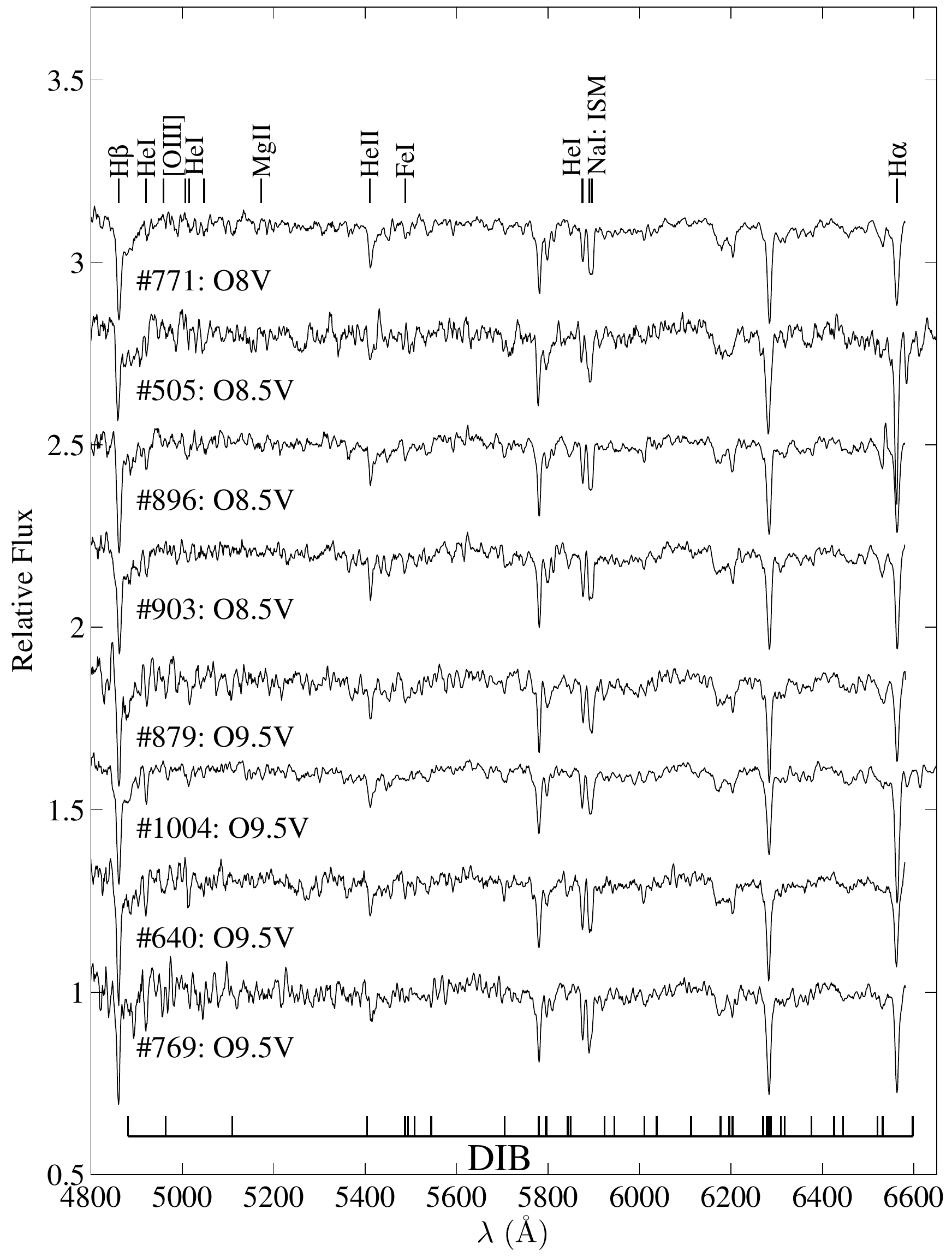}
\caption{Continuation of Figure~\ref{spec1}.\label{spec2}}
\end{figure}


\clearpage

\begin{figure}
\epsscale{0.6}
\plotone{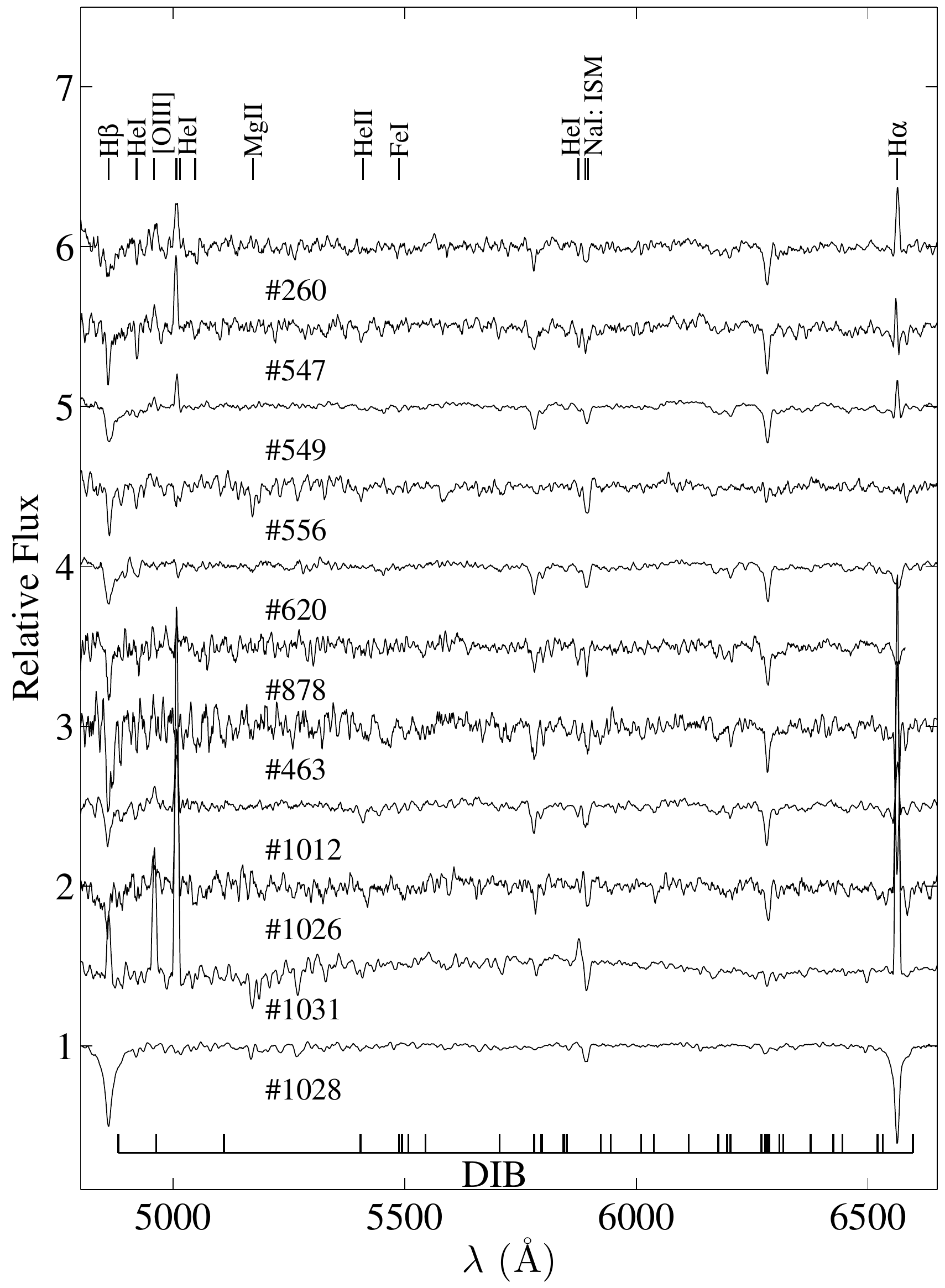}
\caption{Similar to Figure~\ref{spec1} but for the field stars or stars with indeterminate spectral type.  The main stellar absorption lines are labeled at the top of the figure (along with the interstellar \ion{Na}{1} D lines), and the strongest diffuse interstellar band (DIB) lines are identified at the bottom the figure.\label{spec3}}
\end{figure}


\clearpage

\begin{figure}
\epsscale{0.9}
\plotone{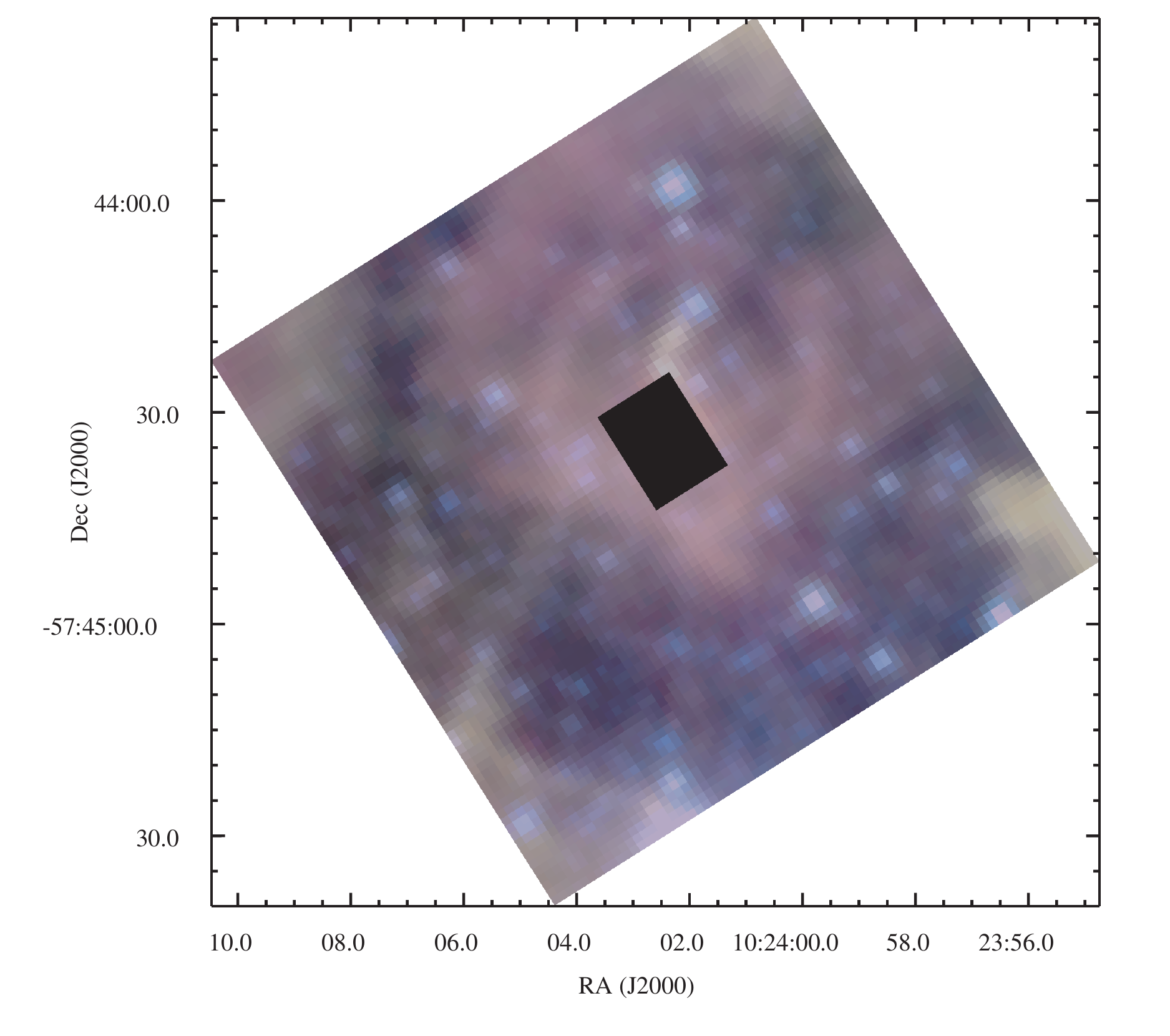}
\caption{\emph{Spitzer} three-color image of the diffuse mid-IR ring surrounding the stellar subgroup centered on \#889 (MSP~18). The image displays 4.5 $\mu m$ in blue, 5.8 $\mu m$ in green, and 8.0 $\mu m$ in red.  MSP~18 has been masked out for easier visualization of the ring.\label{ring}}
\end{figure}


\clearpage

\begin{figure}
\epsscale{0.5}
\plotone{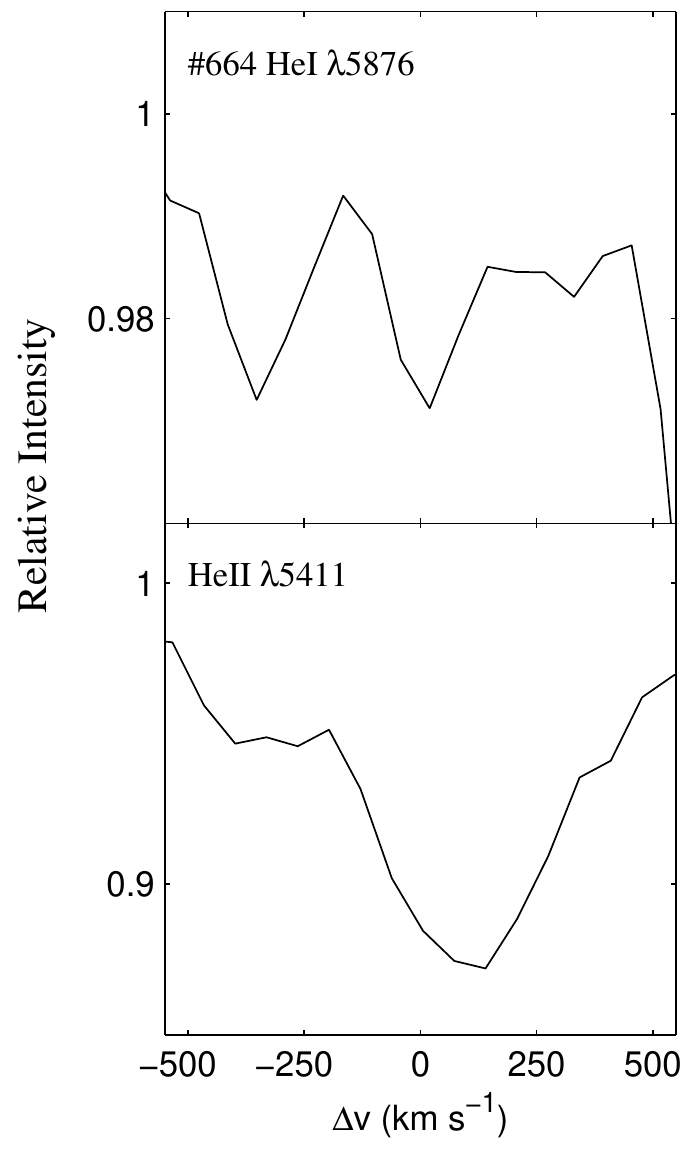}
\caption{Velocity plot of \ion{He}{1} $\lambda$5876 (top panel), and \ion{He}{2} $\lambda$5411 (bottom panel) lines of the probable binary system \#664 showing that each line appears to have two components separated by 360 -- 490 \kms.\label{664vel}}
\end{figure}


\clearpage

\begin{figure}
\epsscale{0.8}
\plotone{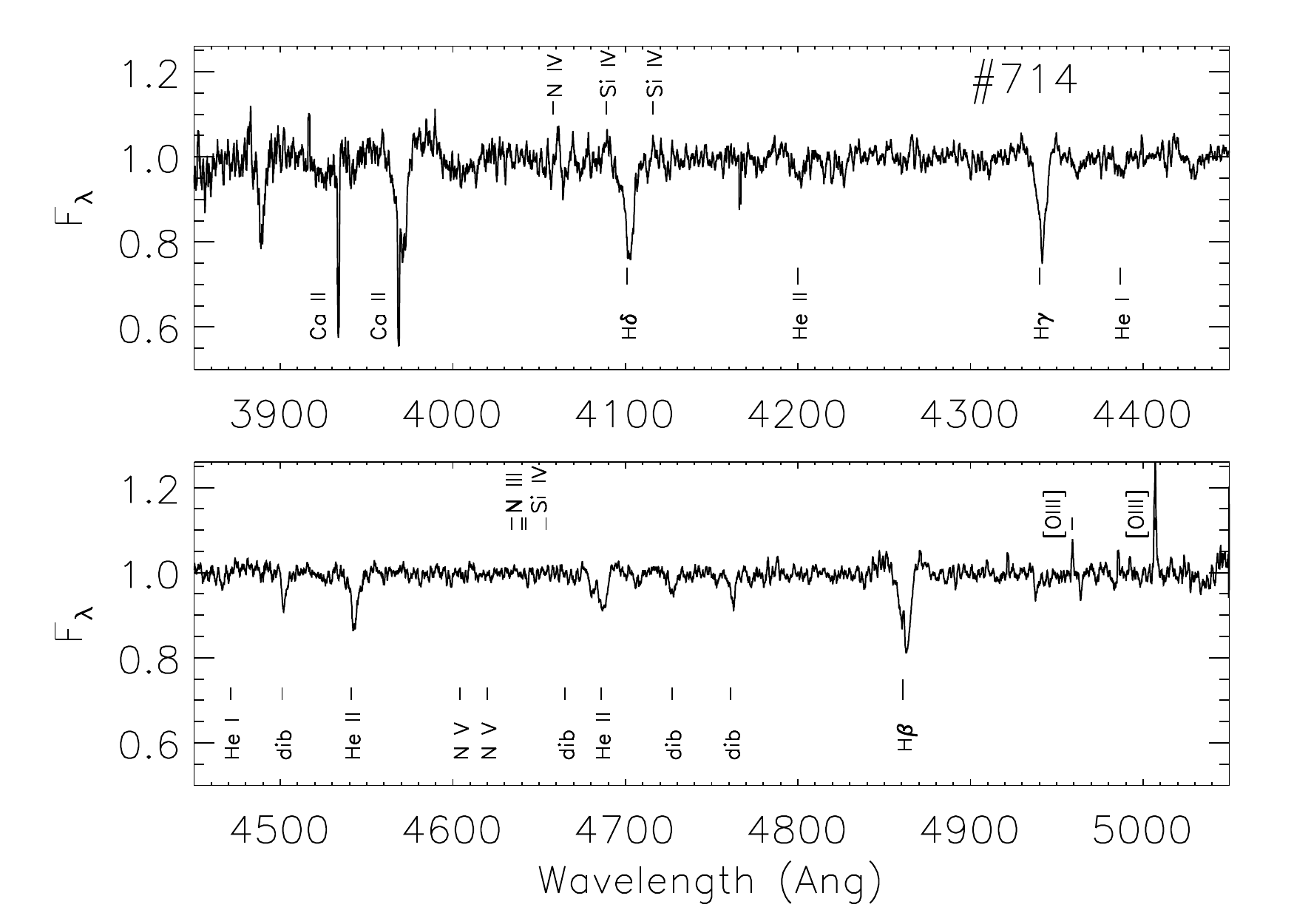}
\caption{MIKE spectrum of \#714 showing the very weak \ion{He}{1}, and strong \ion{He}{2} indicative of a hot star.  Note the residual nebular [\ion{O}{3}] emission.\label{mike714}}
\end{figure}


\clearpage

\begin{figure}
\epsscale{0.8}
\plotone{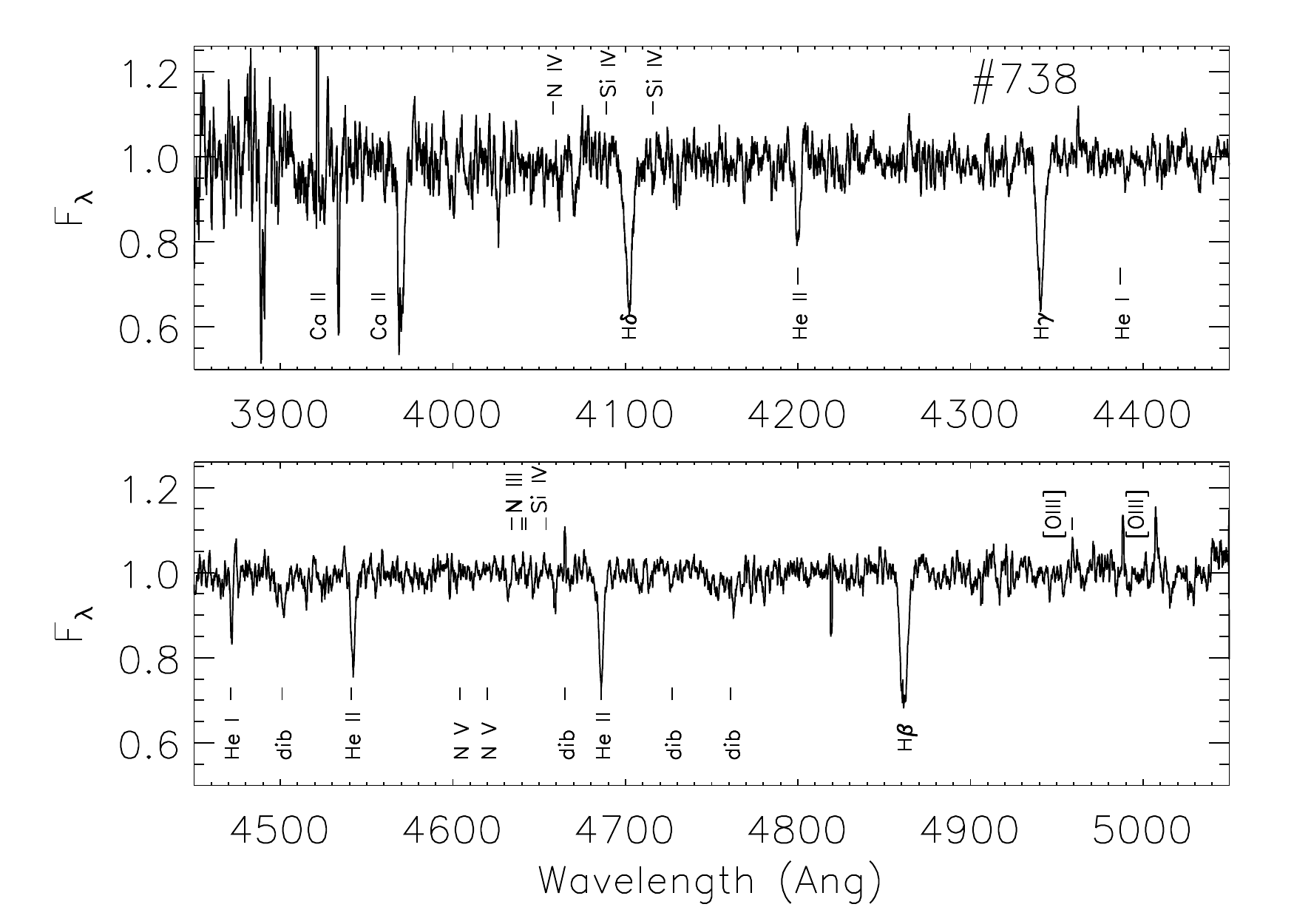}
\caption{MIKE spectrum of \#738, classified as O5.  Note the residual nebular [\ion{O}{3}] emission.\label{mike738}}
\end{figure}


\clearpage
\begin{figure}
\epsscale{0.8}
\plotone{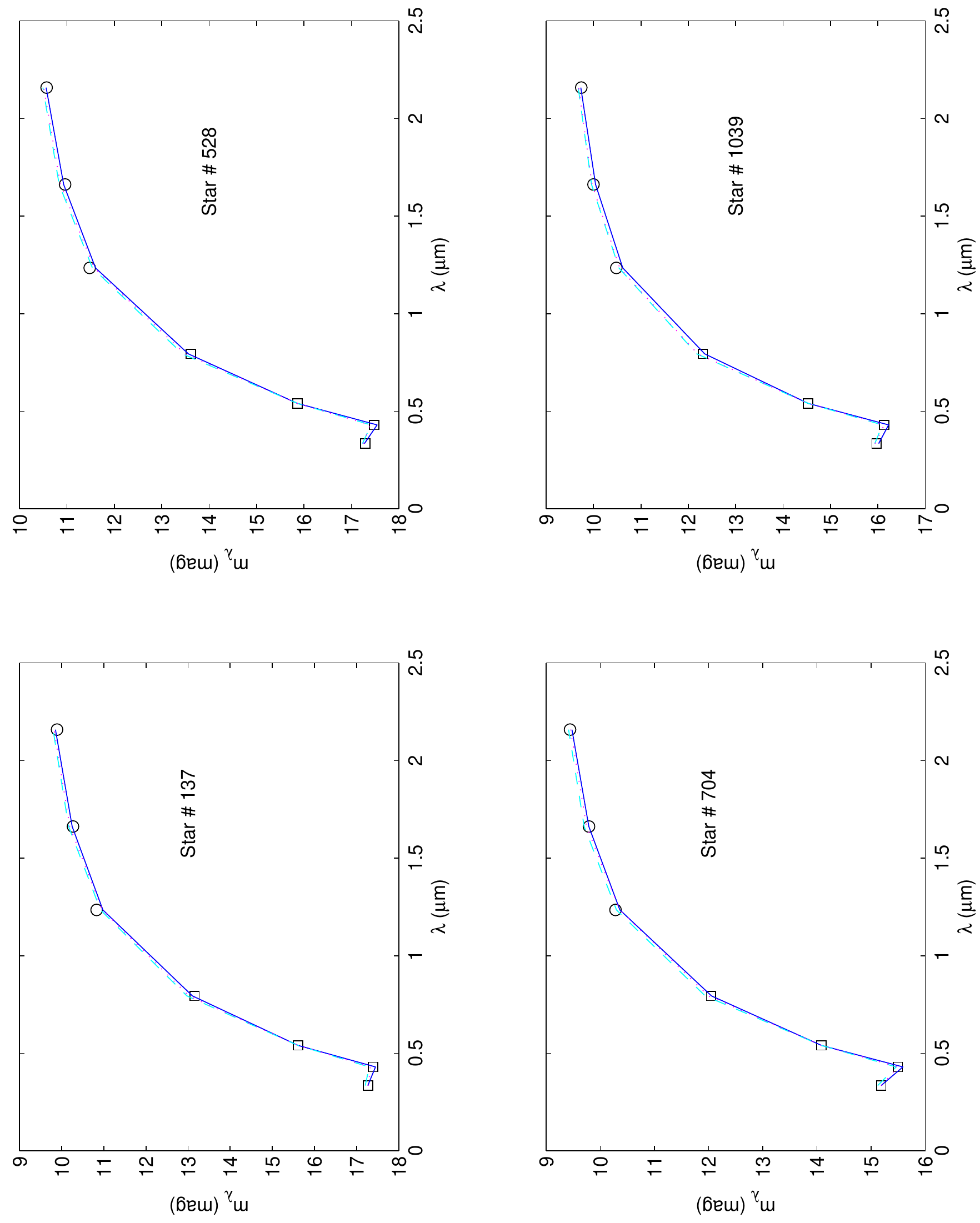}
\caption{Example reddened magnitude SEDs for stars \#137, 528, 704, and 1039.  The solid lines are the reddened magnitudes, the squares are the \emph{HST} flight photometry and the circles are the \citet{asc} IR photometry.  The solid line was reddened with the CCM89 reddening law, the dashed line used F04, and the dotted line used  FM07 (these are colored in blue, cyan, and magenta, respectively, in the electronic edition).\label{sed-plot}}
\end{figure}


\clearpage

\begin{figure}
\epsscale{0.9}
\plotone{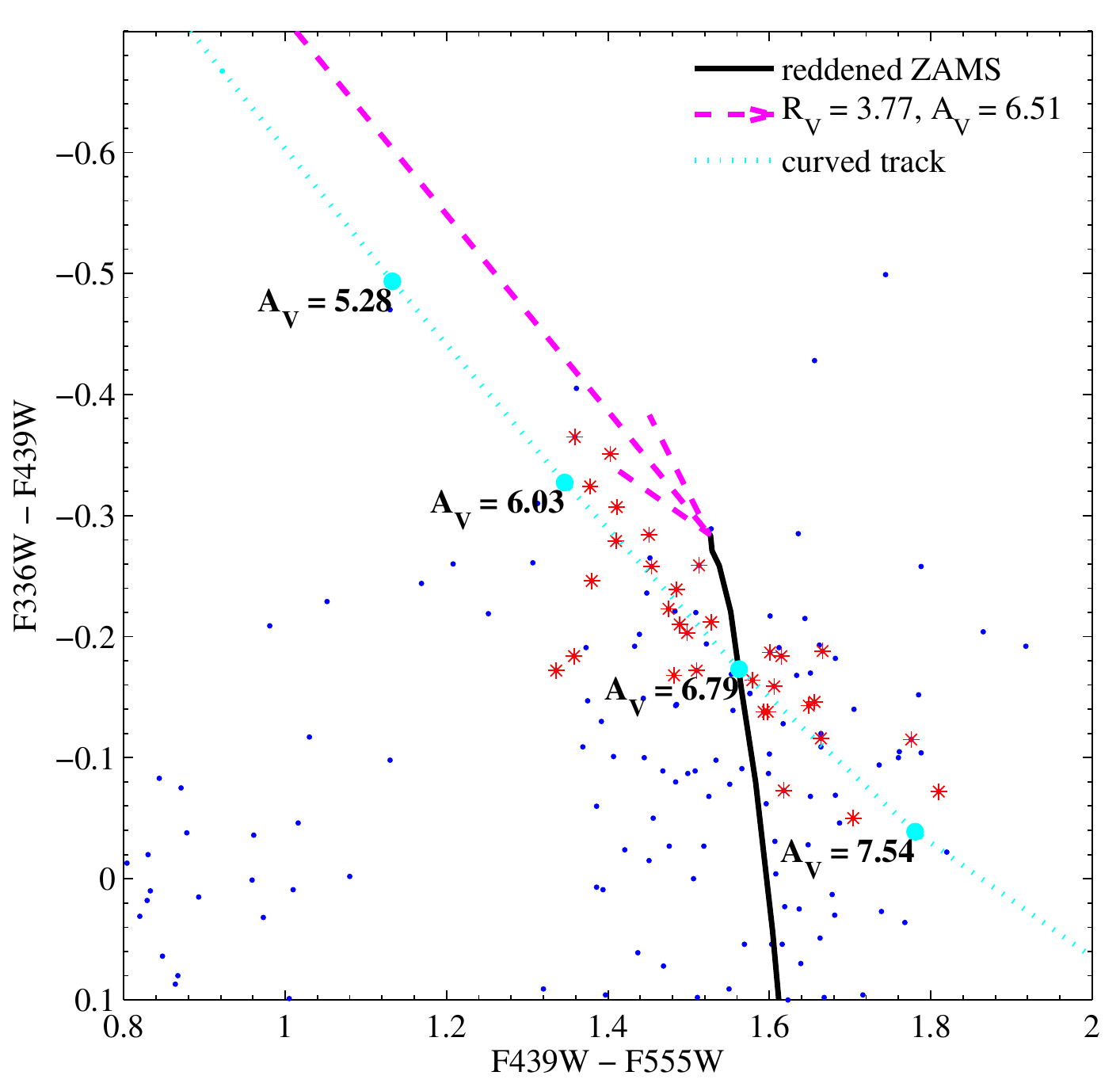}
\caption{CCD showing a section of the reddened ZAMS fit to the cluster Wd2.  The asterisks (red in the electronic edition) denote the stars having spectral types discussed in this work, R07, and R11.  The ZAMS was reddenened by $R_{V} = 3.77$, $A_{V} = 6.51$ with corresponding color excess of $E$(F336W -- F439W) = 1.47, and $E$(F439W -- F555W) = 1.81.  Dashed and dotted lines represent a section of the linear and curved reddening tracks, respectively, for early-type stars (colored in magenta and cyan in the color edition).  Numerical labels mark extinctions, $A_V$, along the curved reddening track.\label{ccd-iso-fit}}
\end{figure}


\clearpage

\begin{figure}
\epsscale{0.8}
\plotone{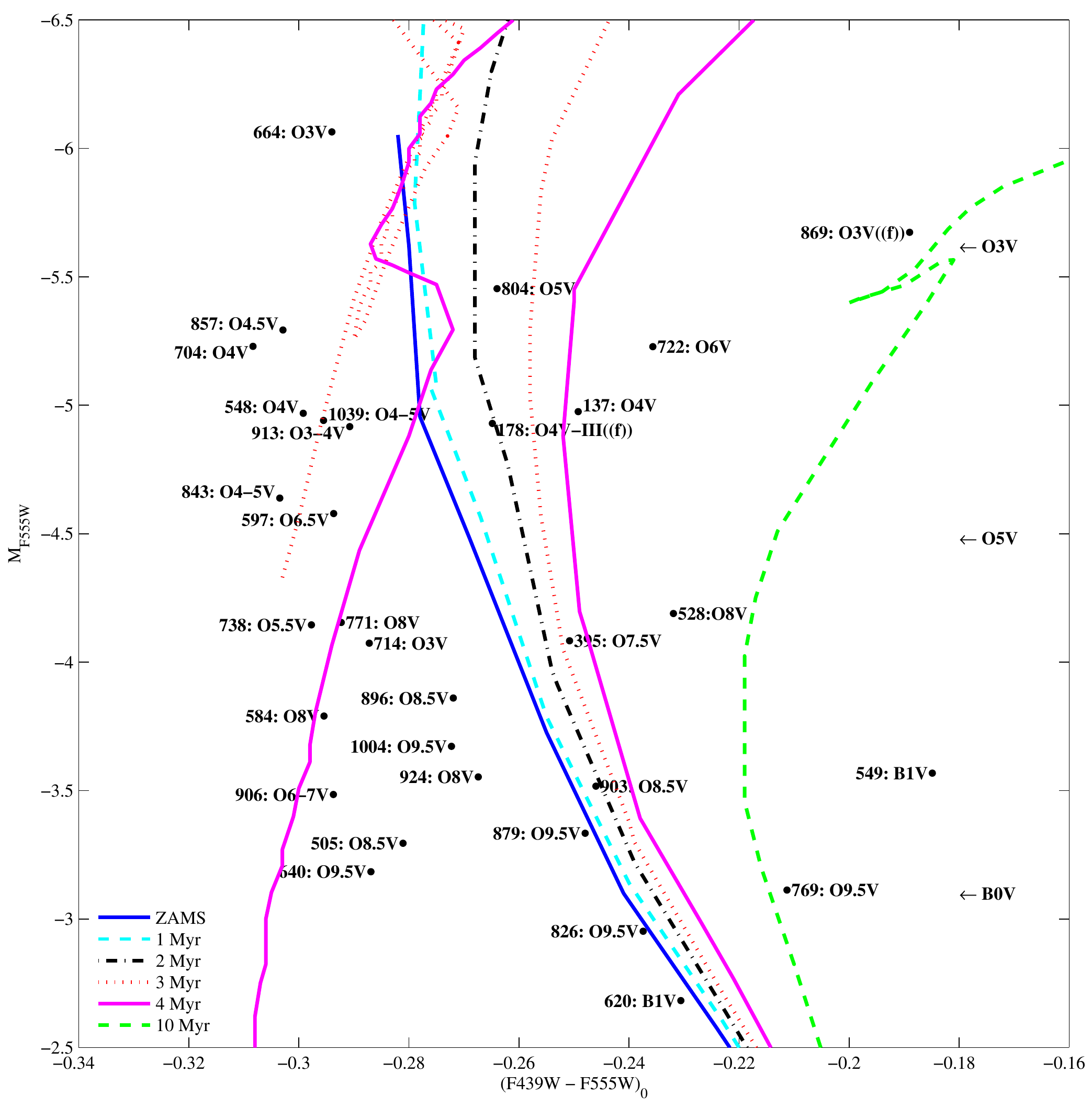}
\caption{Dereddened CMD of Wd2 at the adopted distance modulus of 12.98 mag.  Labels show the IDs and respective spectral types.  Labels at the right of the figure indicate the spectral types according to the ZAMS isochrone.  The curves represent Padova isochrones, with ages as indicated in the legend.\label{unred-cmd}}
\end{figure}


\clearpage
\end{document}